\documentclass[11pt]{article}
\usepackage{jheppub}
\usepackage[usenames,dvipsnames,table]{xcolor}
\usepackage{graphicx,amsmath,amssymb,amsthm}
\usepackage{slashed}
\usepackage{cleveref}
\usepackage{cancel}
\usepackage{relsize}
\usepackage{slashed}

\usepackage{tikz}
\usepackage{pgfplots}
\usetikzlibrary{backgrounds,arrows,shapes,shapes.geometric,shapes.misc,fillbetween,backgrounds,decorations.markings,intersections,patterns,hobby}
\tikzstyle{tikzfig}=[baseline=-0.25em,scale=0.5]
\tikzstyle{none}=[inner sep=0mm]

\pgfdeclarelayer{edgelayer}
\pgfdeclarelayer{nodelayer}
\pgfsetlayers{background,edgelayer,nodelayer,main}

\newcommand{\mcal}[1]{\mathcal{#1}}

\title{\centering Hyperbolic Geometry and  Closed  Bosonic String Field Theory II\\ {\large The Rules for Evaluating the Quantum BV Master Action}}

 \author{Seyed Faroogh Moosavian, Roji Pius}
\affiliation{Perimeter Institute for Theoretical Physics, Waterloo, ON N2L 2Y5, Canada}
\emailAdd{sfmoosavian@perimeterinstitute.ca \qquad rpius@perimeterinstitute.ca}

\abstract{The quantum Batalian-Vilkovisky master action for closed string field theory consists of kinetic term and infinite number of interaction terms. The interaction strengths (coupling constants) are given by integrating  the off-shell string measure over the distinct string diagrams describing the elementary interactions of the closed strings.  In the first paper of this series, it was shown that the string diagrams describing the elementary interactions  can be characterized using  the Riemann surfaces endowed with the hyperbolic metric with constant curvature $-1$. In this paper, we construct the off-shell bosonic-string measure as a function of the Fenchel-Nielsen coordinates  of  the Teichm\"uller space  of hyperbolic Riemann surfaces. We also describe an explicit procedure for integrating the off-shell string measure over the region inside the moduli space corresponding to the elementary interactions of the closed strings.
\\
~
\\
{\it  We dedicate this paper to the memory of Maryam Mirzakhani who tragically passed away recently, and whose seminal ideas about the space of hyperbolic  Riemann surfaces form some of the basic tools that we use in this work.}}

\begin{document} 
\maketitle
\flushbottom

\section{Introduction}
\label{sec:intro}

The basic object in string theory is a one-dimensional extended object, the string. The harmonics of the vibrating string correspond to the elementary particles with  different masses and quantum numbers. String can support infinitely many harmonics, and hence string theory contains infinite number of elementary particles. Therefore, we can consider string theory as a  framework for describing the interactions of infinitely many  elementary particles. String field theory describe the dynamics of this system of infinite number of elementary particles  using the language of quantum field theory \cite{Witten:1985cc,Thorn:1988hm,Zwiebach:1992ie}. Compared to the conventional formulation of string perturbation theory \cite{Friedan:1985ge}, string field theory has the advantage that the latter provide us with the standard tools in quantum field theory for computing the S-matrix elements that are free from infrared divergences  \cite{Pius:2013sca,Pius:2014iaa,Sen:2014pia,Pius:2014gza,Sen:2014dqa,Sen:2015uoa,Sen:2015uaa,Sen:2015hha,deLacroix:2017lif}. Furthermore, the S-matrix  computed using string field  is unitary \cite{Pius:2016jsl,Pius:2018crk,Sen:2016ubf,Sen:2016bwe,Sen:2016uzq}. Since string field theory is based on a Lagrangian, it also has the potential to open the door towards the non-perturbative regime of string theory \cite{Schnabl:2005gv}, even though no one has succeeded in studying the non-perturbative behaviour of  closed strings using closed string field theory yet \cite{Sen:2016qap,Yang:2005rx}. Moreover, string field theory can be used for the first principle construction for the effective actions describing the low energy dynamics of  strings \cite{Sen:2014dqa,Sen:2015hha,Sen:2016qap}.  \par

The gauge transformations of closed string field theory form a complicated infinite dimensional gauge group.  Consequently, the quantization of closed string field theory requires  the sophisticated machinery of  Batalian-Vilkovisky formalism (BV formalism)  \cite{Batalin:1981jr,Batalin:1984jr,Barnich:1994db,Barnich:1994mt,Henneaux:1989jq,Henneaux:1992ig,Gomis:1994he}. The quantum BV master action for closed string field theory can be obtained by solving the quantum BV master equation. The perturbative solution of quantum BV master action for the closed bosonic string field theory has been already constructed \cite{Zwiebach:1992ie}. The striking feature of closed string field theory is that, albeit, the quantum BV master action contains a kinetic term and infinite number of interaction terms, the theory has only one independent parameter, the closed string coupling. The interaction strengths (coupling constants) of the elementary interactions  in closed string field theory are expressed as  integrals over the distinct two dimensional world-sheets describing the elementary interactions of the closed strings.\par

 The collection of  world-sheets describing the elementary interactions of the closed strings are called as the string vertex. A consistent set of string vertices provide a cell decomposition of the moduli space of Riemann surfaces \cite{Zwiebach:1992ie}. The main challenge in constructing string field theory is  to find a consistent set of string vertices that  give rise to a suitable cell decomposition of the moduli spaces of  Riemann surfaces.  In principle, all the string vertices that provide  such a cell decomposition of the moduli space   can be constructed using the Riemann surfaces endowed with the  metric solving the generalized minimal area problem \cite{Zwiebach:1992ie}. Unfortunately, our current understanding of minimal area metrics is insufficient to obtain a calculable formulation of closed string field theory \footnote{Recently, the cubic vertex of heterotic string field theory has constructed by using $SL(2,\mathbb{C})$ local coordinate maps which in turn has been used to construct the one loop tadpole string vertex in heterotic string field theory \cite{Erler:2017pgf}. The cubic string vertex defined this way differ from the cubic string vertex defined by the minimal area metric.}.  However, there exist an  alternate construction of the string vertices using Riemann surfaces endowed with metrics having constant curvature $-1$  \cite{Moosavian:2017fta,Moosavian:2017qsp}.  They can be characterized using the Fenchel-Nielsen coordinates for the Teichm\"uller space and the local coordinates around the punctures on the world-sheets in terms of the hyperbolic metric. They can be used to construct a closed string field theory with approximate gauge invariance.  \par

 The interaction strengths  in closed string field theory are obtained by integrating the off-shell string measure  over the region in the moduli space   that corresponds to the distinct two dimensional world-sheets describing the elementary interactions of the closed strings.  The explicit evaluation of the interaction strength requires:
  \begin{enumerate}
 
 \item A convenient choice of parametrization of the Teichm\"uller space and the conditions on them that specify the region of the moduli space inside the Teichm\"uller space.
 
 \item An explicit description for the region inside moduli space that corresponds to the string vertex and a consistent choice of local coordinates around the punctures on the Riemann surfaces belong to the string vertex.
 
  \item An explicit procedure for constructing the off-shell string measure in terms of the chosen coordinates of the moduli space.
 
 \item Finally, an explicit procedure for integrating the off-shell string measure over the region inside the moduli space that corresponds to the string vertex.
 
 \end{enumerate}
 In this paper,  we provide detailed descriptions for each of them. \\

\noindent{\underline{\bf Summary of the results}}: The main results of this paper are as follows:
 \begin{itemize}
 \item We  explicitly construct the off-shell string measure in terms of the Fenchel-Nielsen coordinates for the Teichm\"uller space using a specific choice of local coordinates that is encoded in the definition of the string vertices.  
 
 \item The interaction strengths  in closed string field theory are obtained by integrating the off-shell string measure, which is an MCG-invariant object,  over the region in the moduli space that corresponds to the Riemann surfaces describing the elementary interactions of the closed strings. The moduli space  is the quotient of the Teichm\"uller space   with the action of the mapping class group (MCG). However, in the generic case, an explicit fundamental region for the action of the MCG inside the Teichm\"uller space is not known.    Therefore, integrating an MCG-invariant function over a region in the moduli space of the Riemann surfaces is not a straightforward operation to perform. In this paper, we  discuss a way to bypass this difficulty and obtain an effective expression for the integral using the prescription for performing the integration over the moduli space  of  hyperbolic Riemann surfaces, parametrized using the Fenchel-Nielsen coordinates, introduced by M.Mirzakhani \cite{Mirzakhani:2006fta}. 
  
 \item  We show that this integration method has an important property when we restrict the integration to a thin region around the boundary of the moduli space. Using this property, we find an effective expression for the integral  of the off-shell string measure  over the region inside the moduli space that corresponds to the string vertex.

 \end{itemize}
 In short, we describe a systematic method for evaluating the quantum BV master action for closed bosonic string field theory. \\
 
  \noindent{\underline{\bf Organization of the paper}}:  This paper is organized as follows. In section \ref{QBVMA}, we briefly review the general construction of the quantum BV master action for closed bosonic string field theory and explain what do we mean by the explicit evaluation of the quantum action.   In section \ref{vertices} we discuss the construction string vertices using hyperbolic Riemann surfaces described in \cite{Moosavian:2017qsp}.  In section \ref{OSM},  we  describe the explicit construction of the off-shell string measure in terms of  the Fenchel-Nielsen coordinates of the Teichm\"uller space. In section \ref{IOMS} we discuss the concept of effective string vertices and the practical procedure of evaluating the corrected interaction vertices. In \ref{disc} we provide a brief summary of the paper and mention some of the future directions.  In appendix \ref{hyperbolic}, we review the theory of hyperbolic Riemann surfaces.   In appendx \ref{MMidentity} and appendix \ref{LuoTan}, we discuss two classes of non-trivial identities satisfied by the lengths of the simple closed geodesics on a hyperbolic Riemann surface.

\section{The quantum BV master action}\label{QBVMA}

The quantum BV master action for  closed string field theory is a functional of the the fields and the antifields in the theory. The {\it fields} and {\it antifields}  are specified by splitting the  string field $|\Psi\rangle$, which is an arbitrary element in the Hilbert space of the worldsheet CFT \cite{Thorn:1988hm}, as
\begin{equation}\label{stringfieldanti}
|\Psi\rangle=|\Psi_-\rangle+|\Psi_+\rangle.
\end{equation}
Both $|\Psi_-\rangle$ and $|\Psi_+\rangle$ are annihilated by $b_0^-$ and $L_0^-$. The string field  $|\Psi_-\rangle$ contains all the fields and the string field $|\Psi_+\rangle$ contains all the antifields.  They can be decomposed as follows
\begin{align}\label{psipmdecomp}
|\Psi_-\rangle&=\sideset{}{'}\sum_{G(\Phi_s)\leq 2}|\Phi_s\rangle \psi^s,\nonumber\\
|\Psi_+\rangle&=\sideset{}{'}\sum_{G(\Phi_s)\leq 2}|\tilde \Phi_s\rangle \psi^*_s,
\end{align}
where  $|\tilde \Phi_s\rangle=b_0^-|\Phi^c_s\rangle$, such that  $\langle\Phi_r^c|\Phi_s\rangle=\delta_{rs}$. The state $\langle\Phi_r^c|$ is the conjugate state of $|\Phi_r\rangle$.  The sum in (\ref{psipmdecomp}) extends over the basis states $|\Phi_s\rangle$ with ghost number less than or equal to two. The prime over the summation sign reminds us that the sum is only over those states that are annihilated by $L_0^-$. The target space field $\psi^*_s$ is the antifield that corresponds to the target space field $\psi^s$. The target space ghost number of the fields $g^t(\psi^s)$ takes  all possible non-negative   values and that of antifields $g^t(\psi^*_s)$ takes all possible negative values. They are related via the following relation 
\begin{equation}\label{ghostnumberaf}
g^t(\psi_s^*)+g^t(\psi^s)=-1.
\end{equation}
Therefore, the statistics of the antifield is opposite to that of the field.  Moreover, it is possible to argue that corresponding to each target space field $\psi^s$ there is a unique antifield $\psi^*_s$ \cite{Thorn:1988hm}.\par

The quantum BV master action must be a solution of  the following quantum BV master equation for  the closed bosonic string theory
\begin{equation}\label{mastereqbcsft}
\frac{\partial _r S}{\partial \psi^s}\frac{\partial _l S}{\partial \psi^*_s}+\hbar\frac{\partial _r }{\partial \psi^s}\frac{\partial _l S}{\partial \psi^*_s}=0,
\end{equation}
 where the target space field $\psi^*_s$ is the antifield corresponding to the field $\psi^s$ and $\partial_r, \partial_l$ denote the right and left derivatives respectively.  The  perturbative solution of this equation in the closed string coupling $g_s$   is given by \cite{Zwiebach:1992ie}:
 \begin{equation}\label{cbstringfieldaction}
 S(\Psi)=g_s^{-2}\left[\frac{1}{2}\langle\Psi|c_0^-Q_B|\Psi \rangle+\sum_{g\geq 0}(\hbar g_s^2)^g\sum_{n\geq 1}\frac{g_s^n}{n!}\{\Psi^n\}_g\right],
 \end{equation} 
 where   $\Psi$ denotes the  string field (\ref{stringfieldanti}) having arbitrary ghost number that is built using target space fields and antifields carrying arbitrary ghost numbers. $\{\Psi^n\}_g$ denotes  the $g$-loop elementary interaction vertex $\{\Psi_1,\cdots,\Psi_n\}_{g}$ for $n$ closed string fields with  $\Psi_i=\Psi$ for $i=1,\cdots,n$.  The $g$-loop elementary interaction vertex $\{\Psi_1,\cdots,\Psi_n\}_{g}$ for $n$ closed string fields  can be defined as the integral of the off-shell string measure $ \Omega^{(g,n)}_{6g-6+2n}\left(|\Psi_1\rangle,\cdots,|\Psi_n\rangle\right)$ over the string vertex $\mathcal{V}_{g,n}$:
\begin{equation}\label{bstringvertex}
\{\Psi_1,\cdots,\Psi_n\}_{g}\equiv\int_{\mathcal{V}_{g,n}} \Omega^{(g,n)}_{6g-6+2n}\left(|\Psi_1\rangle,\cdots,|\Psi_n\rangle\right),
\end{equation}
 where $\Psi_1,\cdots,\Psi_n$ denotes the off-shell  closed string states $|\Psi_1\rangle,\cdots,|\Psi_n\rangle$. The definition of the string vertices and the construction of off-shell measure is discussed  below.
 
 \subsection{The string vertex $ \mathcal{V}_{g,n}$}
  
The string vertex $ \mathcal{V}_{g,n}$ for the closed strings  can be understood as a collection of genus $g$ Riemann surfaces with $n$ punctures that belong to a specific  region inside the compactified moduli space $\overline{\mathcal{M}}_{g,n}$. We can define the string vertices  by stating the properties that they must satisfy \cite{Zwiebach:1992ie}:
\begin{itemize}

\item The string vertices must not contain Riemann surfaces that are  arbitrarily close to the degeneration.

\item  The Riemann surfaces that belong to the string vertices must be  equipped with a specific choice of  local coordinates around each of its punctures. The coordinates  are only defined up to a constant phase and they are defined continuously over the set  $\mathcal{V}_{g,n}$.  
\
\item The  local coordinates around the punctures on the Riemann surfaces that belong to the string vertices  must be  independent of the labeling of the punctures. Moreover, if a Riemann surface $\mathcal{R}$ with labeled punctures is in $ \mathcal{V}_{g,n}$ then copies of $\mathcal{R}$ with all other inequivalent  labelings of the punctures  also must be included in $ \mathcal{V}_{g,n}$.  

\item If a Riemann surface belongs to the string vertex, then  its complex conjugate also must be included in the string vertex.  A complex conjugate Riemann surface of a Riemann surface $\mathcal{R}$ with coordinate $z$ can be obtained by using the anti-conformal map $z\to -\overline{z}$.

\end{itemize}

 The string vertices with the above mentioned properties must  also satisfy the following geometric identity. This identity can be understood as the geometric realization of the quantum BV master equation (\ref{mastereqbcsft}):   
 \begin{equation}\label{bvmastercond}
 \partial \mathcal{V}_{g,n}=-\frac{1}{2}\mathop{\sum_{g_1,g_2}}_{g_1+g_2=g}\mathop{\sum_{n_1,n_2}}_{n_1+n_2=n}\mathbf{ S}[\{ \mathcal{V}_{g_1,n_1}, \mathcal{V}_{g_2,n_2}\}]-\Delta\mathcal{V}_{g-1,n+2},
 \end{equation} 
where $\partial  \mathcal{V}_{g,n}$ denotes the boundary of the string vertex $\mathcal{V}_{g,n}$ and $\mathbf{ S}$ represents the operation of summing over all inequivalent permutations of the external  punctures.  $\{ \mathcal{V}_{g_1,n_1},\mathcal{V}_{g_2,n_2}\}$ denotes the set of Riemann surfaces obtained  by taking a Riemann surface from the string vertex $\mathcal{V}_{g_1,n_1}$ and a Riemann surface from  the string vertex $\mathcal{V}_{g_2,n_2}$ and gluing them by identifying the regions around  one of the puncture from each via the special plumbing fixture relation: 
\begin{equation}\label{specialplumbing}
zw=e^{i\theta},\qquad\qquad0\leq\theta\leq2\pi,
\end{equation}
where $z$ and $w$ denote the local coordinates around the punctures that are being glued. The special plumbing fixture corresponds to the locus $|t|=1$ of the plumbing fixture relation
\begin{equation}\label{plumbing}
zw=t,\qquad\qquad t\in \mathbb{C},\qquad\qquad 0\leq |t|\leq 1,
\end{equation}

The resulting surface has genus $g=g_1+g_2$ and $n=n_1+n_2-2$. $\Delta$ denotes the operation of taking a pair of  punctures on a Riemann surface that belongs to the string vertex  $ \mathcal{V}_{g-1,n+2}$ and gluing them via the special plumbing fixture relation (\ref{specialplumbing}). Therefore, the first term of (\ref{bvmastercond})  represents  the gluing of two distinct surfaces via the special plumbing fixture and the second terms represents  the special plumbing fixture applied to a single surface. \par

{\it The geometric condition (\ref{bvmastercond})  demands that the set of Riemann surfaces that belong to the boundary of a  string vertex having dimension, say $d$, must agree with the set of union of surfaces having dimension $d$  obtained by applying the special plumbing fixture construction  (\ref{specialplumbing})  to the surfaces belong to the lower dimensional string vertices only once, both in  their moduli parameters and in their local coordinates around the punctures}.

\subsection{The off-shell string measure $ \Omega^{(g,n)}_{6g-6+2n}$}\label{offmeasure}

The   off-shell string measure $ \Omega^{(g,n)}_{6g-6+2n}\left(|\Psi_1\rangle,\cdots,|\Psi_n\rangle\right)$ is constructed using $n$ number of vertex operators with arbitrary conformal dimensions. Consequently,  the off-shell string measure depends on the choice of  local coordinates around the punctures on the Riemann surface. Therefore,  the integration measure of an off-shell amplitude is not a  genuine differential form on the moduli space $\mathcal{M}_{g,n}$, because the moduli spaces do not know about the  various  choices of local coordinates around the punctures. Instead, we need to  consider it as a differential form defined on a section of a larger space $\widehat{\mathcal{P}}_{g,n}$. This space  is  defined as a fiber bundle over $\mathcal{M}_{g,n}$. The fiber direction of  the fiber bundle $\pi: \widehat{\mathcal{P}}_{g,n}\to \mathcal{M}_{g,n}$  contains the information about different choices of local coordinates around each of the $n$ punctures that differ only by a phase factor. The section of our interest  corresponds to the choice of a specific  set of  local coordinates around the punctures for each point $\mathcal{R}_{g,n}\in \mathcal{M}_{g,n}$. Therefore, in order to construct a differential form on such a section, we only need to consider  the tangent vectors of $\widehat{\mathcal{P}}_{g,n}$ that are the tangent vectors of  the moduli space of Riemann surfaces equipped with the choice  local coordinates that defines the section. They are given by the Beltrami differentials spanning the tangent space of the moduli space of  Riemann surfaces \cite{Yoichi}.  \par

Let us denote the coordinates of $\mathcal{M}_{g,n}$ by $\left(t_{1},\cdots,t_{6g-6+2n} \right)$. Consider  $B_p$,  an operator-valued $p$-form defined on the section of the  space $\widehat{\mathcal{P}}_{g,n}$. The contraction of $B_p$ with $\left\{V_1,\cdots,V_p\right\}, p$ tangent vectors of the section, is given by 
\begin{equation}\label{opevormbos}
B_p[V_1,\cdots,V_p]=b(V_1)\cdots b(V_p),
\end{equation}
where
 \begin{equation}\label{bvgen2}
b(V_{k})=\int d^2z\Big(b_{zz}\mu_{k\bar z}^z+b_{\bar z\bar z}\mu_{kz}^{\bar z}\Big),
\end{equation} 
 Here $\mu_k$ denotes the Beltrami differential associated with the moduli $t_{k}$ of the Riemann surfaces belong to the  section  of the fiber space $\widehat{\mathcal{P}}_{g,n}$ in which we are interested.  The $p$-form on the section can be obtained by taking the expectation value of  the operator valued $p$-form $B_p$ between the surface state $\langle \mathcal{R}|$ and the state $|\Phi\rangle$: 
 \begin{equation}\label{pformpgnbos}
\Omega_p^{(g,n)}(|\Phi\rangle)=(2\pi \mathrm{i})^{-(3g-3+n)}\langle\mathcal{R}|B_p|\Phi\rangle.
\end{equation}
The state $|\Phi\rangle$ is  the tensor product of external off-shell states $|\Psi_i\rangle,~i=1,\cdots,n$  inserted at the punctures  and the  state $\langle \mathcal{R}|$ is the surface state associated with the surface $\mathcal{R}_{g,n}$. It describes the state that is created on the boundaries of the discs $D_i,~i=1,\cdots,n$ by performing a functional integral over the fields of CFT on $\mathcal{R}-\sum_iD_i$.  The inner product between $\langle\mathcal{R}|$ and a state $|\Psi_1\rangle\otimes\cdots\otimes |\Psi_n\rangle\in\mathcal{H}^{\otimes n}$ 
\begin{equation}\label{innerprcft}
\langle\mathcal{R}|(|\Psi_1\rangle\otimes\cdots\otimes |\Psi_n\rangle),
\end{equation} 
can be understood as the $n$-point correlation function on $\mathcal{R}$ with the vertex operator for $|\Psi_i\rangle$ inserted at the $i^{th}$ puncture using the local coordinate  around that puncture. \par

The path integral representation of $ \Omega^{(g,n)}_{6g-6+2n}\left(|\Psi_1\rangle,\cdots,|\Psi_n\rangle\right)$ is given by 
\begin{align}\label{pathintrep}
& \Omega^{(g,n)}_{6g-6+2n}\left(|\Psi_1\rangle,\cdots,|\Psi_n\rangle\right)\nonumber\\
&=\frac{dt_1\cdots dt_{6g-6+2n}}{(2\pi \mathrm{i})^{(3g-3+n)}}\int \mathcal{D}x^{\mu}\int\mathcal{D}c~\mathcal{D}\overline{c}~\mathcal{D}b~\mathcal{D}\overline{b}~e^{-I_m(x)-I_{gh}(b,c)}\prod_{j=1}^{6g-6+2n} b(V_j)\prod_{i=1}^n\left[c\overline{c}~V_{i}(k_{i})\right]_{w_i},
\end{align}
where $\left[c\overline{c}~V_{i}(k_{i})\right]_{w_i}$ denotes the vertex operator corresponds to the state $|\Psi_i\rangle$ inserted using the local coordinate $w_i$. $I_m(x)$ is the action for matter fields and $I_{gh}(b,c)$ is the actions for ghost fields.  $z$ is the global coordinate on $\mathcal{R}$.

\subsection{The explicit evaluation of the quantum master action}\label{EEQMA}

In this subsection, we explain, what  we mean by the explicit evaluation of the quantum BV master action for the closed string field theory. Let us denote the vertex operator corresponds to the basis state $|\Phi_s\rangle$ by $\mathcal{A}(\Phi_s)$. Then the string field entering in the quantum BV master action can be expressed as
\begin{equation}\label{stringfieldanti1}
|\Psi\rangle=\sideset{}{'}\sum_{G(\Phi_s)\leq 2}\sum_p\psi^s(p)\mathcal{A}(\Phi_s) |\mathbf{1},p\rangle +\sideset{}{'}\sum_{G(\Phi_s)\leq 2}\sum_p\psi^*_s(p)\mathcal{A}(\widetilde{\Phi}_s) |\mathbf{1},p\rangle,
\end{equation}
where $|\mathbf{1},p\rangle $ denotes the $SL(2,\mathbb{C})$ invariant family of vacua for the worldsheet CFT for the closed bosonic string theory, parameterized by $p$. The expression for the quantum BV master action in terms of the target space fields and the target space antifields can be obtained by substituting this expansion of the string field $\Psi$ in the quantum BV master action (\ref{cbstringfieldaction}):
\begin{align}\label{cbstringfieldaction1}
 S(\Psi)&=\frac{1}{2g_s^{2}}\mathop{\sideset{}{'}\sum_{G(\Phi_{s_j})\leq 2}}_{j=1,2}\mathop{\sum_{\phi^{s_i}\in \mathcal{S}_i}}_{i=1,2}\sum_{p_1,p_2}\phi^{s_1}(p_1)\mathbf{P}_{s_1s_2}\left(p_1,p_2\right)\phi^{s_2}(p_2)\nonumber\\
&+\mathop{\sum_{g\geq 0}}_{n\geq 1}\frac{\hbar^g g_s^{2g+n-2}}{n!}\mathop{\sideset{}{'}\sum_{G(\Phi_{s_j})\leq 2}}_{j=1,\cdots,n}\mathop{\sum_{\phi^{s_i}\in \mathcal{S}_i}}_{i=1,\cdots,n}\sum_{p_1,\cdots,p_n}\mathbf{V}_{s_1\cdots s_n}^{g,n}\left(p_1,\cdots,p_n \right)\phi^{s_1}(p_1)\cdots\phi^{s_n}(p_n),
 \end{align} 
where  $\mathcal{S}_i=\left\{ \psi^{s_i}, \psi^*_{s_i}\right\}$ is the set of all fields and antifields of the closed bosonic string field theory spectrum. $\mathbf{P}_{s_1s_2}\left(p_1,p_2\right)$, the inverse of the  propagator, is given by
\begin{equation}\label{propagator}
\mathbf{P}_{s_1s_2}\left(p_1,p_2\right)=\left\langle\mathcal{Y}_{s_1},p_1\left|c_0^-Q_B\right|\mathcal{Y}_{s_2},p_2 \right\rangle,
\end{equation} 
and $\mathbf{V}_{s_1\cdots s_n}^{g,n}\left(p_1,\cdots,p_n \right)$, the  $g$ loop interaction vertex of $n$ target spacetime fields/antifields $\left\{\phi^{s_1}(p_1),\cdots,\phi^{s_n}(p_n)\right\}$, is given by
\begin{equation}\label{interaction}
\mathbf{V}_{s_1\cdots s_n}^{g,n}\left(p_1,\cdots,p_n \right)=\int_{\mathcal{V}_{g,n}} \Omega^{(g,n)}_{6g-6+2n}\left(|\mathcal{Y}_{s_1},p_1\rangle,\cdots,|\mathcal{Y}_{s_n},p_n\rangle\right).
\end{equation} 
 Here, $|\mathcal{Y}_{s_i},p_i\rangle$ is the state associated with the string field/antifield $\phi^{s_i}(p_i)$. The state $|\mathcal{Y}_{s_i},p_i\rangle$ is annihilated by both $b^-_0$ and $L^-_0$. By the explicit evaluation of the quantum master action, we mean the explicit evaluation of  $\mathbf{V}_{s_1\cdots s_n}^{g,n}\left(p_1,\cdots,p_n \right)$. The explicit evaluation requires:
  \begin{enumerate}
 
 \item A convenient choice of parametrization of the Teichm\"uller space and the conditions on them that specify the region of the moduli space inside the Teichm\"uller space.
 
  \item An explicit procedure for constructing the off-shell string measure in terms of the chosen coordinates of the moduli space.
 
 \item An explicit description for the region inside moduli space that corresponds to the string vertex and a consistent choice of local coordinates around the punctures on the Riemann surfaces belong to the string vertex.
 
 \item Finally, an explicit procedure for integrating the off-shell string measure over the region inside moduli space that corresponds to the string vertex.
 
 \end{enumerate}

In the remaining sections of this paper, we provide a detailed description for each of these steps.

\section{The string  vertices using  hyperbolic  metric}\label{vertices}

The main challenge in constructing string field theory is  to find a suitable cell decomposition of the moduli spaces of closed Riemann surfaces.  In principle, the string vertices satisfying the conditions listed in the section (\ref{QBVMA})  that provide  such a cell decomposition of the moduli space   can be constructed using the Riemann surfaces endowed with the  metric solving the generalized minimal area problem \cite{Zwiebach:1992ie}. Unfortunately, the current understanding of minimal area metrics is not sufficient enough to obtain a calculable formulation of closed string field theory. In our previous paper \cite{Moosavian:2017qsp}, we  described an  alternate construction of the string vertices using Riemann surfaces endowed with metric having constant curvature $-1$.   We  briefly review this construction below. For a brief review of the theory of hyperbolic Riemann surfaces, read appendix (\ref{hyperbolic}). \par

A hyperbolic Riemann surface can be represented  as a quotient of the upper half-plane $\mathbb{H} $ by a Fuchsian group.  A puncture on a hyperbolic Riemann surface corresponds to the fixed point of a parabolic element (element having trace $\pm 2$) of the Fuchsian group acting on the upper half-plane $\mathbb{H} $. For a puncture $p$ on a hyperbolic Riemann surface, there is a natural local conformal coordinate $w$ with $w(p) = 0$, induced from the hyperbolic metric. The local expression for the hyperbolic metric around the puncture is given by 
\begin{equation}\label{metriclocal}
ds^2=\left(\frac{|dw|}{|w|\ln|w|}\right)^2.
\end{equation}

We can naively define the string vertices by means of the Riemann surfaces endowed with the hyperbolic metric  as below.\\

\noindent{\bf The naive string vertex $\mathcal{V}^0_{g,n}$}:  {\it Consider a genus-$g$ hyperbolic Riemann surface $\mathcal{R}$ having $n$ punctures  with no simple closed geodesics that has geodesic length $l\leq c_*$, where $c_*$ is some positive real number such that $c_*\ll 1$.  The local coordinates around the punctures on $\mathcal{R}$ are chosen to be  $e^{\frac{\pi^2}{c_*}}w$, where $w$ is the natural local coordinate induced from the hyperbolic metric on  $\mathcal{R}$. The set of all such inequivalent hyperbolic Riemann surfaces forms the  string vertex $\mathcal{V}^0_{g,n}$}. \\

 It was shown in  \cite{Moosavian:2017qsp} that  the string vertices $\mathcal{V}^0_{g,n}$ fails to provide a single cover of the moduli space for any non-vanishing value of $c_*$. The argument goes as follows. We can claim that the string vertex $\mathcal{V}^0_{g,n}$ together with the Feynman diagrams provide  a cell decomposition of  the moduli space only if  the  Fenchel-Nielsen length parameters and the local coordinates around the punctures on the surfaces at the boundary of the string vertex region matches exactly with the  Fenchel-Nielsen length parameters and the local coordinates around the punctures on the surface obtained by the special plumbing fixture construction
\begin{equation}\label{specialplumbing1}
\widetilde z\cdot \widetilde w=e^{i\theta},\qquad 0\leq\theta\leq2\pi,
\end{equation}
where $\widetilde z$ and $\widetilde w$ denote the local coordinates around the punctures that are being glued. However, the metric on the surface obtained by the plumbing fixture of  a set of hyperbolic Riemann surface fails to be  exactly hyperbolic all over the surface \cite{Wolpert2,Wolpert3,Melrose}.

Consider the   Riemann surface $\mathcal{R}_t$, for $t=(t_1,\cdots,t_m)$ obtained via plumbing fixture around $m$ nodes of a hyperbolic surface  $\mathcal{R}_{t=0}\equiv\mathcal{R}_0$ with $m$ nodes. We denote the set of Riemann surfaces obtained by removing the nodes from $\mathcal{R}_0$  by $\hat{\mathcal{R}}$, i.e.,  $\hat{\mathcal{R}}_0=\mathcal{R}_0-\{\mathrm{nodes}\}$. The Riemann surfaces $\hat{\mathcal{R}}_0$ have a pair of punctures $(a_j,b_j)$ in place of the $j$\textsuperscript{th} node of $\mathcal{R}_0,~j=1,\cdots,m$. Assume that $w^{(1)}_j$ and $w^{(2)}_j$ are the local coordinates around the punctures $a_j$ and $b_j$ with the property that $w^{(1)}_j(a_j)=0$ and $w^{(2)}_j(b_j)=0$. Let us choose the local coordinates $w^{(1)}_j$ and $w^{(2)}_j$ such that, in terms of these local coordinates, the hyperbolic metric around the punctures of $\widehat{\mathcal{R}}_0$ has the local expression 
 \begin{equation}
ds^2=\left (\frac{|d\zeta|}{|\zeta|\mathrm{ln}~|\zeta|}\right)^2,\qquad\qquad\zeta=w^{(1)}_j~\mathrm{or}~w^{(2)}_j.
 \end{equation}
 Let us call the metric on the glued surface  $\mathcal{R}_t$ as the the {\it grafted metric} $ds_{\text{graft}}^2$. The grafted metric has curvature $-1$ except at the collar boundaries, where the interpolation leads to a deviation of magnitude $(\mathrm{ln}|t|)^{-2}$ \cite{Wolpert2}. This deviation makes the resulting surface almost hyperbolic except at the boundaries of the plumbing collar.  \par
 
However, we can compute the hyperbolic metric on $\mathcal{R}_t$ by solving the {\it curvature correction equation} \cite{Wolpert2,Wolpert3}. To describe the  curvature correction equation, consider a compact Riemann surface having metric $ds^2$ with Gauss curvature \footnote{In two dimension, the Gaussian curvature is half of the Ricci curvature of the surface.} $\mathbf{ C}$.  Then, another  metric $e^{2f}ds^2$ on this surface has constant curvature $-1$ provided
  \begin{equation}\label{constantcurvatureq}
  Df-e^{2f}=\mathbf{ C},
  \end{equation}
  where $D$ is the Laplace-Beltrami operator on the surface. Therefore, in order to get the  hyperbolic metric on $\mathcal{R}_t$, we need to solve this curvature correction equation perturbatively around the grafted metric by adding a Weyl factor.  Then we can invert this expression for hyperbolic metric on $\mathcal{R}_t$ in terms of the grafted metric to obtain the grafted metric  in terms of the hyperbolic metric.  \par
  
  To the second order the hyperbolic metric on $\mathcal{R}_t$, Riemann surface at the boundary of the string vertex $\mathcal{V}_{g,n}^0$ obtained by the special  plumbing  fixture (\ref{specialplumbing1}) of the hyperbolic Riemann surfaces, is related to the grafted metric as follows
\begin{equation}
ds^2_{\text{hyp}}=ds^2_{\text{graft}}\left(1+\sum_{i=1}^m\frac{c_*^{2}}{3}\left(E^{\dagger}_{i,1}+E^{\dagger}_{i,2}\right)+\mathcal{O}\left(c_*^3\right)\right). \label{graftedhyp2c}
\end{equation}
The functions $E^{\dagger}_{i,1}$ and $E^{\dagger}_{i,2}$ are the melding of the Eisenstein series $E(\cdot;2)$ associated to the pair of cusps plumbed to form the $i$\textsuperscript{th} collar. For the definition of these functions, see \cite{Moosavian:2017qsp}. The details of these functions are not very important for our discussions. \par

Using this relation, we modify the definition of the  string vertices by changing the choice of local coordinates on the surfaces which belong to the boundary region of the string vertices as follows  \cite{Moosavian:2017qsp}. The boundary of the string vertex with $m$ plumbing collar is defined as the locus in the moduli space of the hyperbolic Riemann surfaces with $m$ non-homotopic and disjoint  non trivial simple closed curves having length equal to that of the  length of the simple geodesic  on any plumbing collar of a Riemann surface obtained by gluing $m$ pair of punctures on a set of hyperbolic Riemann surfaces via the special plumbing fixture relation (\ref{specialplumbing1}).  To the second order in $c_*$, there is no correction to the hyperbolic length of the  geodesics on the plumbing collars. Therefore, to second order in $c_*$, we don't have to correct the definition of the region corresponding to the string vertex in the moduli space for the hyperbolic Riemann surfaces parametrized using the Fenchel-Nielsen coordinates. However,  the choice of local coordinates around the punctures must be modified to make it gluing compatible to second order in $c_*$.     In order to modify the assignment of  local coordinates in the string vertex $\mathcal{V}^0_{g,n}$, we divide it into subregions. Let us denote the subregion in the region corresponds to the string vertex $\mathcal{V}^0_{g,n}$ consists of surfaces with $m$ simple closed geodesics (none of them are related to each other by the action of any elements in MCG) of length between $c_*$ and $(1+\delta)c_*$ by $\mathbf{W}^{(m)}_{g,n}$, where $\delta$ is an infinitesimal real number. Then we modify the local coordinates as follows:
 \begin{itemize}
 \item For surfaces belong to the subregion $\mathbf{W}^{(0)}_{g,n}$, we choose the local coordinate around the $j^{th}$ puncture to be $e^{\frac{\pi^2}{c_*}}w_j$. In terms of $w_j$, the hyperbolic metric in the neighbourhood of the puncture takes the following form
\begin{equation}
 \left(\frac{|dw_j|}{|w_j|\ln|w_j|}\right)^2, \qquad  j=1,\cdots,n. 
\end{equation}
\item For surfaces belong to the region $\mathbf{W}^{(m)}_{g,n}$ with $m\ne 0$, we choose the local coordinates around the $j^{th}$ puncture to be  $e^{\frac{\pi^2}{c_*}}\widetilde{w}_{j,m}$, where $\widetilde{w}_{j,m}$, up to a phase ambiguity,  is given by
\begin{equation}
\widetilde{w}_{j,m}=e^{\frac{c_*^2}{6}\sum_{i=1}^mf(l_i)Y_{ij}}w_{j}.
\end{equation}

\end{itemize}

We found $\widetilde w_{j,m}$ by solving the following equation
\begin{equation}
\left(\frac{|d\widetilde{w}_{j,m}|}{|\widetilde{w}_{j,m}|\mathrm{ln}|\widetilde{w}_{j,m}|}\right)^2=\left(\frac{|dw_j|}{|w_j|\text{ln}|w_j|}\right)^2\left\{1-\frac{c_*^2}{3~\text{ln}|w_j|}\sum_{i=1}^mf(l_i)Y_{ij}\right\},
\end{equation}
where $l_i$ denotes the length of the $i^{th}$ degenerating simple closed geodesic and the function $f(l_i)$ is an arbitrary smooth real function of the geodesic length $l_i$ defined in the interval $\left(c_*,c_*+\delta c_*\right)$, such that $f(c_*)=1$ and $f(c_*+\delta c_*)=0$. The coefficient $Y_{ij}$ is given by
\begin{align}\label{yij}
Y_{ij}&=\sum_{q=1}^2\sum_{c_i^q,d_i^q}\pi^{2}\frac{\epsilon(j,q)}{|c_i^q|^4}\nonumber\\ c_i^q>0 \qquad &d_i^q~\text{mod}~c_i^q \qquad\left(\begin{array}{cc}* & * \\c_i^q & d_i^q\end{array}\right)\in \quad (\sigma_i^q)^{-1}\Gamma_{i}^{q}\sigma_j
\end{align} 
Here, $\Gamma_i^q$ denotes the Fuchsian group for the component Riemann surface with the cusp  denoted by the index $q$ that is being glued via plumbing fixture to obtain the $i^{th}$ collar.  The transformation $\sigma_j^{-1}$ maps the cusp corresponding to the $j^{th}$ cusp to $\infty$ and $(\sigma_j^q)^{-1}$ maps the cusp denoted by the index $q$ that is being glued via plumbing fixture to obtain the $i^{th}$ collar to $\infty$. The factor $\epsilon(j,q)$ is one if both the $j^{th}$ cusp and he cusp denoted by the index $q$ that is being glued via plumbing fixture to obtain the $i^{th}$ collar belong to the same component surface other wise $\epsilon(j,q)$ is zero.

The string vertices corrected in this way are denoted as $\mathcal{V}^{2}_{g,n}$. They provide an  improved approximate cell decomposition of the moduli space that has no mismatch up to the order  $c_*^2$. 

\section{The off-shell string measure and Fenchel-Nielsen parameters}\label{OSM}

 In this section, we  describe the explicit construction of the off-shell string measure in terms of  the Fenchel-Nielsen coordinates of the Teichm\"uller space. As explained in  subsection (\ref{offmeasure}), the off-shell string measure can be defined using a specific choice of local coordinates, that is encoded in the definition of the string vertices, and  the Beltrami differentials associated with the moduli parameters. \par


 A flow in   $\mathcal{T}_{g,n}$, the Teichm\"uller space of  $\mathcal{R}$, hyperbolic Riemann surfaces with $g$ handles and $n$ borders, can be generated by a twist field  defined with respect a simple closed curve on the Riemann surface \cite{Kerckhoff,Wolpert5,Wolpert6,Wolpert7}. The twist field $t_{\alpha}$, where $\alpha$ is a simple closed geodesic, generates a flow in $\mathcal{T}_{g,n}$ that can be understood as the {\it Fenchel-Nielsen deformation} of $\mathcal{R}$ with respect to $\alpha$.  The Fenchel-Nielsen deformation is the operation of  cutting the hyperbolic surface along $\alpha$ and attaching the boundaries after rotating one boundary relative to the other by some amount $\delta$. The magnitude $\delta$ parametrizes the flow on  $\mathcal{T}_{g,n}$. \par

Assume that  $\mathcal{R}$ is uniformized  as $\mathbb{H}/\Gamma$. Suppose that  the element of $\Gamma$ that corresponds to a simple closed geodesic $\alpha$  is the matrix
$$A=\left(\begin{array}{cc}a & b \\c & d\end{array}\right).$$
Then, the  Beltrami differential corresponds to the twist vector field $t_{\alpha}$ is given by \cite{Wolpert5}  
\begin{equation}\label{beltramitwist}
\mathbf{t }_{\alpha}=\frac{\mathrm{i}}{\pi}(\mathrm{Im}z)^2\overline{ \Theta}_{\alpha}.
\end{equation}
 $\Theta_{\alpha}$ is the following relative Poincar\'e series 
\begin{equation}\label{poicares}
\Theta_{\alpha}=\sum_{B\in \langle A\rangle \backslash  \Gamma}\omega_{B^{-1}AB},
\end{equation}
where   $\langle A\rangle$ denote  the infinite cyclic group  generated by the element $A$, and   $\omega_A$ is given by
\begin{equation}
\omega_A=\frac{(a+d)^2-4}{\left(cz^2+(d-a)z-b\right)^2}.
\end{equation}
Consider  the Fenchel-Nielsen  coordinates of the Teichm\"uller space $\left(\tau_i,\ell_i\right),~ i=1,\cdots,3g-3+n$ defined with respect to the  pants decomposition $ \mathcal{P}=\left\{C_1,\cdots,C_{3g-3+n}\right\}$, where $C_i$ denotes a simple geodesic on $\mathcal{R}$. By definition, for  $i\neq j$, the curves $C_i$ and $C_j$ are disjoint and non-homotopic to each other.  The tangent space at a point in the Teichm\"uller space is spanned by the Fenchel-Nielsen  coordinate vector fields  $ \left\{\frac{\partial}{\partial \tau_i},\frac{\partial }{\partial \ell_i}\right\},~ i=1,\cdots,3g-3+n$.   The Fenchel-Nielsen  coordinate vector field $\frac{\partial}{\partial \tau_i}$ can be identified with  the  twist vector field $t_{C_i}$ defined with respect to  the curve $C_i$. Hence, the  {\it  Beltrami differential corresponds to the Fenchel-Nielsen  coordinate vector field $\frac{\partial}{\partial \tau_i}$ is given by $\mathbf{ t}_{C_i}$}. The Beltrami differential for the Fenchel-Nielsen  coordinate vector field $\frac{\partial}{\partial l_i}$ can also be constructed by noting the that with respect to the WP symplectic form $\frac{\partial}{\partial l_i}$ is dual to the twist vector field $\frac{\partial}{\partial \tau_i}$ \cite{Wolpert6}. We denote the  Beltrami differential for the Fenchel-Nielsen  coordinate vector field $\frac{\partial}{\partial l_i}$ as $\mathbf{ l}_{C_i}$.\par
   
Putting these together, the off-shell bosonic-string measure can be written as 
\begin{align}\label{eq:the bosonic-string measure}
	&\Omega_{6g-6+2n}^{(g,n)}(|\Psi_1\rangle\otimes\cdots\otimes|\Psi_{n}\rangle)\nonumber\\
	&=\frac{\prod_{i=1}^{3g-3+n}d\ell_id\tau_i}{(2\pi \mathrm{i})^{(3g-3+n)}}\int \mathcal{D}x^{\mu}\int\mathcal{D}c~\mathcal{D}\overline{c}~\mathcal{D}b~\mathcal{D}\overline{b}~e^{-I_m(x)-I_{gh}(b,c)}\prod_{j=1}^{3g-3+n} b(\mathbf{t}_{C_j})b(\mathbf{l}_{ C_j})\prod_{i=1}^n\left[c\overline{c}~V_{i}(k_{i})\right]_{w_i},
\end{align}
where $[c\overline{c}V_i(k_i)]_{w_i}$ denote the vertex operator for the state $|\Psi_i\rangle$ inserted at $i^{\text{th}}$ puncture using the local coordinate $w_i$ and 
\begin{alignat}{1}\label{eq:the expressions for b(v)}
	b(\mathbf{t}_{C_i})&=\int_{\mathcal{F}} d^2z\left(b_{zz}\mathbf{t}_{C_i}+b_{\bar z\bar z}\overline{\mathbf{t}}_{C_i}\right),\nonumber
	\\
	b(\mathbf{l}_{C_i})&=\int_{\mathcal{F}} d^2z\left(b_{zz}\mathbf{l}_{C_i}+b_{\bar z\bar z}\overline{\mathbf{l}}_{C_i}\right).
\end{alignat}
Here $\mathcal{F}$ denotes the fundamental domain in the upper half-plane for the action of the Fuchsian group $\Gamma$ that corresponds to  $\mathcal{R}$.  Here, we assumed  that  $\mathcal{R}$ belongs to  the string vertex $\mathcal{V}_{g,n}$. Remember that, the Riemann surfaces belong to the string vertices carry a specific choice of local coordinates  around its punctures which is consistent with the geometrical identity (\ref{bvmastercond}). \par

Assume that the vertex operator $V_i(k_i)$ has conformal dimension $h_i$ with no ghost fields in it, for $i=1,\cdots,n$. Then we have 
\begin{align}\label{eq:the bosonic-string measure1}
	&\Omega_{6g-6+2n}^{(g,n)}(|\Psi_1\rangle\otimes\cdots\otimes|\Psi_{n}\rangle)\nonumber\\
	&=\frac{\prod_{i=1}^{3g-3+n}d\ell_id\tau_i}{(2\pi \mathrm{i})^{(3g-3+n)}}z\left|\frac{\partial z}{\partial w_i}\right|^{2h_i-2}\sqrt{\text{det}' P_1^{\dagger}P_1}\left(\frac{2\pi^2}{\int d^2z~\sqrt{g}}\text{det}'\Delta \right)^{-13}\int \mathcal{D}x^{\mu}~e^{-I_m(x)}\prod_{i=1}^nV_{i}(k_{i}),
\end{align}
 where, $\Delta$ is the Laplacian acting on scalars defined on $\mathcal{R}$ a genus $g$ hyperbolic Riemann surface with $n$ punctures. The prime indicates that we do not include contributions from zero modes while computing the determinant of $\Delta$. The operator $P_1=\nabla^1_z\oplus\nabla^z_{-1}$ and $P_1^{\dagger}=-\left(\nabla^2_z\oplus\nabla^z_{-2}\right)$. Operators $\nabla^n_z$  and  $\nabla^z_n$ are defined by their action on $T(dz)^n$, which is given by
\begin{align}
\nabla^n_z \left(T(dz)^n\right)&=(g_{z\overline{z}})^n\frac{\partial}{\partial z}\left((g^{z\overline{z}})^nT\right)(dz)^{n+1},\nonumber\\
\nabla^z_n \left(T(dz)^n\right)&=g^{z\overline{z}}\frac{\partial}{\partial\overline{z}}T(dz)^{n-1}.
\end{align}
Interestingly, the determinant $\text{det}' P_1^{\dagger}P_1$ and $\text{det}'\Delta$ can be evaluated on any hyperbolic Riemann surface in terms of Selberg zeta functions \cite{Sarnak, Dhokerdet, Bolte, Hejhal, DHoker:1985een}. For instance, $\text{det}'\Delta$  on a genus $g$ hyperbolic Riemann surface with $n$ punctures can be expressed as follows \cite{LPT}
\begin{align}\label{selbergdet}
\text{det}'\Delta=2^{\frac{n}{2}+\frac{1}{2}\text{tr}\Phi(\frac{1}{2})}(2\pi)^{g-1+\frac{n}{2}}e^{(2g-2+n)\left(2\zeta'(-1)-\frac{1}{4} \right)}\frac{d}{ds}Z(s)\Big|_{s=1}
\end{align}
where $\zeta(s)$ is the Riemann zeta function and $\Phi(s)=\left(\phi_{ij}(s)\right)_{1\leq i, j \leq n}$. The elements $\phi_{ij}$ can be found by expanding the Eisenstein series defined with respect to the $i^{\text{th}}$ puncture around the  $j^{\text{th}}$ puncture. The expansion can be obtained by taking the limit $(y=\text{Im}(z))\to \infty$
\begin{equation}\label{phiij}
E_i(\sigma_jz,s)=\delta_{ij}y^s+\phi_{ij}(s)y^{1-s}+\cdots,
\end{equation}
where $\sigma_i^{-1}\kappa_i\sigma_i=\left(\begin{array}{cc}1 & 1 \\0 & 1\end{array}\right)$. $\kappa_i$ is the parabolic generator associated with the $i^{\text{th}}$ puncture. Finally $Z(s)$ is the Selberg zeta function given by
\begin{equation}\label{selbergzeta}
Z(s)=\prod_{\gamma\in\text{S}}\prod_{k=1}^{\infty}\left[1-e^{-(s+k)\ell_{\gamma}} \right],
\end{equation}
where $\gamma$ is a simple closed geodesic on $\mathcal{R}$ and $S$ is the set of all simple closed geodesics on $\mathcal{R}$. A simple closed geodesic on $\mathcal{R}$ corresponds to a primitive element in the Fuchsian group $\Gamma$. A hyperbolic element of $\Gamma$ is said to be a primitive element if it can not be written as a power of any hyperbolic element in $\Gamma$. However, a primitive element can be an inverse of another primitive element in $\Gamma$. If $g\in \Gamma$ represents the simple closed geodesic $\gamma$, then the length of $\gamma$ is given by 
\begin{equation}\label{lgamma}
\ell_{\gamma}=\text{cosh}^{-1}\left(\frac{1}{2}\left|\text{tr} ~g\right|\right).
\end{equation}
Therefore the Selberg zeta function can be expressed as a product over all the primitive elements in $\Gamma$. The $\text{det} P_1^{\dagger}P_1$ on $\mathcal{R}$ also can be similarly expressed in terms of the Selberg zeta functions.\par 

The matter sector path integral can be expressed in terms of the Green's function $G$  for the Laplacian acting on the scalars on $\mathcal{R}$. To demonstrate this, let us consider the case where all the external states are tachyons, ie. $V_i(k_i)=e^{\text{i} k_i\cdot X_i}$. Then we have 
\begin{align}\label{eq:the bosonic-string measure2}
	&\Omega_{6g-6+2n}^{(g,n)}(|T_1\rangle\otimes\cdots\otimes|T_{n}\rangle)\nonumber\\
	&=\frac{\prod_{i=1}^{3g-3+n}d\ell_id\tau_i}{(2\pi \mathrm{i})^{(3g-3+n)}}\left|\frac{\partial z}{\partial w_i}\right|^{2h_i-2}\sqrt{\text{det}' P_1^{\dagger}P_1}\left(\frac{2\pi^2}{\int d^2z~\sqrt{g}}\text{det}'\Delta \right)^{-13}e^{\frac{1}{2}\sum_{i,j}k_i\cdot k_jG(x_i,x_j)}(2\pi)^{26}\delta(k_1+\cdots+k_n),
\end{align}
where $x_i$ denotes the fixed point corresponds to the $i^{\text{th}}$ puncture. The Green's function on $\mathcal{R}$ can be constructed by first constructing the Green's function on $\mathbb{H}$ and then by considering the sum over all the elements of $\Gamma$, which is same as the idea of method of images  \cite{Hejhal}. \par

Assume that the hyperbolic Riemann surface $\mathcal{R}$ corresponds to a point in the Teichm\"uller space with coordinate $(\ell_1,\tau_1,\cdots,\ell_{3g-3+n},\tau_{3g-3+n})$. Then by following the general algorithm described in \cite{Maskit}, it is possible to express the matrix elements of the generators of $\Gamma$ as functions of $(\ell_1,\tau_1,\cdots,\ell_{3g-3+n},\tau_{3g-3+n})$. Using these generators it is in principle possible to construct all the primitive elements of $\Gamma$. Therefore we can express the determinants of the Laplacians and the Green's functions on $\mathcal{R}$ as functions of the Fenchel-Nielsen coordinates. Finally we get an expression of the off-shell string measure in terms of the Fenchel-Nielsen coordinates.   \par

\section{The effective string vertices} \label{IOMS}

The interaction vertices in closed string field theory is obtained by integrating the off-shell bosonic string measure constructed in the previous section over the region in the compactified moduli space $\overline{\mathcal{M}}_{g,n}$  that corresponds to the string vertex $\mathcal{V}_{g,n}$, which is denoted as $\mathcal{W}_{g,n}$. The modification of the local coordinates requires dividing  $\mathcal{W}_{g,n}$ into different sub-regions.   The moduli space $\mathcal{M}_{g,n}$ can be understood as the quotient of the Teichm\"uller space  $\mathcal{T}_{g,n}$ with the action of the MCG (mapping class group). Unfortunately, in generic cases, an explicit fundamental region for the action of  MCG is not known in terms of the Fenchel-Nielsen coordinates. This is due to the fact that the form of the action of   MCG on  the Fenchel-Nielsen coordinates is not yet known \cite{Thurs1,Hatch1}.   Therefore, modifying the naive string vertex, to make it consistent to $\mathcal{O}(c_*^2)$, appears to be impractical. In this section, we discuss a way to overcome this difficulty by following  the prescription for performing intgrations in the moduli space  introduced by M.Mirzakhani \cite{Mirzakhani:2006fta}.  \par

	\subsection{The effective calculations}
	
	 Consider the space $\mathcal{M}$ with a covering space $\mathcal{N}$. The covering map is given by 
	$$\pi: \mathcal{N}\to \mathcal{M}.$$
	 If $dv_{\mathcal{M}}$ is a volume form for $\mathcal{M}$, then
	$$ dv_{\mathcal{N}}\equiv\pi^{-1}*(dv_{\mathcal{M}}),$$
	   defines the volume form for the covering space $\mathcal{N}$.  Assume that $h$ is a smooth function defined in the space $\mathcal{N}$. Then the push forward of  the function $h$ at a point $x$ in the space $\mathcal{M}$, which is denoted by $\pi_*h(x)$,  can be obtained by the summation over the values of the function $h$ at all points in the fiber of the point $x$ in $\mathcal{N}$:
	\begin{equation}
	(\pi_*h)(x)\equiv\sum_{y\in \pi^{-1}\{x\}}h(y). \label{coveri}
	\end{equation}
	 This relation defines a smooth function on the space $\mathcal{M}$. As a result, the integral of this pull-back function over the space $\mathcal{M}$ can be lifted to the covering space $\mathcal{N}$ as follows: 
		\begin{equation}
		\int_{\mathcal{M}}dv_{\mathcal{M}}~(\pi_*h)~=\int_{\mathcal{N}}dv_{\mathcal{N}}~h. \label{covint}
		\end{equation}

	  \noindent{{\bf Integration over $\mathbb{S}^1$ as an integration over $\mathbb{R}$:}} In order to elucidate the basic logic  behind the integration method, let us  discuss a simple and explicit example. Consider the real  line $\mathbb{R}=(-\infty,\infty)$ as the covering space of circle $\mathbb{S}^1=[0,1)$. We  denote the covering map  by $$\pi: \mathbb{R}\to \mathbb{S}^1.$$
	Assume that $f(x)$ is a function living in $\mathbb{S}^1$, i.e. $f(x+k)=f(x),~k\in \mathbb{Z}$. Then we can convert the integration over $\mathbb{S}^1$ into an integration over $\mathbb{R}$ with the help of the identity 
	\begin{equation}\label{identity}
	1=\sum_{k=-\infty}^{\infty}\frac{\text{sin}^2\left(\pi [x-k] \right)}{\pi^2\left( x-k\right)^2},
	\end{equation}
	 as follows:
	\begin{align}\label{covercircle}
	\int_{0}^1dx~f(x)&= \int_0^1 dx~ \left(\sum_{k=-\infty}^{\infty}\frac{\text{sin}^2\left(\pi [x-k] \right)}{\pi^2\left( x-k\right)^2}\right)f(x)\nonumber\\
	&= \int_{0}^{1} dx~ \sum_{k=-\infty}^{\infty}\left(\frac{\text{sin}^2\left(\pi [x-k] \right)}{\pi^2\left( x-k\right)^2} f(x-k)\right)\nonumber\\
	&= \sum_{k=-\infty}^{\infty} \int_{0}^{1} dx~\frac{\text{sin}^2\left(\pi [x-k] \right)}{\pi^2\left( x-k\right)^2}f(x-k)\nonumber\\
	&= \int_{-\infty}^{\infty} dx~ \frac{\text{sin}^2\left(\pi x \right)}{ \pi^2x^2}f(x).
	\end{align}
	In the last step, we absorbed the summation over $k$ and converted the integration over $\mathbb{S}^1$ to the integration over $\mathbb{R}$. For instance, choosing $f(x)$ to be the ione, gives the following well-known result $$1= \int_{-\infty}^{\infty} dx~ \frac{\text{sin}^2\left(\pi x \right)}{ \pi^2x^2}.$$ \par

\subsection{Effective regions in the Teichm\"uller spaces}

The discussion in the previous subsection suggest that, if we have a region in the Teichm\"uller space that can be identified as a covering space of a region in the moduli space, then  the integration of a differential form defined in the moduli space can be performed by expressing the differential form as a push-forward of a differential form in the Teichm\"uller space using the covering map. In the remaining part of this section, we shall explain that it is indeed possible to find such a covering map and express the off-shell string measure as a push-forward of a differential form defined in the Teichm\"uller space. \\

		  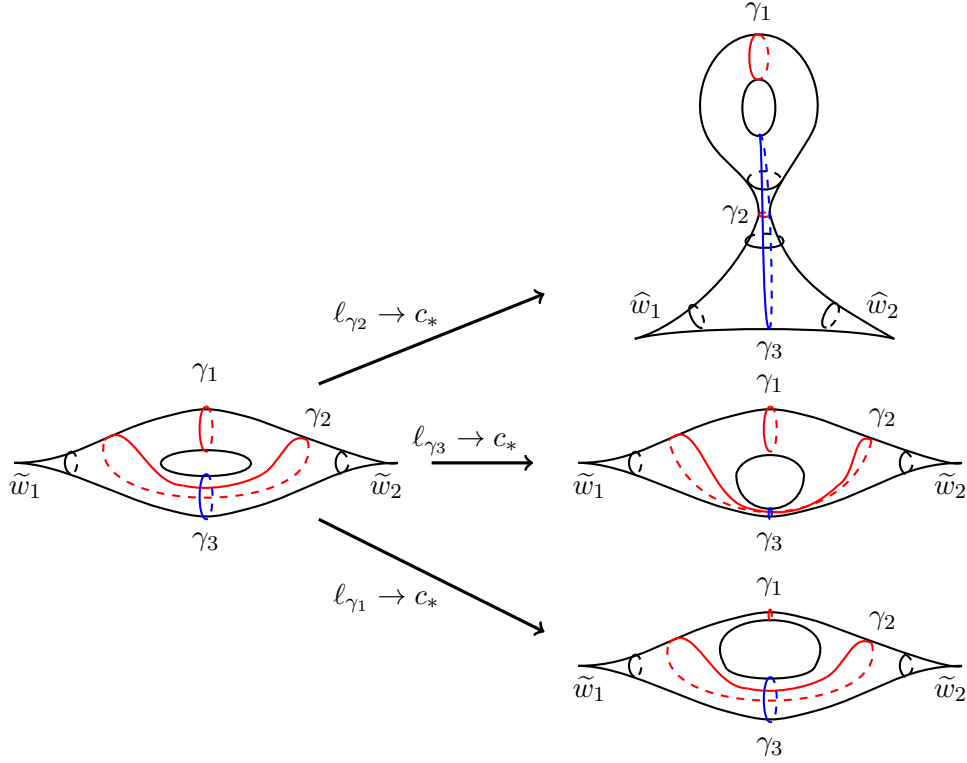
\begin{figure}
\begin{center}
\usetikzlibrary{backgrounds}
\begin{tikzpicture}[scale=.6]
\begin{pgfonlayer}{nodelayer}
		\node [style=none] (0) at (-15, -2.5) {};
		\node [style=none] (1) at (-11.75, -1.5) {};
		\node [style=none] (2) at (-9.75, -1.5) {};
		\node [style=none] (3) at (-6.5, -2.5) {};
		\node [style=none] (4) at (-9.75, -3.5) {};
		\node [style=none] (5) at (-11.75, -3.5) {};
		\node [style=none] (6) at (-11.75, -2.5) {};
		\node [style=none] (7) at (-9.75, -2.5) {};
		\node [style=none] (8) at (-13.75, -2.25) {};
		\node [style=none] (9) at (-13.75, -2.75) {};
		\node [style=none] (10) at (-7.75, -2.75) {};
		\node [style=none] (11) at (-7.75, -2.25) {};
		\node [style=none] (12) at (-10.75, -1.25) {};
		\node [style=none] (13) at (-8.5, -2) {};
		\node [style=none] (14) at (-10.75, -2.25) {};
		\node [style=none] (15) at (-13, -2) {};
		\node [style=none] (16) at (-8.25, -0.75) {};
		\node [style=none] (17) at (-3.25, 1.25) {};
		\node [style=none] (18) at (-1.25, 0.25) {};
		\node [style=none] (19) at (4.5, 0.25) {};
		\node [style=none] (20) at (1.5, 3) {};
		\node [style=none] (21) at (1.5, 7) {};
		\node [style=none] (22) at (2.75, 5) {};
		\node [style=none] (23) at (0.25, 5) {};
		\node [style=none] (24) at (1.5, 6) {};
		\node [style=none] (25) at (1.5, 4.75) {};
		\node [style=none] (26) at (1.25, 3.75) {};
		\node [style=none] (27) at (2, 3.75) {};
		\node [style=none] (28) at (1.25, 2.5) {};
		\node [style=none] (29) at (2, 2.5) {};
		\node [style=none] (30) at (0, 1) {};
		\node [style=none] (31) at (0.25, 0.5) {};
		\node [style=none] (32) at (3.25, 1) {};
		\node [style=none] (33) at (3, 0.5) {};
		\node [style=none] (35) at (-10.75, -0.5) {$\gamma_1$};
		\node [style=none] (37) at (-8.25, -1.5) {$\gamma_2$};
		\node [style=none] (38) at (1.5, 7.5) {$\gamma_1$};
		\node [style=none] (40) at (1, 3) {$\gamma_2$};
		\node [style=none] (41) at (1.75, 3) {};
		\node [style=none] (42) at (-14.75, -3) {};
		\node [style=none] (43) at (-14.75, -3) {$\widetilde{w}_1$};
		\node [style=none] (44) at (-6.75, -3) {$\widetilde{w}_2$};
		\node [style=none] (45) at (-1, 1) {$\widehat{w}_1$};
		\node [style=none] (47) at (4.25, 1) {$\widehat{w}_2$};
		\node [style=none] (48) at (-11.25, -3) {};
		\node [style=none] (49) at (-9.5, -2.75) {};
		\node [style=none] (50) at (-10.75, -2.75) {};
		\node [style=none] (51) at (-10.75, -3.75) {};
		\node [style=none] (52) at (-10.75, -2.25) {};
		\node [style=none] (53) at (-10.75, -4.25) {$\gamma_3$};
		\node [style=none] (54) at (1.5, 4.75) {};
		\node [style=none] (55) at (1.75, 0.5) {};
		\node [style=none] (56) at (1.5, 5.25) {};
		\node [style=none] (57) at (1.75, 0) {$\gamma_3$};
		\node [style=none] (58) at (-2.5, -2.5) {};
		\node [style=none] (59) at (0.75, -1.5) {};
		\node [style=none] (60) at (2.75, -1.5) {};
		\node [style=none] (61) at (6, -2.5) {};
		\node [style=none] (62) at (2.75, -3.5) {};
		\node [style=none] (63) at (0.75, -3.5) {};
		\node [style=none] (64) at (1, -2.75) {};
		\node [style=none] (65) at (2.5, -2.75) {};
		\node [style=none] (66) at (-1.25, -2.25) {};
		\node [style=none] (67) at (-1.25, -2.75) {};
		\node [style=none] (68) at (4.75, -2.75) {};
		\node [style=none] (69) at (4.75, -2.25) {};
		\node [style=none] (70) at (1.75, -1.25) {};
		\node [style=none] (71) at (4, -2) {};
		\node [style=none] (72) at (1.75, -2.25) {};
		\node [style=none] (73) at (-0.5, -2) {};
		\node [style=none] (74) at (-5.75, -2.5) {};
		\node [style=none] (75) at (-3.5, -2.5) {};
		\node [style=none] (76) at (1.75, -0.75) {$\gamma_1$};
		\node [style=none] (77) at (4.25, -1.5) {$\gamma_2$};
		\node [style=none] (78) at (-2.25, -3) {};
		\node [style=none] (79) at (-2.25, -3) {$\widetilde{w}_1$};
		\node [style=none] (80) at (5.75, -3) {$\widetilde{w}_2$};
		\node [style=none] (81) at (1.25, -3.5) {};
		\node [style=none] (82) at (3.25, -3) {};
		\node [style=none] (83) at (1.75, -3.5) {};
		\node [style=none] (84) at (1.75, -3.75) {};
		\node [style=none] (85) at (1.75, -2.25) {};
		\node [style=none] (86) at (1.75, -4.25) {$\gamma_3$};
		\node [style=none] (87) at (1.25, -3.5) {};
		\node [style=none] (105) at (-8.25, -3.75) {};
		\node [style=none] (106) at (-3.25, -6.25) {};
		\node [style=none] (107) at (-2.5, -7) {};
		\node [style=none] (108) at (0.75, -6) {};
		\node [style=none] (109) at (2.75, -6) {};
		\node [style=none] (110) at (6, -7) {};
		\node [style=none] (111) at (2.75, -8) {};
		\node [style=none] (112) at (0.75, -8) {};
		\node [style=none] (113) at (0.75, -7) {};
		\node [style=none] (114) at (2.75, -7) {};
		\node [style=none] (115) at (-1.25, -6.75) {};
		\node [style=none] (116) at (-1.25, -7.25) {};
		\node [style=none] (117) at (4.75, -7.25) {};
		\node [style=none] (118) at (4.75, -6.75) {};
		\node [style=none] (119) at (1.75, -5.75) {};
		\node [style=none] (120) at (4, -6.5) {};
		\node [style=none] (121) at (1.75, -6) {};
		\node [style=none] (122) at (-0.5, -6.5) {};
		\node [style=none] (124) at (1.75, -5.25) {$\gamma_1$};
		\node [style=none] (125) at (4.25, -6) {$\gamma_2$};
			\node [style=none] (125) at (1.75, -8.8) {$\gamma_3$};
		\node [style=none] (126) at (-2.25, -7.5) {};
		\node [style=none] (127) at (-2.25, -7.5) {$\widetilde{w}_1$};
		\node [style=none] (128) at (5.75, -7.5) {$\widetilde{w}_2$};
		\node [style=none] (129) at (1.25, -7.5) {};
		\node [style=none] (130) at (3, -7.25) {};
		\node [style=none] (131) at (1.75, -7.25) {};
		\node [style=none] (132) at (1.75, -8.25) {};
		\node [style=none] (134) at (4.25, -8.25) {};
		\node [style=none] (135) at (-6.75, 0.75) {$\ell_{\gamma_2}\to c_*$};
		\node [style=none] (136) at (-5, -2) {$\ell_{\gamma_3}\to c_*$};
		\node [style=none] (137) at (-6.75, -5.5) {$\ell_{\gamma_1}\to c_*$};
	\end{pgfonlayer}
	\begin{pgfonlayer}{edgelayer}
		\draw [thick, in=-165, out=0] (0.center) to (1.center);
		\draw [thick, in=165, out=15, looseness=1.25] (1.center) to (2.center);
		\draw [thick, in=-150, out=-15, looseness=0.25] (2.center) to (3.center);
		\draw [thick, in=15, out=-180] (3.center) to (4.center);
		\draw [thick, in=-15, out=-165, looseness=1.25] (4.center) to (5.center);
		\draw [thick, in=0, out=165, looseness=0.75] (5.center) to (0.center);
		\draw [thick, bend right=90, looseness=0.50] (6.center) to (7.center);
		\draw [thick, bend left=285, looseness=0.50] (7.center) to (6.center);
		\draw [thick, bend right=60] (8.center) to (9.center);
		\draw [thick, bend right=60] (11.center) to (10.center);
		\draw [thick, style=dashed, bend left=60, looseness=1.25] (11.center) to (10.center);
		\draw [thick, color=red, bend right=75, looseness=0.50] (12.center) to (14.center);
		\draw [thick, color=red, style=dashed, bend left=90, looseness=0.50] (12.center) to (14.center);
		\draw [thick, color=red, style=dashed, in=-105, out=-75] (13.center) to (15.center);
		\draw [very thick, ->] (16.center) to (17.center);
		\draw [thick, in=255, out=15, looseness=0.75] (18.center) to (20.center);
		\draw [thick, in=165, out=15, looseness=0.50] (18.center) to (19.center);
		\draw [thick, in=0, out=75] (22.center) to (21.center);
		\draw [thick, in=-255, out=-180] (21.center) to (23.center);
		\draw [thick, in=90, out=-75] (23.center) to (20.center);
		\draw [thick, bend left=90] (24.center) to (25.center);
		\draw [thick, bend right=90] (24.center) to (25.center);
		\draw [thick, style=dashed, bend left, looseness=0.75] (28.center) to (29.center);
		\draw [thick, bend right=135, looseness=1.50] (28.center) to (29.center);
		\draw [thick, bend right=60] (26.center) to (27.center);
		\draw [thick, style=dashed, bend left=90] (26.center) to (27.center);
		\draw [thick, bend right=90, looseness=0.75] (30.center) to (31.center);
		\draw [thick, style=dashed, bend left=75, looseness=0.75] (30.center) to (31.center);
		\draw [thick, style=dashed, bend left=60, looseness=0.75] (32.center) to (33.center);
		\draw [thick, bend left=240, looseness=1.25] (32.center) to (33.center);
		\draw [thick, color=red, style=dashed, bend left=75, looseness=0.75] (21.center) to (24.center);
		\draw [thick, color=red, bend left=255, looseness=0.75] (21.center) to (24.center);
		\draw [thick, style=dashed, bend left=90] (8.center) to (9.center);
		\draw [thick, in=-105, out=105, looseness=0.50] (41.center) to (22.center);
		\draw [thick, in=150, out=-75] (41.center) to (19.center);
		\draw [thick, color=red, bend right=60, looseness=0.75] (20.center) to (41.center);
		\draw [thick, color=red, in=180, out=45] (15.center) to (48.center);
		\draw [thick, color=red, in=-150, out=-15, looseness=0.75] (48.center) to (49.center);
		\draw [thick, color=red, in=150, out=30] (49.center) to (13.center);
		\draw [thick, color=red, style=dashed, bend left=90, looseness=0.75] (20.center) to (41.center);
		\draw [thick, color=blue, bend right=75, looseness=0.50] (50.center) to (51.center);
		\draw [thick, color=blue, style=dashed, bend left=90, looseness=0.50] (50.center) to (51.center);
		\draw [thick, color=blue, in=-120, out=75, looseness=0.25] (54.center) to (55.center);
		\draw [thick, color=blue, style=dashed, in=75, out=60, looseness=0.25] (54.center) to (55.center);
		\draw [thick, in=-165, out=0] (58.center) to (59.center);
		\draw [thick, in=165, out=15, looseness=1.25] (59.center) to (60.center);
		\draw [thick, in=-150, out=-15, looseness=0.25] (60.center) to (61.center);
		\draw [thick, in=15, out=-180] (61.center) to (62.center);
		\draw [thick, in=-15, out=-165, looseness=1.25] (62.center) to (63.center);
		\draw [thick, in=0, out=165, looseness=0.75] (63.center) to (58.center);
		\draw [thick, bend right=90, looseness=1.75] (64.center) to (65.center);
		\draw [thick, bend left=285] (65.center) to (64.center);
		\draw [thick, bend right=60] (66.center) to (67.center);
		\draw [thick, bend right=60] (69.center) to (68.center);
		\draw [thick, style=dashed, bend left=60, looseness=1.25] (69.center) to (68.center);
		\draw [thick, color=red, bend right=75, looseness=0.50] (70.center) to (72.center);
		\draw [thick, color=red, style=dashed, bend left=90, looseness=0.50] (70.center) to (72.center);
		\draw [thick, color=red, style=dashed, in=-75, out=-105, looseness=1.25] (71.center) to (73.center);
		\draw [very thick, ->] (74.center) to (75.center);
		\draw [thick, style=dashed, bend left=90] (66.center) to (67.center);
		\draw [thick, color=red, in=165, out=45] (73.center) to (81.center);
		\draw [thick, color=red, bend right] (81.center) to (82.center);
		\draw [thick, color=red, in=150, out=30] (82.center) to (71.center);
		\draw [thick, color=blue, bend right=75, looseness=0.50] (83.center) to (84.center);
		\draw [thick, color=blue, style=dashed, bend left=90, looseness=0.50] (83.center) to (84.center);
		\draw [very thick, ->] (105.center) to (106.center);
		\draw [thick, in=-165, out=0] (107.center) to (108.center);
		\draw [thick, in=165, out=15, looseness=1.25] (108.center) to (109.center);
		\draw [thick, in=-150, out=-15, looseness=0.25] (109.center) to (110.center);
		\draw [thick, in=15, out=-180] (110.center) to (111.center);
		\draw [thick, in=-15, out=-165, looseness=1.25] (111.center) to (112.center);
		\draw [thick, in=0, out=165, looseness=0.75] (112.center) to (107.center);
		\draw [thick, bend right=75, looseness=0.50] (113.center) to (114.center);
		\draw [thick, bend right=120, looseness=2.00] (114.center) to (113.center);
		\draw [thick, bend right=60] (115.center) to (116.center);
		\draw [thick, bend right=60] (118.center) to (117.center);
		\draw [thick, style=dashed, bend left=60, looseness=1.25] (118.center) to (117.center);
		\draw [thick, color=red, bend right=75, looseness=0.50] (119.center) to (121.center);
		\draw [thick, color=red, style=dashed, bend left=90, looseness=0.50] (119.center) to (121.center);
		\draw [thick, color=red, style=dashed, in=-105, out=-75] (120.center) to (122.center);
		\draw [thick, style=dashed, bend left=90] (115.center) to (116.center);
		\draw [thick, color=red, in=180, out=45] (122.center) to (129.center);
		\draw [thick, color=red, in=-150, out=-15, looseness=0.75] (129.center) to (130.center);
		\draw [thick, color=red, in=150, out=30] (130.center) to (120.center);
		\draw [thick, color=blue, bend right=75, looseness=0.50] (131.center) to (132.center);
		\draw [thick, color=blue, style=dashed, bend left=90, looseness=0.50] (131.center) to (132.center);
	\end{pgfonlayer}
\end{tikzpicture}
\end{center}

\caption{Curves $\gamma_1,\gamma_2,\gamma_3$ are different non-self intersecting closed geodesics on twice-punctured torus. By shrinking these curves we can reach the boundaries of the string vertex $\mathcal{V}_{1,2}$.}
\label{twiceptorus}
\end{figure}

 \noindent{\underline{\bf Naive interaction vertex $\mathbf{S}_{1,2}$}}: Let us start by constructing the naive one-loop interaction vertex $\mathbf{S}_{1,2}$ with two external states external states represented by the unintegrated vertex operators $V_1$ and $V_2$.   It is given by
\begin{equation}\label{eq:the bosonic-string amplitude}
	\mathbf{S}_{1,2} =(2\pi \mathrm{i})^{-2} \int_{\mathcal{W}_{1,2}} d\ell_{\gamma_1}d\tau_{\gamma_1}d\ell_{\gamma_2}d\tau_{\gamma_2} ~\langle\mathcal{R}_{1,2}|b(\mathbf{t}_{\gamma_1})b(\mathbf{l}_{ \gamma_1}) b(\mathbf{t}_{\gamma_{2}})b(\mathbf{l}_{ \gamma_{2}}) |V_1\rangle_{w_1}\otimes|V_2\rangle_{w_2},
\end{equation}
where $|\mathcal{R}_{1,2}\rangle$ is the surface state associated with the twice-punctured torus, and $|V_i\rangle_{w_i}$ denotes the state inserted at the $i^{\text{th}}$ puncture of the torus using the coordinate $e^{\frac{\pi^2}{c_*}}w_i$ induced from the hyperbolic metric on $\mathcal{R}_{1,2}$. The parameters $(\tau_{\gamma_j},\ell_{\gamma_j}),~j=1,2$ denote the Fenchel-Nielsen coordinates for the Teichm\"uller space $\mathcal{T}_{1,2}$ of twice-punctured tori defined with respect to the curves $\gamma_1$ and $\gamma_2$, see figure \ref{twiceptorus}. And 
\begin{alignat}{1}\label{eq:the expressions for b(v)}
	b(\mathbf{t}_{\gamma_i})&=\int_{\mathcal{F}} d^2z\left(b_{zz}\mathbf{t}_{\gamma_i}+b_{\bar z\bar z}\overline{\mathbf{t}}_{\gamma_i}\right),\nonumber
	\\
	b(\mathbf{l}_{\gamma_i})&=\int_{\mathcal{F}} d^2z\left(b_{zz}\mathbf{l}_{\gamma_i}+b_{\bar z\bar z}\overline{\mathbf{l}}_{\gamma_i}\right),
\end{alignat}
where $\mathcal{F}$ denotes the fundamental domain  of the action of  $\Gamma_{1,2}$, the Fuchsian group associated with $\mathcal{R}_{1,2}$, in $\mathbb{H}$. $\mathbf{t}_{\gamma_i}$ and $\mathbf{l}_{\gamma_i}$ are the Beltrami differentials associated with the Fenchel-Nielsen coordinates $(\tau_{\gamma_i},\ell_{\gamma_i})$. Finally, $\mathcal{W}_{1,2}$ is the region covered by the naive string vertex $\mathcal{V}_{1,2}^0$ in the moduli space. Although a copy of $\mathcal{W}_{1,2}$ is a subspace in $\mathcal{T}_{1,2}$, it has no simple descriptionin terms of the Fenchel-Nielsen coordinates. \par

 In order to evaluate $\mathbf{S}_{1,2}$ we must specify $\mathcal{W}_{1,2}$ in terms of the Fenchel-Nielsen cordinates. This seems impossible, since there is no simple description of $\mathcal{W}_{1,2}$ or even $\mathcal{M}_{1,2}$  in terms of $(\tau_{\gamma_1},\ell_{\gamma_1},\tau_{\gamma_2},\ell_{\gamma_2})$. However, there is an interesting to resolution to this issue. The lengths of the non-self intersecting closed geodesics on $\mathcal{R}_{1,2}$ satisfy the following curious identity \cite{McShane1}:
   \begin{equation}\label{gmidentityp}
\sum_{g_1\in \text{MCG}(\mathcal{R}_{1,2},\gamma_1+\gamma_3)}\frac{2}{1+e^{\frac{\ell_{g_1\cdot\gamma_1}+\ell_{g_1\cdot\gamma_3}}{2}}}+\sum_{g_2\in \text{MCG}(\mathcal{R}_{1,2},\gamma_2)}\frac{2}{1+e^{\frac{l_{g_2\cdot\gamma_2}}{2}}}=1,
  \end{equation}
where $\gamma_1,\gamma_2$ and $\gamma_3$ are the non-self intersecting closed geodesics on $\mathcal{R}_{1,2}$ as shown in figure \ref{twiceptorus}, and $\ell_{\gamma_i}$ denotes the hyperbolic length of $\gamma_i$. $\text{MCG}(\mathcal{R}_{1,2},\gamma_1+\gamma_3)$ denotes the subgroup of mapping class group (MCG) of $\mathcal{R}_{1,2}$ that acts non-trivially only on the curve $\gamma_1+\gamma_3$. Similarly, $\text{MCG}(\mathcal{R}_{1,2},\gamma_2)$ denotes the subgroup of MCG of $\mathcal{R}_{1,2}$ that acts non-trivially only on the curve $\gamma_2$. \par

The  MCG group $\text{MCG}(\mathcal{R}_{1,2})$ can be factorized in different ways as follows:
\begin{align}\label{MCGfactR12}
\text{MCG}(\mathcal{R}_{1,2})&=\text{MCG}(\mathcal{R}_{1,2},\gamma_1+\gamma_3)\times \text{Dehn}(\gamma_1)\times \text{Dehn}(\gamma_3),\nonumber\\
\text{MCG}(\mathcal{R}_{1,2})&=\text{MCG}(\mathcal{R}_{1,2},\gamma_2)\times \text{Dehn}^*(\gamma_2)\times \text{MCG}(\mathcal{R}_{1,1}(\ell_{\gamma_2})),
\end{align}
where $ \text{MCG}(\mathcal{R}_{1,1}(\ell_{\gamma_2}))$ denotes the MCG of the torus  $\mathcal{R}_{1,1}(\ell_{\gamma_2})$ with a border having length $\ell_{\gamma_2}$. $\text{Dehn}(\gamma_i)$ denotes the group generated by the Dehn twist $\tau_{\gamma_i}\to \tau_{\gamma_i}+\ell_{\gamma_i}$ and $\text{Dehn}^*(\gamma_i)$ denotes the group generated by the half Dehn twist $\tau_{\gamma_i}\to \tau_{\gamma_i}+\frac{1}{2}\ell_{\gamma_i}$. Interestingly, the lengths of the non-self intersecting closed geodesics on $\mathcal{R}_{1,1}(\ell_{\gamma_2})$  also satisfy an identity of the kind (\ref{gmidentityp}) \cite{Mirzakhani:2006fta}:
\begin{equation}\label{torusMidentity}
\sum_{g\in \text{MCG}(\mathcal{R}_{1,1}(\ell_{\gamma_2}))}\left[1-\frac{1}{\ell_{\gamma_2}}\mathrm{ln}\left(\frac{\mathrm{cosh}(\frac{\ell_{g\cdot\gamma_1}}{2})+\mathrm{cosh}(\frac{\ell_{\gamma_2}+\ell_{g\cdot\gamma_1}}{2})}{\mathrm{cosh}(\frac{\ell_{g\cdot\gamma_1}}{2})+\text{cosh}(\frac{\ell_{\gamma_2}-\ell_{g\cdot\gamma_1}}{2})}\right)\right]=1.
\end{equation}
We also have an identity that involves the sum over all images of the elements in the group $\text{Dehn}(\gamma_i)$, and is given by
\begin{equation}\label{Dehn}
\sum_{g\in \text{Dehn}(\gamma_i)}\text{sinc}^2\left(\frac{\tau_{g\cdot\gamma_i}}{l_{g\cdot\gamma_i}}\right)= \sum_{g\in \text{Dehn}_*(\gamma_i)}\text{sinc}^2\left(\frac{2\tau_{g\cdot\gamma_i}}{l_{g\cdot\gamma_i}}\right)=1,
\end{equation}
where $\text{sinc}(x)=\frac{\text{sin}\pi x}{\pi x}$. The identity (\ref{Dehn}) can be verified using the following well known identity
  \begin{equation}\label{identity}
	\sum_{k=-\infty}^{\infty}\text{sinc}^2\left(x-k \right)=1\qquad x\in \mathbb{R}.
	\end{equation}
	Combining the identities (\ref{gmidentityp},\ref{torusMidentity}, \ref{Dehn}) give the following identity
	  \begin{equation}\label{gmidentityp1}
\sum_{g\in \text{MCG}(\mathcal{R}_{1,2})}G_1(\ell_{g\cdot\gamma_1},\tau_{g\cdot\gamma_1},\ell_{g\cdot\gamma_3},\tau_{g\cdot\gamma_3})+\sum_{g\in \text{MCG}(\mathcal{R}_{1,2})}G_1(\ell_{g\cdot\gamma_1},\tau_{g\cdot\gamma_1},\ell_{g\cdot\gamma_2},\tau_{g\cdot\gamma_2})=1,
  \end{equation}
	where $G_1$ and $G_2$ are given by
	  \begin{align}\label{gmidentityp2}
G_1(\ell_{\gamma_1},\tau_{\gamma_1},\ell_{\gamma_3},\tau_{\gamma_3})&=\frac{2~\text{sinc}^2\left(\frac{\tau_{\gamma_1}}{\ell_{\gamma_1}}\right)\text{sinc}^2\left(\frac{\tau_{\gamma_3}}{l_{\gamma_3}}\right)}{1+e^{\frac{\ell_{\gamma_1}+\ell_{\gamma_3}}{2}}}, \nonumber\\
G_2(\ell_{\gamma_1},\tau_{\gamma_1},\ell_{\gamma_2},\tau_{\gamma_2})&=\frac{2~\text{sinc}^2\left(\frac{2\tau_{\gamma_2}}{l_{\gamma_2}}\right)\left[1-\frac{1}{\ell_{\gamma_2}}\mathrm{ln}\left(\frac{\mathrm{cosh}(\frac{\ell_{\gamma_1}}{2})+\mathrm{cosh}(\frac{\ell_{\gamma_2}+\ell_{\gamma_1}}{2})}{\mathrm{cosh}(\frac{\ell_{\gamma_1}}{2})+\text{cosh}(\frac{\ell_{\gamma_2}-\ell_{\gamma_1}}{2})}\right)\right]}{1+e^{\frac{l_{\gamma_2}}{2}}}.
  \end{align}
  Notice that the functions $G_1$ and $G_2$ have the following decaying property
  \begin{equation}\label{decay}
\lim_{\ell_{\gamma_3\to \frac{1}{c_*}}}G_1(\ell_{\gamma_1},\tau_{\gamma_1},\ell_{\gamma_3},\tau_{\gamma_3})=\mathcal{O}(e^{-1/c_*}),\qquad \lim_{\ell_{\gamma_2\to \frac{1}{c_*}}}G_2(\ell_{\gamma_1},\tau_{\gamma_1},\ell_{\gamma_2},\tau_{\gamma_2})=\mathcal{O}(e^{-1/c_*}).
  \end{equation}
  Using the identity (\ref{gmidentityp1}) we can express the amplitude $\mathbf{S}_{1,2}$ as an integral over the Teichm\"uller space $\mathcal{T}_{1,2}$ of twice-punctured tori as follows:
  \begin{align}\label{eq:the bosonic-string amplitude1}
	\mathbf{S}_{1,2}&=(2\pi \mathrm{i})^{-2} \int_{\mathcal{TW}^{\mathbf{P}_1}_{1,2}} d\ell_{\gamma_1}d\tau_{\gamma_1}d\ell_{\gamma_2}d\tau_{\gamma_2} ~G_1(\ell_{\gamma_1},\tau_{\gamma_1},\ell_{\gamma_2},\tau_{\gamma_2})\langle\mathcal{R}_{1,2}|b(\mathbf{t}_{\gamma_1})b(\mathbf{l}_{ \gamma_1}) b(\mathbf{t}_{\gamma_{2}})b(\mathbf{l}_{ \gamma_{2}}) |{V}_1\rangle_{w_1}\otimes|{V}_2\rangle_{w_2}\nonumber\\
	&+(2\pi \mathrm{i})^{-2} \int_{\mathcal{TW}^{\mathbf{P}_2}_{1,2}} d\ell_{\gamma_1}d\tau_{\gamma_1}d\ell_{\gamma_3}d\tau_{\gamma_3} ~G_2(\ell_{\gamma_1},\tau_{\gamma_1},\ell_{\gamma_3},\tau_{\gamma_3})\langle\mathcal{R}_{1,2}|b(\mathbf{t}_{\gamma_1})b(\mathbf{l}_{ \gamma_1}) b(\mathbf{t}_{\gamma_{3}})b(\mathbf{l}_{ \gamma_{3}}) |{V}_1\rangle_{w_1}\otimes|{V}_2\rangle_{w_2},
\end{align}
where $\mathcal{TW}^{\mathbf{P}_1}_{1,2}$ is the image of $\mathcal{W}_{1,2}$ in the Teichm\"uller space defined with respect to the pair of pants decomposition $\mathbf{P}_1$ given by the curves $\gamma_1$ and $\gamma_2$.   $\mathcal{TW}^{\mathbf{P}_2}_{1,2}$ is the union of all the images of $\mathcal{W}_{1,2}$ in the Teichm\"uller space defined with respect to the pair of pants decomposition $\mathbf{P}_2$ given by the curves $\gamma_1$ and $\gamma_3$.  Although $\mathcal{TW}^{\mathbf{P}_1}_{1,2}$ and $\mathcal{TW}^{\mathbf{P}_2}_{1,2}$ do not have a nice description, the decay behaviour of  the functions  $G_1$ and $G_2$ (\ref{decay})  allows us to replace them with the effective regions  $E\mathcal{W}^{\mathbf{P}_1}_{1,2}$ and  $E\mathcal{W}^{\mathbf{P}_2}_{1,2}$  without changing the value of $\mathbf{S}_{1,2}$. The string vertex region $\mathcal{W}_{1,2}$ has the property that it does not contain any hyperbolic Riemann surface having simple closed geodesics with length less than $c_*$. Consequently, $\mathbf{S}_{1,2}$ computed by integrating the off-shell string measure over  $\mathcal{W}_{1,2}$ does not receive any contribution from  hyperbolic Riemann surfaces having a simple closed geodesics with length less than $c_*$.  Therefore, $\mathbf{S}_{1,2}$ computed by integrating the differential form in $\mathcal{T}_{1,2}$ over $E\mathcal{W}^{\mathbf{P}_1}_{1,2}$ must also not receive any finite contribution from such surfaces. This is true if we identify $E\mathcal{W}^{\mathbf{P}_1}_{1,2}$ with the following region in $\mathcal{T}_{1,2}$ 
\begin{equation}
E\mathcal{W}^{\mathbf{P}_1}_{1,2}:\qquad \ell_{\gamma_1}\in [c_*,\infty) \qquad  \ell_{\gamma_2}\in [c_*,\infty), \qquad \tau_{\gamma_1}\in(-\infty,\infty), \qquad \tau_{\gamma_2}\in (-\infty,\infty),
\end{equation}
and $E\mathcal{W}^{\mathbf{P}_2}_{1,2}$ with the following region 
\begin{equation}
E\mathcal{W}^{\mathbf{P}_2}_{1,2}:\qquad \ell_{\gamma_1}\in [c_*,\infty) \qquad  \ell_{\gamma_3}\in [c_*,\infty), \qquad \tau_{\gamma_1}\in(-\infty,\infty), \qquad \tau_{\gamma_3}\in (-\infty,\infty).
\end{equation}
 Notice that the region $E\mathcal{W}^{\mathbf{P}_1}_{1,2}$  includes hyperbolic Riemann surfaces with simple closed geodesic $\gamma_3$ having length less than $c_*$. Interestingly, when $\ell_{\gamma_3}\to c_*$ the length of $\gamma_2$ decay very fast and the function $G_2$ exponentially decays. As a result, the integration over region  $E\mathcal{W}^{\mathbf{P}_1}_{1,2}$ does not include any finite contribution  from  hyperbolic Riemann surfaces with simple closed geodesic $\gamma_3$ having length less than $c_*$. Similar statement is true for the integration over $E\mathcal{W}^{\mathbf{P}_2}_{1,2}$. Then we can write $\mathbf{S}_{1,2}$ as
  \begin{align}\label{eq:the bosonic-string amplitude1}
	\mathbf{S}_{1,2}&=(2\pi \mathrm{i})^{-2} \int_{E\mathcal{W}^{\mathbf{P}_1}_{1,2}} d\ell_{\gamma_1}d\tau_{\gamma_1}d\ell_{\gamma_2}d\tau_{\gamma_2} ~G_1(\ell_{\gamma_1},\tau_{\gamma_1},\ell_{\gamma_2},\tau_{\gamma_2})\langle\mathcal{R}_{1,2}|b(\mathbf{t}_{\gamma_1})b(\mathbf{l}_{ \gamma_1}) b(\mathbf{t}_{\gamma_{2}})b(\mathbf{l}_{ \gamma_{2}}) |{V}_1\rangle_{w_1}\otimes|{V}_2\rangle_{w_2}\nonumber\\
	&+(2\pi \mathrm{i})^{-2} \int_{E\mathcal{W}^{\mathbf{P}_2}_{1,2}} d\ell_{\gamma_1}d\tau_{\gamma_1}d\ell_{\gamma_3}d\tau_{\gamma_3} ~G_2(\ell_{\gamma_1},\tau_{\gamma_1},\ell_{\gamma_3},\tau_{\gamma_3})\langle\mathcal{R}_{1,2}|b(\mathbf{t}_{\gamma_1})b(\mathbf{l}_{ \gamma_1}) b(\mathbf{t}_{\gamma_{3}})b(\mathbf{l}_{ \gamma_{3}}) |{V}_1\rangle_{w_1}\otimes|{V}_2\rangle_{w_2},
\end{align}

 \noindent{\underline{\bf Corrected interaction vertex $\widetilde{\mathbf{S}}_{1,2}$}}: The naive interaction vertex  $\mathbf{S}_{1,2}$ must be modified to make them suitable for constructing a string field theory with approximate gauge invariance. The modification can be implemented once we specify the subregions $\mathbf{W}^{(0)}_{1,2}, \mathbf{W}^{(1)}_{1,2}$ and $\mathbf{W}^{(2)}_{1,2}$ inside $\mathcal{W}_{1,2}$. \par
 
 The subregion $\mathbf{W}^{(0)}_{1,2}$ has the property that it does not include any hyperbolic Riemann surface with one or more simple closed geodesic having length less than $c_*(1+\delta)$. Let us denote the union of all the images of $\mathbf{W}^{(0)}_{1,2}$ in $\mathcal{T}_{1,2}$ defined with respect to the pants decomposition $\mathbf{P}_1$ as $\mathcal{T}\mathbf{W}^{\mathbf{P_1},(0)}_{1,2}$.  For $\mathcal{T}_{1,2}$ defined with respect to the pants decomposition $\mathbf{P}_2$,  the union of all images of $\mathbf{W}^{(0)}_{1,2}$ is denoted as $\mathcal{T}\mathbf{W}^{\mathbf{P_2},(0)}_{1,2}$. Then by repeating the arguments in the previous paragraph we can identify the effective region $E\mathbf{W}^{\mathbf{P}^1,(0)}_{1,2}$ in $\mathcal{T}_{1,2}$ that corresponds to $\mathcal{T}\mathbf{W}^{\mathbf{P}^1,(0)}_{1,2}$ with the following region 
 \begin{equation}
 E\mathbf{W}^{\mathbf{P}^1,(0)}_{1,2}: \quad \ell_{\gamma_1}\in [c_*(1+\delta),\infty), \quad  \ell_{\gamma_2}\in [c_*(1+\delta),\infty), \quad \tau_{\gamma_1}\in (-\infty,\infty), \quad \tau_{\gamma_2}(-\infty,\infty).
 \end{equation}
 Similarly, we can identify the effective region $E\mathbf{W}^{\mathbf{P}^2,(0)}_{1,2}$ that corresponds to $\mathcal{T}\mathbf{W}^{\mathbf{P}^2,(0)}_{1,2}$ with the following region 
  \begin{equation}
 E\mathbf{W}^{\mathbf{P}^2,(0)}_{1,2}: \quad \ell_{\gamma_1}\in [c_*(1+\delta),\infty),\quad  \ell_{\gamma_3}\in [c_*(1+\delta),\infty), \quad \tau_{\gamma_1}\in (-\infty,\infty), \quad \tau_{\gamma_3}(-\infty,\infty).
 \end{equation}

Now let us analyze the subregion $\mathbf{W}^{(1)}_{1,2}$. It has the property that any hyperbolic Riemann surface in this region has only one simple closed geodesic having length between $c_*$ and $c_*(1+\delta)$.   Let us denote the union of all the images of $\mathbf{W}^{(1)}_{1,2}$ in $\mathcal{T}_{1,2}$ defined with respect to the pants decomposition $\mathbf{P}_1$ as $\mathcal{T}\mathbf{W}^{\mathbf{P_1},(1)}_{1,2}$ and that defined with respect to the pants decomposition $\mathbf{P}_2$ as $\mathcal{T}\mathbf{W}^{\mathbf{P_2},(1)}_{1,2}$. We can identify the effective regions correspond to 
$\mathcal{T}\mathbf{W}^{\mathbf{P_1},(1)}_{1,2}$ and $\mathcal{T}\mathbf{W}^{\mathbf{P_2},(1)}_{1,2}$ as follows:
\begin{align}
 E\mathbf{W}^{\mathbf{P}^1,(1)}_{1,2}&= E\mathbf{W}^{\mathbf{P}^1,\gamma_1}_{1,2}\cup E\mathbf{W}^{\mathbf{P}^1,\gamma_2}_{1,2}\nonumber\\
 E\mathbf{W}^{\mathbf{P}^2,(1)}_{1,2} &= E\mathbf{W}^{\mathbf{P}^2,\gamma_1}_{1,2}\cup E\mathbf{W}^{\mathbf{P}^2,\gamma_3}_{1,2}
 \end{align}
 where
 \begin{align}
 E\mathbf{W}^{\mathbf{P}^1,\gamma_1}_{1,2}&=  \ell_{\gamma_1}\in [c_*,c_*(1+\delta)),\quad  \ell_{\gamma_2}\in [c_*(1+\delta),\infty), \quad \tau_{\gamma_1}\in (-\infty,\infty), \quad \tau_{\gamma_2}(-\infty,\infty), \nonumber\\
  E\mathbf{W}^{\mathbf{P}^1,\gamma_2}_{1,2}&=\ell_{\gamma_1}\in [c_*(1+\delta),\infty),\quad  \ell_{\gamma_2}\in [c_*,c_*(1+\delta)), \quad \tau_{\gamma_1}\in (-\infty,\infty), \quad \tau_{\gamma_2}(-\infty,\infty), \nonumber\\
   E\mathbf{W}^{\mathbf{P}^2,\gamma_1}_{1,2}&=  \ell_{\gamma_1}\in [c_*,c_*(1+\delta)),\quad  \ell_{\gamma_3}\in [c_*(1+\delta),\infty), \quad \tau_{\gamma_1}\in (-\infty,\infty), \quad \tau_{\gamma_3}(-\infty,\infty), \nonumber\\
  E\mathbf{W}^{\mathbf{P}^2,\gamma_3}_{1,2}&=\ell_{\gamma_1}\in [c_*(1+\delta),\infty),\quad  \ell_{\gamma_3}\in [c_*,c_*(1+\delta)), \quad \tau_{\gamma_1}\in (-\infty,\infty), \quad \tau_{\gamma_3}(-\infty,\infty),
\end{align}

Finally, the effective regions $E\mathbf{W}^{\mathbf{P}^1, \gamma_1,\gamma_2}_{1,2}$ and $E\mathbf{W}^{\mathbf{P}^2, \gamma_1,\gamma_3}_{1,2}$ for the subregion $\mathbf{W}^{(2)}_{1,2}$  in $\mathcal{T}_{1,2}$ defined with respect to the pants decomposition $\mathbf{P}^1$ and $\mathbf{P}^2$ respectively are given by
\begin{align}
 E\mathbf{W}^{\mathbf{P}^1,\gamma_1,\gamma_2}_{1,2}&=  \ell_{\gamma_1}\in [c_*,c_*(1+\delta)),\quad  \ell_{\gamma_2}\in [c_*c_*(1+\delta)), \quad \tau_{\gamma_1}\in (-\infty,\infty), \quad \tau_{\gamma_2}(-\infty,\infty), \nonumber\\
  E\mathbf{W}^{\mathbf{P}^2,\gamma_1,\gamma_3}_{1,2}&=\ell_{\gamma_1}\in [c_*,c_*(1+\delta)),\quad  \ell_{\gamma_3}\in [c_*c_*(1+\delta)), \quad \tau_{\gamma_1}\in (-\infty,\infty), \quad \tau_{\gamma_3}(-\infty,\infty).
\end{align}
  Given, that we have identified the effective regions for the subregions in $\mathcal{W}_{1,2}$, let us construct the corrected interaction vertex $\widetilde{\mathbf{S}}_{1,2}$. It is given by 
  \begin{align}\label{eq:the bosonic-string amplitude2}
	\widetilde{\mathbf{S}}_{1,2}&= \int_{ E\mathbf{W}^{\mathbf{P}^1,(0)}_{1,2}} \frac{d\ell_{\gamma_i}d\tau_{\gamma_i}d\ell_{\gamma_2}d\tau_{\gamma_2}}{(2\pi \mathrm{i})^{2}} ~G_1(\ell_{\gamma_1},\tau_{\gamma_1},\ell_{\gamma_2},\tau_{\gamma_2})\langle\mathcal{R}_{1,2}|b(\mathbf{t}_{\gamma_1})b(\mathbf{l}_{ \gamma_1}) b(\mathbf{t}_{\gamma_{2}})b(\mathbf{l}_{ \gamma_{2}}) |{V}_1\rangle_{\widetilde{w}_1}\otimes|{V}_2\rangle_{\widetilde{w}_2}\nonumber\\
	&+ \int_{ E\mathbf{W}^{\mathbf{P}^2,(0)}_{1,2}} \frac{d\ell_{\gamma_1}d\tau_{\gamma_1}d\ell_{\gamma_3}d\tau_{\gamma_3}}{(2\pi \mathrm{i})^{2}} ~G_2(\ell_{\gamma_1},\tau_{\gamma_1},\ell_{\gamma_3},\tau_{\gamma_3})\langle\mathcal{R}_{1,2}|b(\mathbf{t}_{\gamma_1})b(\mathbf{l}_{ \gamma_1}) b(\mathbf{t}_{\gamma_{3}})b(\mathbf{l}_{ \gamma_{3}}) |{V}_1\rangle_{{\widetilde{w}}_1}\otimes|{V}_2\rangle_{{\widetilde{w}}_2}\nonumber\\
	&+ \int_{ E\mathbf{W}^{\mathbf{P}^1,\gamma_1}_{1,2}} \frac{d\ell_{\gamma_1}d\tau_{\gamma_1}d\ell_{\gamma_2}d\tau_{\gamma_2}}{(2\pi \mathrm{i})^{2}} ~G_1(\ell_{\gamma_1},\tau_{\gamma_1},\ell_{\gamma_2},\tau_{\gamma_2})\langle\mathcal{R}_{1,2}|b(\mathbf{t}_{\gamma_1})b(\mathbf{l}_{ \gamma_1}) b(\mathbf{t}_{\gamma_{2}})b(\mathbf{l}_{ \gamma_{2}}) |{V}_1\rangle_{{\widetilde{w}}^{\gamma_1}_1}\otimes|{V}_2\rangle_{{\widetilde{w}}^{\gamma_1}_2}\nonumber\\
	&+ \int_{ E\mathbf{W}^{\mathbf{P}^1,\gamma_2}_{1,2}} \frac{d\ell_{\gamma_1}d\tau_{\gamma_1}d\ell_{\gamma_2}d\tau_{\gamma_2}}{(2\pi \mathrm{i})^{2}} ~G_1(\ell_{\gamma_1},\tau_{\gamma_1},\ell_{\gamma_2},\tau_{\gamma_2})\langle\mathcal{R}_{1,2}|b(\mathbf{t}_{\gamma_1})b(\mathbf{l}_{ \gamma_1}) b(\mathbf{t}_{\gamma_{2}})b(\mathbf{l}_{ \gamma_{2}}) |{V}_1\rangle_{{\widetilde{w}}^{\gamma_2}_1}\otimes|{V}_2\rangle_{{\widetilde{w}}^{\gamma_2}_2}\nonumber\\
	&+ \int_{ E\mathbf{W}^{\mathbf{P}^2,\gamma_1}_{1,2}} \frac{d\ell_{\gamma_1}d\tau_{\gamma_1}d\ell_{\gamma_3}d\tau_{\gamma_3}}{(2\pi \mathrm{i})^{2}} ~G_2(\ell_{\gamma_1},\tau_{\gamma_1},\ell_{\gamma_3},\tau_{\gamma_3})\langle\mathcal{R}_{1,2}|b(\mathbf{t}_{\gamma_1})b(\mathbf{l}_{ \gamma_1}) b(\mathbf{t}_{\gamma_{3}})b(\mathbf{l}_{ \gamma_{3}}) |{V}_1\rangle_{{\widetilde{w}}^{\gamma_1}_1}\otimes|{V}_2\rangle_{{\widetilde{w}}^{\gamma_1}_2}\nonumber\\
	&+\int_{ E\mathbf{W}^{\mathbf{P}^2,\gamma_3}_{1,2}} \frac{d\ell_{\gamma_1}d\tau_{\gamma_1}d\ell_{\gamma_3}d\tau_{\gamma_3}}{(2\pi \mathrm{i})^{2}} ~G_2(\ell_{\gamma_1},\tau_{\gamma_1},\ell_{\gamma_3},\tau_{\gamma_3})\langle\mathcal{R}_{1,2}|b(\mathbf{t}_{\gamma_1})b(\mathbf{l}_{ \gamma_1}) b(\mathbf{t}_{\gamma_{2}})b(\mathbf{l}_{ \gamma_{2}}) |{V}_1\rangle_{{\widetilde{w}}^{\gamma_3}_1}\otimes|{V}_2\rangle_{{\widetilde{w}}^{\gamma_3}_2}\nonumber\\
&+ \int_{ E\mathbf{W}^{\mathbf{P}^1,\gamma_1,\gamma_2}_{1,2}} \frac{d\ell_{\gamma_1}d\tau_{\gamma_1}d\ell_{\gamma_2}d\tau_{\gamma_2}}{(2\pi \mathrm{i})^{2}} ~G_1(\ell_{\gamma_1},\tau_{\gamma_1},\ell_{\gamma_2},\tau_{\gamma_2})\langle\mathcal{R}_{1,2}|b(\mathbf{t}_{\gamma_1})b(\mathbf{l}_{ \gamma_1}) b(\mathbf{t}_{\gamma_{2}})b(\mathbf{l}_{ \gamma_{2}}) |{V}_1\rangle_{{\widetilde{w}}^{\gamma_1\gamma_2}_1}\otimes|{V}_2\rangle_{{\widetilde{w}}^{\gamma_1\gamma_2}_2}\nonumber\\
	&+ \int_{ E\mathbf{W}^{\mathbf{P}^2,\gamma_1,\gamma_3}_{1,2}} \frac{d\ell_{\gamma_1}d\tau_{\gamma_1}d\ell_{\gamma_3}d\tau_{\gamma_3}}{(2\pi \mathrm{i})^{2}} ~G_2(\ell_{\gamma_1},\tau_{\gamma_1},\ell_{\gamma_3},\tau_{\gamma_3})\langle\mathcal{R}_{1,2}|b(\mathbf{t}_{\gamma_1})b(\mathbf{l}_{ \gamma_1}) b(\mathbf{t}_{\gamma_{2}})b(\mathbf{l}_{ \gamma_{2}}) |{V}_1\rangle_{{\widetilde{w}}^{\gamma_1\gamma_3}_1}\otimes|{V}_2\rangle_{{\widetilde{w}}^{\gamma_1\gamma_3}_2}.
\end{align}
 The local coordinates are as follows
 \begin{align}
  \widetilde{w}_j&=e^{\frac{\pi^2}{c_*}}w_j,\nonumber\\
   \widetilde{w}^{\gamma_1}_j&=e^{\frac{c_*^2}{6}f(\ell_{\gamma_1})Y^{(1)}_{1j}}e^{\frac{\pi^2}{c_*}}w_{j},\nonumber\\
     \widetilde{w}^{\gamma_2}_j&=e^{\frac{c_*^2}{6}f(\ell_{\gamma_2})Y^{(1)}_{2j}}e^{\frac{\pi^2}{c_*}}w_{j},\nonumber\\
          \widetilde{w}^{\gamma_3}_j&=e^{\frac{c_*^2}{6}f(\ell_{\gamma_3})Y^{(1)}_{3j}}e^{\frac{\pi^2}{c_*}}w_{j},\nonumber\\
      \widetilde{w}^{\gamma_1\gamma_2}_j&=e^{\frac{c_*^2}{6}\left[f(\ell_{\gamma_1})Y_{1j}^{(2)}+f(\ell_{\gamma_2})Y^{(2)}_{2j}\right]}e^{\frac{\pi^2}{c_*}}w_{j}\nonumber\\
         \widetilde{w}^{\gamma_1\gamma_3}_j&=e^{\frac{c_*^2}{6}\left[f(\ell_{\gamma_1})Y_{1j}^{(2)}+f(\ell_{\gamma_3})Y^{(2)}_{3j}\right]}e^{\frac{\pi^2}{c_*}}w_{j}\nonumber
  \end{align}

where $f$ is an arbitrary smooth real function of the geodesic length defined in the interval $\left(c_*,c_*+\delta c_*\right)$, such that $f(c_*)=1$ and $f(c_*+\delta c_*)=0$. The coefficient $Y^{(1)}_{1j}$ is given by
\begin{align}
Y^{(1)}_{1j}&=\sum_{q=1}^2\sum_{c_i^q,d_i^q}\frac{\pi^2}{|c_i^q|^4}\nonumber\\ 
c_i^q>0 \qquad &d_i^q~\text{mod}~c_i^q \qquad\left(\begin{array}{cc}* & * \\c_i^q & d_i^q\end{array}\right)\in \quad (\sigma_{\gamma_1}^q)^{-1}\Gamma_{0,4}\sigma_j.
\end{align} 
where the transformation $\sigma_j^{-1}$ maps the cusp corresponding to the $j^{th}$ puncture to $\infty$ and $(\sigma_{\gamma_1}^q)^{-1}$ maps the cusp corresponds to the  one of the two punctures, marked as $q$, obtained by degenerating the curve $\gamma_1$, to  $\infty$.  $\Gamma_{0,4}$ is the Fuchisan group of a four punctured hyperbolic Riemann surface with Fenchel-Nielsen parameters $(\ell_{\gamma_1},\tau_{\gamma_1},\ell_{\gamma_2},\tau_{\gamma_2})$.$Y^{(1)}_{1j}$ is given by
\begin{align}
Y^{(1)}_{2j}&=\sum_{q=1}^2\sum_{c_i^q,d_i^q}\frac{\pi^2}{|c_i^q|^4}\nonumber\\ 
c_i^q>0 \qquad &d_i^q~\text{mod}~c_i^q \qquad\left(\begin{array}{cc}* & * \\c_i^q & d_i^q\end{array}\right)\in \quad (\sigma_{\gamma_2}^q)^{-1}\Gamma_{0,4}\sigma_j.
\end{align} 
where  $(\sigma_{\gamma_2}^q)^{-1}$ maps the cusp corresponds to the  one of the two punctures, marked as $q$, obtained by degenerating the curve $\gamma_2$, to  $\infty$.   $Y^{(1)}_{3j}$ is given by
\begin{align}
Y^{(1)}_{3j}&=\sum_{q=1}^2\sum_{c_i^q,d_i^q}\frac{\pi^2}{|c_i^q|^4}\nonumber\\ 
c_i^q>0 \qquad &d_i^q~\text{mod}~c_i^q \qquad\left(\begin{array}{cc}* & * \\c_i^q & d_i^q\end{array}\right)\in \quad (\sigma_{\gamma_3}^q)^{-1}\Gamma_{0,4}\sigma_j,
\end{align} 
where   $(\sigma_{\gamma_3}^q)^{-1}$ maps the cusp corresponds to the  one of the two punctures, marked as $q$, obtained by degenerating the curve $\gamma_3$, to  $\infty$.  $Y^{(2)}_{1j}$ is given by
\begin{align}
Y^{(2)}_{1j}&=\sum_{q=1}^2\sum_{c_i^q,d_i^q}\pi^{2}\frac{\epsilon(j,q)}{|c_i^q|^4}\nonumber\\ 
c_i^q>0 \qquad &d_i^q~\text{mod}~c_i^q \qquad\left(\begin{array}{cc}* & * \\c_i^q & d_i^q\end{array}\right)\in \quad (\sigma_{\gamma_1}^q)^{-1}\Gamma_{0,3}\sigma_j,
\end{align} 
  where $\Gamma_{0,3}$ is the Fuchsian group of a thrice punctured hyperbolic sphere.  The factor $\epsilon(j,q)$ is one if both the $j^{th}$ puncture and the puncture denoted by the index $q$ obtained by degenerating the curve $\gamma_1$ belong to the same thrice punctured sphere, other wise $\epsilon(j,q)$ is zero.  $Y^{(2)}_{2j}$ is given by
\begin{align}
Y^{(2)}_{2j}&=\sum_{q=1}^2\sum_{c_i^q,d_i^q}\pi^{2}\frac{\epsilon(j,q)}{|c_i^q|^4}\nonumber\\ 
c_i^q>0 \qquad &d_i^q~\text{mod}~c_i^q \qquad\left(\begin{array}{cc}* & * \\c_i^q & d_i^q\end{array}\right)\in \quad (\sigma_{\gamma_2}^q)^{-1}\Gamma_{0,3}\sigma_j,
\end{align} 
and  $Y^{(2)}_{3j}$ is given by
\begin{align}
Y^{(2)}_{3j}&=\sum_{q=1}^2\sum_{c_i^q,d_i^q}\pi^{2}\frac{\epsilon(j,q)}{|c_i^q|^4}\nonumber\\ 
c_i^q>0 \qquad &d_i^q~\text{mod}~c_i^q \qquad\left(\begin{array}{cc}* & * \\c_i^q & d_i^q\end{array}\right)\in \quad (\sigma_{\gamma_3}^q)^{-1}\Gamma_{0,3}\sigma_j.
\end{align} 
The last ingredient that one need for computing of the corrected interaction vertex $\widetilde{S}_{1,2}$ is the explicit form of the generators of the Fuchsian groups $\Gamma_{0,3}$, associated with the thrice punctured hyperbolic sphere and $\Gamma_{0,4}$, associated with the four punctured hyperbolic Riemann surface with specific Fenchel-Nielsen parameters.  Interestingly, following the algorithm given in \cite{Maskit}, it is possible to construct the  Fuchsian group of any hyperbolic Riemann surface having specific Fenchel-Nielsen parameters. For example, the group $\Gamma_{0,3}$ is generated by the transformations $$z\to \frac{z}{2z+1}\qquad \qquad z\to z+2.$$  The Fuchsian group $\Gamma_{0,4}(\ell,\tau)$ that produces a four punctured sphere with Fenchel-Nielsen parameter $(\ell,\tau)$ can be generated using the following three elements
\begin{align}\label{Gamma04lt}
a_1&=\left(\begin{array}{cc}1+\beta & -\beta \\  \beta  & 1-\beta \end{array}\right)  \nonumber\\ 
    a_2&=\left(\begin{array}{cc}\left(1-\beta\right)& -\beta e^{2\tau}\\  \beta e^{-2\tau} &  \left(1+\beta\right)\end{array}\right) \nonumber\\
     a_3&=-\left(\begin{array}{cc}(1+\beta )e^{\ell}& \beta e^{-\ell+2\tau}\\  -\beta e^{\ell-2\tau}& (1 -\beta) e^{-\ell}\end{array}\right),
\end{align}
where $\beta=-\frac{\text{cosh}\ell+1}{\text{sinh}\ell}$. \par

\noindent{\underline{\bf Arbitrary interaction vertex}}:  It is straightforward to generalize this discussion to the case of a general interaction vertex in closed string field theory.  This is because the lengths of simple closed geodesics on a general hyperbolic Riemann surface with borders also satisfies identities of the kind (\ref{gmidentityp}).  Assume that $\mathcal{R}(L_1,\cdots,L_n)$ is  a Riemann surface with $g$ handles and $n$ boundaries $\beta_1,\cdots,\beta_n$ having hyperbolic lengths $L_1,\cdots,L_n$. In the limit $L_i\to 0,~i=1,\cdots,n$, the bordered surface $\mathcal{R}(L_1,\cdots,L_n)$ becomes $\mathcal{R}$, a genus $g$ Riemann surface with $n$ punctures.  The lengths of the non-self intersecting closed geodesics on $\mathcal{R}(L_1,\cdots,L_n)$ satisfy the following identity  \cite{Mirzakhani:2006fta}: 
  \begin{equation}\label{gmidentitya}
\sum_{i=1}^n\sum_k\sum_{g\in \text{MCG}(\mathcal{R}(L_1,\cdots,L_n),\mathcal{C}^k_i)} \mathcal{Q}_i(L_1,L_i,\ell_{g\cdot\mathcal{C}^k_i})=1,
  \end{equation}
  where  
   \begin{align}\label{rinterval}
 \mathcal{Q}_i(L_1,L_i,\ell_{\mathcal{C}^k_i}) &= \delta_{1i}~\mathcal{D}(L_1,\ell_{\alpha^k_1},\ell_{\alpha^k_2})+(1-\delta_{i1})~\mathcal{E}(L_1,L_i,\ell_{\mathcal{C}^i_1})\nonumber\\
    \mathcal{D}(x_1,x_2,x_3)&=1-\frac{1}{x_1}\mathrm{ln}\left(\frac{\mathrm{cosh}(\frac{x_2}{2})+\mathrm{cosh}(\frac{x_1+x_3}{2})}{\mathrm{cosh}(\frac{x_2}{2})+\mathrm{cosh}(\frac{x_1-x_3}{2})}\right),\nonumber\\
    \mathcal{E}(x_1,x_2,x_3)&=\frac{2}{x_1}\mathrm{ln}\left(\frac{e^{\frac{x_1}{2}}+e^{\frac{x_2+x_3}{2}}}{e^{-\frac{x_1}{2}}+e^{\frac{x_2+x_3}{2}}}\right).\nonumber
    \end{align}
 $\mathcal{C}^k_1$ is the multi-curve $\alpha^k_1+\alpha^k_2$, where the simple closed geodesics $\alpha^k_1$ and $\alpha^k_2$ together with $\beta_1$  bounds a pair of pants, see figure \ref{cutting1}.  $\mathcal{C}^k_i,~i\in\{2,\cdots,n\}$ is a simple closed geodesic $\gamma^k_i$ which together with $\beta_1$ and $\beta_i$ bounds a pair of pants. The index $k$ distinguishes curves that are not related to each other via the action of elements in $\text{MCG}(\mathcal{R}(L_1,\cdots,L_n))$, the mapping class group of $\mathcal{R}(L_1,\cdots,L_n)$. The summation over $k$ add contributions from all such distinct classes of curves. By $\ell_{C_1^k}$ we mean the pair $(\ell_{\alpha^k_1},\ell_{\alpha^k_2})$. $ \text{MCG}(\mathcal{R}(L_1,\cdots,L_n),\mathcal{C}^k_i)$ is the subgroup of $\text{MCG}(\mathcal{R}(L_1,\cdots,L_n))$ that acts non-trivially only on the curve $\mathcal{C}^k_i$.  Remember that a Dehn twist performed with respect to $\mathcal{C}^k_i$ is not an element of $ \text{MCG}(\mathcal{R}(L_1,\cdots,L_n),\mathcal{C}_i)$. We also have an identity for the  group Dehn twists, and is given by
   \begin{equation}\label{dehnidentity}
\sum_{g\in \text{Dehn}(\mathcal{C}^k_i)} \mathcal{Y}_i(\ell_{g\cdot\mathcal{C}^k_i},\tau_{g\cdot\mathcal{C}^k_i})=1,
  \end{equation}
 where $\text{Dehn}(\mathcal{C}^k_1)$ denotes the product group $\text{Dehn}(\alpha^k_1)\times\text{Dehn}(\alpha^k_2)$, and
 \begin{align}\label{dehnidentity1}
 \mathcal{Y}_i(\ell_{\mathcal{C}^k_i},\tau_{\mathcal{C}^k_i})&= \delta_{i1}~\text{sinc}^2\left(\frac{\tau_{\alpha_1^k}}{\ell_{\alpha_1^k}}\right)\text{sinc}^2\left(\frac{\tau_{\alpha_2^k}}{\ell_{\alpha_2^k}}\right)+(1-\delta_{i1})~\text{sinc}^2\left(\frac{\tau_{\gamma_i^k}}{\ell_{\gamma_i^k}}\right).
  \end{align}

  Combining the identity  (\ref{gmidentitya}) with the identity (\ref{dehnidentity} )gives us the following identity
  \begin{equation}\label{gmidentityc}
\sum_{i=1}^n\sum_k\sum_{g\in \text{MCG}(\mathcal{R}(L_1,\cdots,L_n),\mathcal{C}^k_i)\times \text{Dehn}(\mathcal{C}^k_i)} \mathcal{Z}_i(L_1,L_i,\ell_{g\cdot\mathcal{C}^k_i},\tau_{g\cdot\mathcal{C}^k_i})=1.
  \end{equation}
 where
   \begin{equation}\label{gmidentityc1}
\mathcal{Z}_i(L_1,L_i,\ell_{\mathcal{C}^k_i},\tau_{\mathcal{C}^k_i})=\mathcal{Q}_i(L_1,L_i,\ell_{\mathcal{C}^k_i}) \mathcal{Y}_i(\ell_{\mathcal{C}^k_i},\tau_{\mathcal{C}^k_i}).
  \end{equation}
  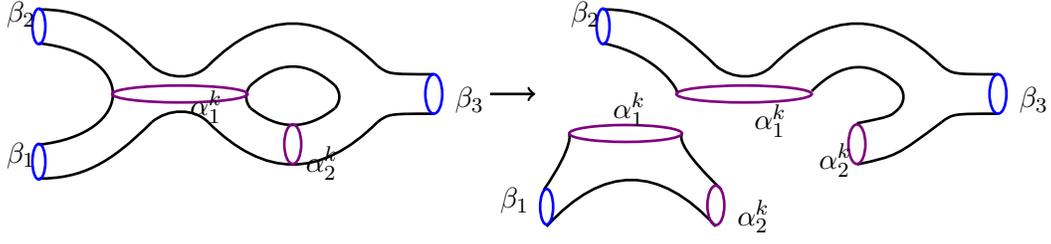
\begin{figure}
	\begin{center}
		\usetikzlibrary{backgrounds}
		\begin{tikzpicture}[scale=.75]
		\draw[line width=1pt] (1,1) .. controls (1.75,1) and (2.25,.75)  ..(3,0);
		\draw[line width=1pt] (1,-2) .. controls(1.75,-2) and (2.25,-1.75)  ..(3,-1);
		\draw[line width=1pt] (1,.4) .. controls(2.75,.2)  and (2.75,-1.2) ..(1,-1.4);
		\draw[blue, line width=1pt] (1,.7) ellipse (.115 and .315);
		\draw[blue, line width=1pt] (1,-1.7) ellipse (.115 and .315);
		\draw[line width=1pt] (3,0) .. controls(3.3,-.25) and (3.75,-.25) ..(4,0);
		\draw[line width=1pt] (3,-1) .. controls(3.3,-.75) and (3.75,-.75) ..(4,-1);
		\draw[line width=1pt] (4,0) .. controls(5,1) and (6,1) ..(7,0);
		\draw[line width=1pt] (4,-1) .. controls(5,-2) and (6,-2) ..(7,-1);
		\draw[line width=1pt] (7,0) .. controls(7.2,-.15)  ..(8,-.15);
		\draw[line width=1pt] (7,-1) .. controls(7.2,-.85)  ..(8,-.85);
		\draw[blue, line width=1pt] (8,-.5) ellipse (.15 and .35);
		\draw[line width=1pt] (4.75,-.75) .. controls(5.25,-1.15) and (5.75,-1.15) ..(6.25,-.75);
		\draw[line width=1pt] (4.75,-.35) .. controls(5.25,0.1) and (5.75,0.1) ..(6.25,-.35);
		\draw[line width=1pt] (4.75,-.35) .. controls(4.65,-.475) and (4.65,-.625) ..(4.75,-.75);
		\draw[line width=1pt] (6.25,-.35) .. controls(6.35,-.475) and (6.35,-.625) ..(6.25,-.75);
		\draw[line width =1pt, color=violet] (3.5,-.5) ellipse (1.2 and .15);
		\draw[line width =1pt, color=violet] (5.5,-1.39) ellipse (.15 and .35);
		\draw (0.25,.5) node[above right] {$\beta_2$}  (0.25,-2) node[above right] {$\beta_1$} (8.2,-.2)node [below right ] {$\beta_3$}  (3.5,-.25) node  [below right ] {$\alpha^k_1$} (5.55,-1.25) node  [below right ] {$\alpha^k_2$};
		\draw[->,line width =1pt] (9,-.5)--(9.8,-.5);
		\draw[line width=1pt] (11,1) .. controls (11.75,1) and (12.25,.75)  ..(13,0);
		\draw[line width=1pt] (11,.4) .. controls(12,.2)  and (12.4,-.5) ..(12.3,-.5);
		\draw[line width=1pt] (9.95,-2.2) .. controls(10.5,-1.4)  and (10.4,-1.2) ..(10.4,-1.2);
		\draw[line width=1pt] (12.4,-1.2) .. controls(12.3,-1.4)  and (12.8,-1.9) ..(13.1,-2.2);
		\draw[line width=1pt] (10,-2.85) .. controls(11,-1.75) and (12,-1.75)  ..(13,-2.85);
		\draw[blue, line width=1pt] (11,.7) ellipse (.115 and .315);
		\draw[blue, line width=1pt] (10,-2.5) ellipse (.115 and .315);
		\draw[line width=1pt] (13,0) .. controls(13.3,-.25) and (13.75,-.25) ..(14,0);
		\draw[line width=1pt] (14,0) .. controls(15,1) and (16,1) ..(17,0);
		\draw[line width=1pt] (15.5,-1.75) .. controls(16.2,-1.6)  ..(17,-1);
		\draw[line width=1pt] (17,0) .. controls(17.2,-.15)  ..(18,-.15);
		\draw[line width=1pt] (17,-1) .. controls(17.2,-.85)  ..(18,-.85);
		\draw[blue, line width=1pt] (18,-.5) ellipse (.15 and .35);
		\draw[line width=1pt] (15.5,-1.02) .. controls(16.1,-.9) ..(16.25,-.75);
		\draw[line width=1pt] (14.7,-.45) .. controls(15.25,0.1) and (15.75,0.1) ..(16.25,-.35);
		\draw[line width=1pt] (16.25,-.35) .. controls(16.35,-.475) and (16.35,-.625) ..(16.25,-.75);
		\draw[line width =1pt, color=violet] (13.5,-.5) ellipse (1.2 and .15);
		\draw[line width =1pt, color=violet] (11.4,-1.2) ellipse (1 and .15);
		\draw[line width =1pt, color=violet] (15.5,-1.39) ellipse (.15 and .35);
		\draw[line width =1pt, color=violet] (13,-2.475) ellipse (.15 and .35);
		\draw (10.25,.5) node[above right] {$\beta_2$}  (9,-2.8) node[above right] {$\beta_1$} (18.2,-.2)node [below right ] {$\beta_3$}  (13.5,-.5) node  [below right ] {$\alpha^k_1$} (14.65,-1.2) node  [below right ] {$\alpha^k_2$} (11,-.3) node  [below right ] {$\alpha^k_1$} (13.2,-2.2) node  [below right ] {$\alpha^k_2$};
		\end{tikzpicture}
	\end{center}
	
	\caption{Cutting a surface along a curve $\mathcal{C}_1^k=\alpha^k_1+\alpha^k_2$ produces a surface with borders.}
	\label{cutting1}
\end{figure}
  Now consider cutting $\mathcal{R}(L_1,\cdots,L_n)$ along $\mathcal{C}_i^k$. Let us denote the surface obtained as a result of this cutting by $\mathcal{R}(L_1,\cdots,L_n; \ell_{\mathcal{C}^k_i})$. Notice that the group $\text{MCG}(\mathcal{R}(L_1,\cdots,L_n); \ell_{\mathcal{C}^k_i})\times \text{Dehn}(\mathcal{C}^k_i)$ has no non-trivial  action on  $\mathcal{R}(L_1,\cdots,L_n;\ell_{\mathcal{C}^k_i})$. Therefore, we can repeat the whole exercise by considering the identity  (\ref{gmidentitya}) on $\mathcal{R}(L_1,\cdots,L_n; \ell_{\mathcal{C}^k_i})$.  At the end we obtain an identity of the following kind
     \begin{equation}\label{MCGIIdentity}
  \sum_{g\in\text{MCG}(\mcal{R}(L_1,\cdots,L_n))}\sum_s\mathcal{G}_s(\ell_{g\cdot\gamma_s},\tau_{g\cdot\gamma_s})=1.
  \end{equation}
  where  $\mathcal{G}_s$s are functions of the Fenchel-Nielsen coordinates of $\mathcal{R}(L_1,\cdots,L_n)$ defined with respect to a multi-curves $\gamma_s=\sum_{i=1}^{3g-3+n}\gamma_s^i$. The collection of curves $\left\{\gamma^1_s,\cdots,\gamma_s^{3g-3+n} \right\}$ form a system of non-self intersecting geodesics that define a pair of pants decomposition $\mathbf{P}^s$ of $\mathcal{R}(L_1,\cdots,L_n)$. The sum over $s$ represents the sum over pair of pants decompositions that are not related to each other via any MCG transformation. \par
  
  The function $\mathcal{G}_s$ has an important property. To demonstrate this property a non-self intersecting closed geodesic $\gamma$ on $\mcal{R}(L_1,\cdots,L_n)$ that can not be mapped to any element in the set $\left\{\gamma^1_s,\cdots,\gamma_s^{3g-3+n} \right\}$ by the action of any element in $\text{MCG}(\mcal{R}(L_1,\cdots,L_n))$. The hyperbolic metric on $\mcal{R}(L_1,\cdots,L_n)$ has the property  that if $\ell_{\gamma}\to c_*$, then the length of at least one of the curve in the set $\left\{\gamma^1_s,\cdots,\gamma_s^{3g-3+n} \right\}$ will have length of the order $e^{\frac{1}{c_*}}$.  Moreover, the function $\mathcal{G}_s$ is a function of all the curves  in the set $\left\{\gamma^1_s,\cdots,\gamma_s^{3g-3+n} \right\}$ constructed by multplying the functions $\mathcal{D}(x,y,z)$ and $\mathcal{E}(x,y,z)$. Note that the function $\mathcal{D}(x,y,z)$ appearing in the Mirzakhani-McShane identity (\ref{gmidentity}) given by 
\begin{equation}\label{DR1}
\mathcal{D}(x,y,z)=2~\ln\left( \frac{e^{\frac{x}{2}}+e^{\frac{y+z}{2}}}{e^{\frac{-x}{2}}+e^{\frac{y+z}{2}}}\right),
\end{equation}
vanishes in the limits $y\to \infty$ keeping $x,z$ fixed and $z\to \infty$ keeping $x,y$ fixed :
\begin{equation}\label{Dlimit}
\lim_{y,z\to\infty} \mathcal{D}(x,y,z)= \lim_{y,z\to\infty} \mathcal{O}\left( {\text{e}}^{-\frac{y+z}{2}}\right).
\end{equation}
$\mathcal{E}(x,y,z)$ given by 
\begin{equation} \label{DR2}
\mathcal{E}(x,y,z)=x-\ln\left( \frac{\cosh\left(\frac{y}{2}\right)+\cosh(\frac{x+z}{2})}{\cosh\left(\frac{y}{2}\right)+\cosh\left(\frac{x-z}{2}\right)}\right).
\end{equation}
becomes 0 in the limit $z\to \infty$ keeping $x,y$ fixed:
\begin{equation}\label{Elimit}
\lim_{z\to\infty} \mathcal{E}(x,y,z)=  \lim_{z\to\infty} \mathcal{O}\left( {\text{e}}^{-\frac{z}{2}}\right).
\end{equation}
This can be easily verified by using the following relation
\begin{equation}\label{DE}
\mathcal{E}(x,y,z)=\frac{\mathcal{D}(x,y,z)+\mathcal{D}(x,-y,z)}{2}.
\end{equation}
 Combining these observations suggests that the function $\mathcal{G}_s$ has the following  property 
  \begin{equation}\label{decay1}
  \lim_{\ell_{\gamma}\to c_*} \mathcal{G}_s=\mathcal{O}(e^{-1/c_*}).
  \end{equation}

   \noindent{\underline{\bf Naive interaction vertex $\mathbf{S}_{g,n}$}}: The naive $g$-loop $n$-point interaction vertex $\mathbf{S}_{g,n}$ for $n$ external off-shell states $|{V}_1\rangle,\cdots,|{V}_n\rangle$ represented by the vertex  operators ${V}_1,\cdots,{V}_n$ constructed using the naive string vertex   $\mathcal{V}^0_{g,n}$ is given by
\begin{equation}\label{eq:the bosonic-string amplitude2}
	\mathbf{S}_{g,n} =\int_{\mathcal{W}_{g,n}} \frac{d\ell_{\gamma^s_1}d\tau_{\gamma^s_1}\cdots d\ell_{\gamma_{Q}^s}d\tau_{\gamma_{Q}^s}}{ (2\pi \mathrm{i})^{Q}}~\langle\mathcal{R}|b(\mathbf{t}_{\gamma^s_1})b(\mathbf{l}_{ \gamma^s_1})\cdots b(\mathbf{t}_{\gamma^s_{Q}})b(\mathbf{l}_{ \gamma^s_{Q}})|{V}_1\rangle_{w_1}\otimes\cdots\otimes|{V}_n\rangle_{w_n},
\end{equation}
 where $Q=3g-3+n$ and the states $|{V}_1\rangle,\cdots,|{V}_n\rangle$ are inserted on the Riemann surface $\mathcal{R}$ using the set of local coordinates $(e^{\frac{\pi^2}{c_*}}w_1,\cdots,e^{\frac{\pi^2}{c_*}} w_n)$ induced from the hyperbolic metric. $|\mathcal{R}\rangle$ is the surface state associated with the Riemann surface $\mathcal{R}$.  $(\tau_{\gamma_j^s},\ell_{\gamma^s_j}),~1\leq j\leq Q$ denote the Fenchel-Nielsen coordinates for the Teichm\"uller space $\mathcal{T}_{g,n}$ defined with respect to the pants decomposition $\mathbf{P}_s$ of $\mathcal{R}$. Using the identity (\ref{MCGIIdentity}), we can decompose (\ref{eq:the bosonic-string amplitude2}) into sum over all possible distinct pants decompositions of $\mathcal{R}$ with each term expressed as an integral over $\mathcal{T}_{g,n}$:
 \begin{equation}\label{eq:the bosonic-string amplitude3}
	\mathbf{S}_{g,n} = \sum_s\int_{\mathcal{TW}_{g,n}}  \frac{d\ell_{\gamma^s_1}d\tau_{\gamma^s_1}\cdots d\ell_{\gamma_{Q}^s}d\tau_{\gamma_{Q}^s}}{(2\pi \mathrm{i})^{Q}}~\mathcal{G}_s\langle\mathcal{R}|b(\mathbf{t}_{\gamma^s_1})b(\mathbf{l}_{ \gamma^s_1})\cdots b(\mathbf{t}_{\gamma^s_{Q}})b(\mathbf{l}_{ \gamma^s_{Q}})|{V}_1\rangle_{w_1}\otimes\cdots\otimes|{V}_n\rangle_{w_n},
\end{equation}
  where $\mathcal{TW}_{g,n}$ is the union of all images $\mathcal{W}_{g,n}$ in $\mathcal{T}_{g,n}$. Since the set local coordinates induced from hyperbolic metric do not satisfy the geometrical identity induced form the quantum BV master equation, the closed string field theory action constructed using the naive interaction vertex $\mathbf{S}_{g,n}$  is not gauge invariant.\\

   \noindent{\underline{\bf Corrected interaction vertex $\widetilde{\mathbf{S}}_{g,n}$}}: In order to correct the interaction vertex $\widetilde{\mathbf{S}}_{g,n}$  the set of local coordinates on the world-sheets in $\mathcal{W}_{g,n}$ induced from the hyperbolic metric used to construct $\mathbf{S}_{g,n}$ must be modified. The set of local coordinates has to be modified if  $\mathcal{R}$ belongs to the regions $\mathbf{W}^{(m)}_{g,n}$ for $m\neq 0$. Although the regions inside $\mathcal{M}_{g,n}$ where we need to modify the local coordinates have a simple description in terms of the length of the simple closed geodesics, it is impossible to specify them as explicit regions inside $\mathcal{T}_{g,n}$ using the Fenchel-Nielsen coordinates. This is due to the fact that there are infinitely many simple closed geodesics on a Riemann surface. \par
   
Interestingly,  the effective expression (\ref{eq:the bosonic-string amplitude3}) has a noteworthy feature due to the decay property of the weight factors $\mathcal{G}_s$s (\ref{decay1}). Assume that the length of a simple closed geodesic $\alpha$ which does not belong to the set of curves $\left\{\gamma_s^1,\cdots,\gamma_s^{3g-3+n} \right\}$ associated with the pants decomposition $\mathbf{P}^s$ becomes $c_*$. Then the weight factor $\mathcal{G}_s$ decays to $\mathcal{O}(e^{-1/c_*})$. As a result, by correcting interaction vertex by modifying the  local coordinates within each term in the effective expression (\ref{eq:the bosonic-string amplitude3}) independently it is possible to make them approximately satisfy the quantum BV master equation. \par

Consider the $s^{\text{th}}$ term in the effective expression. The effective region $E\mathcal{W}^{\mathbf{P}^s}_{g,n}$ for $\mathcal{TW}_{g,n}$ is given by
\begin{equation}
E\mathcal{W}^{\mathbf{P}^s}_{g,n}:~\ell_{\gamma_s^1}\in [c_*,\infty)\quad \cdots \ell_{\gamma_s^Q}\in [c_*,\infty)\quad \tau_1\in (-\infty,\infty)\cdots \tau_Q\in (-\infty,\infty).
\end{equation}
In order to modify the local coordinates we must divide $E\mathcal{W}^{\mathbf{P}^s}_{g,n}$ into subregions $E\mathbf{W}^{\mathbf{P}^s,(m)}_{g,n},~m=0,\cdots, Q$.  Divide the subregion $E\mathbf{W}^{\mathbf{P}^s,(m)}_{g,n}$ further into $\frac{Q!}{m!(Q-m)!}$ number of regions $E\mathbf{W}^{\mathbf{P}^s,\gamma_{j_i}\cdots \gamma_{j_m}}_{g,n}$, where $i_1,\cdots, i_m\in \{1,\cdots,Q\}$. The number $\frac{Q!}{m!(Q-m)!}$ counts the inequivalent ways of choosing $m$ curves from the set $\left\{\gamma_s^1,\cdots,\gamma_s^{Q} \right\}$. For surfaces that belong to the region $E\mathbf{W}^{\mathbf{P}^s,\gamma_s^{i_1}\cdots\gamma_s^{i_m}}_{g,n}$ with $m\ne 0$, we choose the local coordinates around the $j^{th}$ puncture, up to a phase ambiguity,  is given by
\begin{equation}
\widetilde{w}_{j}^{\gamma_{s}^{i_1}\cdots\gamma_s^{i_m}}=e^{\frac{c_*^2}{6}\sum_{k=1}^mf(\ell_{\gamma_{i_k}})Y^{\gamma_{i_1}\cdots\gamma_{i_m}}_{i_kj}}e^{\frac{\pi^2}{c_*}} w_{j}.
\end{equation}
where
\begin{align}
Y^{\gamma_{i_1}\cdots\gamma_{i_m}}_{i_kj}&=\sum_{q=1}^2\sum_{c_j^q,d_j^q}\pi^{2}\frac{\epsilon(j,q)}{|c_j^q|^4}\nonumber\\ c_j^q>0 \qquad &d_j^q~\text{mod}~c_j^q \qquad\left(\begin{array}{cc}* & * \\c_j^q & d_j^q\end{array}\right)\in \quad (\sigma_i^q)^{-1}\Gamma_{{\gamma_{i_1}\cdots\gamma_{i_m}}}^{jq}\sigma_j
\end{align} 
Here, $\Gamma_{{\gamma_{i_1}\cdots\gamma_{i_m}}}^{jq}$ denotes the Fuchsian group for the component Riemann surface obtained from $\mathcal{R}$ by degenerating the curves $\gamma_{i_1},\cdots,\gamma_{i_m}$ carrying the $j^{\text{th}}$ puncture and the puncture  denoted by the index $q$ which is obtained by degenerating the curve $\gamma_{i_k}$.  The transformation $\sigma_j^{-1}$ maps the cusp corresponding to the $j^{th}$ cusp to $\infty$ and $(\sigma_j^q)^{-1}$ maps the puncture denoted by the index $q$ obtained by degenerating the curve $\gamma_{i_k}$ to $\infty$. The factor $\epsilon(j,q)$ is one if both the $j^{th}$ puncture and the puncture denoted by the index $q$  belong to the same component surface, other wise $\epsilon(j,q)$ is zero. \par

Then the corrected interaction vertex $\widetilde{\mathbf{S}}_{g,n}$ is given by
\begin{align}
\widetilde{\mathbf{S}}_{g,n}&=\sum_s\sum_{m=0}^{Q}\sum_{\left\{i_1,\cdots,i_m\right\}}\int_{E\mathbf{W}^{\mathbf{P}^s,\gamma_s^{i_1}\cdots\gamma_s^{i_m}}_{g,n}}  \frac{\prod_{j=1}^Qd\ell_{\gamma^s_j}d\tau_{\gamma^s_j}}{(2\pi \mathrm{i})^{Q}}\mathcal{G}_s\nonumber\\
&\times \langle\mathcal{R}|b(\mathbf{t}_{\gamma^s_1})b(\mathbf{l}_{ \gamma^s_1})\cdots b(\mathbf{t}_{\gamma^s_{Q}})b(\mathbf{l}_{ \gamma^s_{Q}})|{V}_1\rangle_{\widetilde{w}_{1}^{\gamma_{s}^{i_1}\cdots\gamma_s^{i_m}}}\otimes\cdots\otimes|{V}_n\rangle_{\widetilde{w}_{n}^{\gamma_{s}^{i_1}\cdots\gamma_s^{i_m}}},
\end{align}
where the sum over the sets $\{i_1,\cdots,i_m\}$ denotes the sum of $\frac{Q!}{m!(Q-m)!}$  inequivalent ways of choosing $m$ curves from the set $\left\{\gamma_s^1,\cdots,\gamma_s^{Q} \right\}$. The expression for corrected interaction vertex  $\widetilde{\mathbf{S}}_{g,n}$  is true for any values of $g$ and $n$ such that $3g-3+n\geq 0$, and the closed string field theory master action constructed using these corrected interaction vertices will have approximate gauge invariance. 

     \section{Discussion}\label{disc}
In this paper we completed the construction of quantum closed string field theory with approximate gauge invariance by exploring the hyperbolic geometry of Riemann surfaces initiated in \cite{Moosavian:2017qsp}. In  \cite{Moosavian:2017qsp}  it was shown that although the string vertices constructed using Riemann surfaces with local coordinates induced from hyperbolic Riemann surfaces $-1$ constant curvature fails to provide gauge invariant quantum closed string field theory, the corrected string vertices obtained by modifying these local coordinates on Riemann surfaces belongs to the boundary region of string vertices give rise to quantum closed string field theory with approximate gauge invariance. Unfortunately, due to the complicated action of mapping class group on the Fenchel-Nielsen coordinates the implementing the suggested modification seemed to be impractical. However, in this paper we argued that by using the non-trivial identities satisfied by the lengths of simple closed geodesics on hyperbolic Riemann surfaces it is indeed possible to implement the modifications in very convenient fashion. The identities that we explored in this paper are due to McShane and Mirzakhani \cite{McShane1,Mirzakhani:2006fta}.  Although they are very convenient to use, they have a very important drawback. They are applicable only for the case of hyperbolic Riemann surfaces with at least one border or puncture. For instance, for computing the contributions from vacuum graphs to the string field theory action, we can not use them. Interestingly, there exists another class of such identities due to Luo and Tan \cite{LT01,HT01} that are applicable for kinds of hyperbolic Riemann surfaces with no elliptic fixed points. For a quick introduction read appendix \ref{LuoTan}. But they have one disadvantage, the functions  involved in these identities are significantly more complicated than the functions appearing in the identities due to McShane and Mirzakhani.  \par
There are many interesting direction that deserve further study.  It would be very useful to check whether it is possible to construct the string vertices in closed superstring field theory that avoids the occurrence of any unphysical singularities due to the picture changing operators by exploring hyperbolic geometry. It might be worth exploring hyperbolic geometry of super Riemann surfaces to construct closed superstring field theory using the supergeometric formulation of superstring theory. This is particularly interesting due to the fact that there exist generalization of McShane-Mirzakhani identities for the case of super Riemann surfaces \cite{Stanford:2019vob}.  Another interesting direction is to use the formalism discussed in this paper to systematically compute the field theory limit of  string amplitudes. We hope to report on this in the near future. 

\bigskip

{\bf Acknowledgement:} It is our pleasure to thank Davide Gaiotto and Ashoke Sen for important comments and detailed discussions. We thank  Thiago Fleury, Greg McShane, Scott Wolpert and Barton Zwiebach  for helpful discussions.  Research at Perimeter Institute is supported by the Government of Canada through Industry Canada and by the Province of Ontario through the Ministry of Research \& Innovation.

\appendix

\section{Brief review of hyperbolic geometry}\label{hyperbolic}

A Riemann surface is a one-dimensional complex manifold. The map carrying the structure of the complex plane to the Riemann surface is called a chart. The transition maps between two overlapping charts are required to be holomorphic. The charts together with the transition functions between them define a complex structure on the Riemann surface \cite{Kra}. The topological classification of the compact Riemann surfaces is done with the pair $(g,n)$, where $g$ denotes the genus  and $n$ denotes the number of boundaries of the Riemann surface. However, within each topological class determined by $(g,n)$, there are different surfaces endowed with different complex structures.  Two such complex structure are equivalent if there is a conformal (i.e complex analytic) map between them. The set of all such equivalent complex structures define a {\it conformal class} in the set of all complex structures on the surface. \par

 The uniformization theorem \cite{Abi1} asserts that for a Riemann surface $\mathcal{R}$:
\begin{equation}
\left\{\substack{\text{\fontsize{11}{11}\selectfont Hyperbolic Structure}\\ \text{\fontsize{11}{11}\selectfont on $\mathcal{R}$}}\right\}\,\, \leftrightarrow \,\, \left\{\substack{\text{\fontsize{11}{11}\selectfont Complex Structure}\\ \text{\fontsize{11}{11}\selectfont on $\mathcal{R}$}}\right\}
\end{equation}
Therefore, the space of all  conformal classes is the same as the classification space of all inequivalent hyperbolic structures.  A hyperbolic structure on a Riemann surface $\mathcal{S}$ is a diffemorphism $\phi:\mathcal{S}\longrightarrow \mathcal{R}$, where $\mathcal{R}$ is a surface with finite-area hyperbolic metric and geodesic boundary components.  Here, hyperbolic metric on a Riemann surface refers to the metric having constant curvature $-1$ all over the surface.  Two hyperbolic structures on a Riemann surface $\mathcal{S}$, given by  $\phi_1:\mathcal{S}\longrightarrow \mathcal{R}_1$ and $\phi_2:\mathcal{S}\longrightarrow \mathcal{R}_2$, are equivalent if there is an isometry $I:\mathcal{R}_1\longrightarrow \mathcal{R}_2$ such that the maps $I\circ \phi_1 : \mathcal{S} \longrightarrow \mathcal{R}_1$ and $\phi_2 : \mathcal{S}\longrightarrow \mathcal{R}_2$ are homotopic i.e. the map $I\circ \mathcal{R}_1$ can be continuously deformed into $\phi_2$ by a homotopy map.  \par

The classification space of the homotopy classes of the hyperbolic structures on a Riemann surface of a given topological type i.e. the space of the hyperbolic structures up to the homotopies,  is called {\it the Teichm\"uller space of Riemann surfaces}  of a given topological type $(g,n)$ and it is denoted by $\mathcal{T}_{g,n}(\mathcal{R})$ \cite{Yoichi}:
\begin{equation}
\mathcal{T}_{g,n}(\mathcal{R})=\left\{\text{hyperbolic structures on $\mathcal{R}$}\right\}/\text{homotopy}.
\end{equation}

\subsection{Hyperbolic Riemann surfaces and the Fuchsian uniformization}

 A hyperbolic Riemann surface is a Riemann surface with  a metric having constant curvature $-1$ defined on it.  According to  the uniformization theorem,  any Riemann surface with negative Euler characteristic can be made hyperbolic. The hyperbolic Riemann surface $\mathcal{R}$ can be represented as a quotient of the upper half-plane $\mathbb{H}$ by a Fuchsian group $\Gamma$, which is a subgroup  of the isometries of  $\mathbb{H}$ endowed with hyperbolic metric on it.  The hyperbolic metric on  the upper half-plane
\begin{equation}
\mathbb{H}=\{z:~\text{Im}~z>0\}.
\end{equation}
 is given by 
\begin{equation}\label{mDH}
 ds^2_{\text{hyp}}=\frac{dzd\bar z}{(\text{Im}z)^2}.
\end{equation}
Each hyperbolic Riemann surface $\mathcal{R}$ inherits, by projection from $\mathbb{H}$, its own hyperbolic geometry. \par
 
The isometries of the hyperbolic metric on  $\mathbb{H}$    form the  $PSL(2,\mathbb{R})$ group:
\begin{equation}\label{MH}
z\to \frac{az+b}{cz+d},\qquad\qquad a,b,c,d\in \mathbb{R}, \qquad ad-bc=1.
\end{equation}
The hyperbolic distance, $\rho(z,w)$, between two points $z$ and $w$ is defined to be the length of the geodesic in $\mathbb{H}$ that join $z$ to $w$.  A geodesic on the hyperbolic upper half-plane is a segment of the half-circle with origin on the real axis or the line segment parallel to the imaginary axis  which passes through $z$ and $w$ that is orthogonal to the boundary of $\mathbb{H}$.  This distance is given by 
\begin{equation}\label{hdis}
\rho(z,w)=\mathrm{ln}\left(\frac{1+\tau(z,w)}{1-\tau(z,w)}\right)=2~\mathrm{tanh}^{-1}\left(\tau\left(z,w\right)\right).
\end{equation}
where $\tau(z,w)=\left| \frac{z-w}{\bar z- w}\right| $. An open set $\mathcal{F}$ of the upper half plane $\mathbb{H}$ satisfies the following three conditions is called the fundamental domain for $\Gamma$ in $\mathbb{H}$:

\begin{enumerate}
\item $\gamma(\mathcal{F})\cap \mathcal{F}=\emptyset$ for every $\gamma\in \Gamma$ with $\gamma\not=id$. 
\item If $\bar{\mathcal{F}}$ is the closure of $\mathcal{F}$ in $\mathbb{H}$, then $\mathbb{H}=\bigcup_{\gamma\in \Gamma}\gamma(\bar{\mathcal{F}})$.
\item The relative boundary $\partial\mathcal{ F}$ of $\mathcal{F}$ in $\mathbb{H}$ has measure zero with respect to the two dimensional Lebsegue measure.
\end{enumerate}

   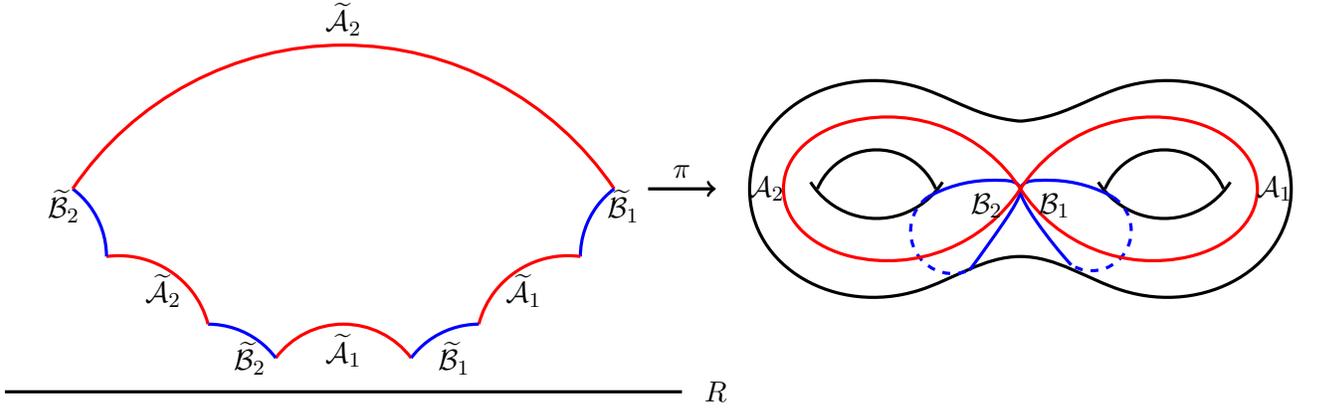
\begin{figure}
\begin{center}
\usetikzlibrary{backgrounds}
\begin{tikzpicture}[scale=.45]
\draw[black,very  thick] (0,2) to[curve through={(1,2.2)..(3,3)..(8,0) .. (3,-3) ..(0,-2)  .. (-3,-3)..(-8,0)..(-3,3)..(-1,2.2)}] (0,2);
\draw[black, very thick] (2.5,0) to[curve through={(3.5,1)..(5,1)}] (6,0);
\draw[black, very thick] (2.3,.2) to[curve through={(3.5,-.75)..(5,-.75)}] (6.2,.2);
\draw[black,very  thick] (-2.5,0) to[curve through={(-3.5,1)..(-5,1)}] (-6,0);
\draw[black,very  thick] (-2.3,.2) to[curve through={(-3.5,-.75)..(-5,-.75)}] (-6.2,.2);
\draw[blue,very thick] (1.5,-2.25) to[curve through={(.5,-1)..(.05,-.25)..(0,0)..(.5,.25)..(1,.25)..}] (2.75,-.25);
\draw[blue,very thick, style=dashed] (2.75,-.25) to[curve through={(3,-2)}] (1.5,-2.25);
\draw[blue,very thick] (-1.5,-2.4) to[curve through={(-.5,-1)..(-.05,-.25)..(0,0)..(-.5,.25)..(-1,.25)..}] (-2.75,-.25);
\draw[blue,very thick, style=dashed] (-2.75,-.25) to[curve through={(-3,-2)}] (-1.5,-2.4);
\draw[red,very thick] (0,0) to[curve through={(5,2)..(7,0)..(5,-2)}] (0,0) ;
\draw[red,very thick] (0,0) to[curve through={(-5,2)..(-7,0)..(-5,-2)}] (0,0);
\draw node at (7.5,0) {$\mathcal{A}_1$};
\draw node at (1,-.5) {$\mathcal{B}_1$};
\draw node at (-7.5,0) {$\mathcal{A}_2$};
\draw node at (-1,-.5) {$\mathcal{B}_2$};
\draw[very thick,->] (-11,0)--(-9,0);
\draw node at (-10,.5) {$\pi$};
\draw[very thick] (-30,-6)--(-10,-6);
\draw node at (-9,-6) {$R$};
\draw[red, very thick] (-22,-5) to[curve through={(-20,-4)}] (-18,-5);
\draw[blue, very thick] (-24,-4) to[curve through={(-22.5,-4.5)}] (-22,-5);
\draw[red, very thick] (-27,-2) to[curve through={(-24.5,-3)}] (-24,-4);
\draw[blue, very thick] (-28,0)to[curve through={(-27.5,-.5)}] (-27,-2) ;
\draw[red, very thick] (-28,0)to[curve through={(-20,4.25)}] (-12,0) ;
\draw[blue, very thick] (-13,-2)to[curve through={(-12.5,-.5)}] (-12,0) ;
\draw[red, very thick] (-13,-2)to[curve through={(-15.5,-3)}] (-16,-4) ;
\draw[blue,very  thick] (-16,-4) to[curve through={(-17.5,-4.5)}] (-18,-5);
\draw node [above] at (-20,4.25) {$\widetilde{\mathcal{A}}_2$};
\draw node [below,right] at (-15.5,-3) {$\widetilde{\mathcal{A}}_1$};
\draw node [below,left] at (-24.5,-3) {$\widetilde{\mathcal{A}}_2$};
\draw node [below,right ] at (-12.5,-.5) {$\widetilde{\mathcal{B}}_1$};
\draw node  [below] at(-20,-4) {$\widetilde{\mathcal{A}}_1$};
\draw node [below,left ] at (-27.5,-.5) {$\widetilde{\mathcal{B}}_2$};
\draw node [below,right] at (-17.5,-5) {$\widetilde{\mathcal{B}}_1$};
\draw node [below,left ] at (-22,-5) {$\widetilde{\mathcal{B}}_2$};
\end{tikzpicture}
\end{center}

\caption{The fundamental domain of Fuchsian uniformization corresponding to the genus 2 surface. }
\label{Fuchsian Fundamental domain}
\end{figure}

The Riemann surface $\mathcal{R}=\mathbb{H}/\Gamma$ is represented as $\bar{\mathcal{F}}$ with points in $\partial\mathcal{ F}$ identified under the action of elements in the group $\Gamma$. Let $\pi:\mathbb{H}\to \mathcal{R}$ be the map that  projects  $\mathbb{H}$ onto $\mathcal{R}=\mathbb{H}/\Gamma$. Since  $ds^2_{\text{hyp}}$ is invariant under the action of elements in $\Gamma$, we obtain a Riemannian metric $ds^2_{\mathcal{R}}$ on $\mathcal{R}$. The metric $ds^2_{\mathcal{R}}$ is the hyperbolic metric on $\mathcal{R}$. Moreover, every $\gamma\in \Gamma$ corresponds to an element $[C_{\gamma}]$ of the fundamental group $\pi_1(\mathcal{R})$ of $\mathcal{R}$. In particular, $\gamma$ determines the free homotopy class of $C_{\gamma}$, where $C_{\gamma}$ is a representative of the class $[C_{\gamma}]$. For hyperbolic element $\gamma\in \Gamma$, i.e. $\left(\text{tr}\left(\gamma\right)\right)^2>4$, the closed curve $L_{\gamma}=A_{\gamma}/\langle \gamma \rangle$, the image on $\mathcal{R}$ of the axis $A_{\gamma}$ by $\pi$, is the unique geodesic with respect to the hyperbolic metric on $\mathcal{R}$ belonging to the same homotopy class of $C_{\gamma}$. The axis of a hyperbolic  element $\gamma$ is the geodesic on $\mathbb{H}$ that connects the fixed points of $\gamma$. Let 
\begin{equation}\label{gammaele}
\gamma= \left(\begin{array}{cc}a & b \\c & d\end{array}\right),\qquad\qquad a,b,c,d\in \mathbb{R}, \qquad ad-bc=1.
\end{equation}
 be a hyperbolic element of $\Gamma$ and $L_{\gamma}$ be the closed geodesic corresponding to $\gamma$. Then the hyperbolic length $l(L_{\gamma})$ of $L_{\gamma}$  satisfies following relation
\begin{equation}\label{lLgamma}
\mathrm{tr}^2(\gamma)=(a+d)^2=4\mathrm{cosh}^2\left(\frac{l\left(L_{\gamma}\right)}{2}\right).
\end{equation}

\subsection{The Fenchel-Nielsen coordinates for the Teichm\"uller space}

Let $\mathcal{R}$ be a hyperbolic Riemann surface. Consider cutting $\mathcal{R}$ along mutually disjoint simple closed geodesics with respect to the hyperbolic metric $ds_{\mathcal{R}}^2$ on $\mathcal{R}$. If there are no more closed geodesics of $\mathcal{R}$ contained in the remaining open set that are non-homotopic to the boundaries of the open set, then every piece should be a pair of pants of $\mathcal{R}$. The complex structure of each pair of pants on $\mathcal{R}$ is uniquely determined by a triple of the hyperbolic lengths of boundary geodesics of it.  To see this, consider a pair of pants $P$ with  boundary components $L_1,L_2$ and $L_3$. Assume that $\Gamma_P$ is the Fuchsian group associated with $P$. Then $\Gamma_P$ is a free group generated by two hyperbolic transformations $\gamma_1$ and $\gamma_2$.  The action of $\gamma_1$ and $\gamma_2$ on $\mathbb{H}$ is given by 
\begin{align}\label{g1g2action}
\gamma_1(z)&=\lambda^2z, &0<\lambda<1, \nonumber\\ 
\gamma_2(z)&=\frac{az+b}{cz+d}, \qquad  ad-bc=1\qquad &0<c.
\end{align}
 We  assume that 1 is the attractive fixed point of $\gamma_2$, or equivalently, $a+b=c+d,~0<-\frac{b}{c}<1$.  We also assume that 
 \begin{equation}
 (\gamma_3)^{-1}(z)=\gamma_2\circ\gamma_1(z)=\frac{\tilde az+\tilde b}{\tilde cz+\tilde d},\qquad\qquad\tilde a\tilde d-\tilde b\tilde c=1.
 \end{equation}
 Then using (\ref{lLgamma}) we obtain the following relations
 \begin{align}
 (\lambda+\frac{1}{\lambda})^2&=4\mathrm{cosh}^2\left(\frac{a_1}{2}\right).\nonumber\\
 (a+d)^2&=4\mathrm{cosh}^2\left(\frac{a_2}{2}\right),\nonumber\\
 (\tilde a+\tilde d)^2&=4\mathrm{cosh}^2\left(\frac{a_3}{2}\right).
 \end{align} 
 This means that the action of the generators of the group $\Gamma_P,~\gamma_1$ and $\gamma_2$ are uniquely determined by the  hyperbolic lengths of the boundary components. \par

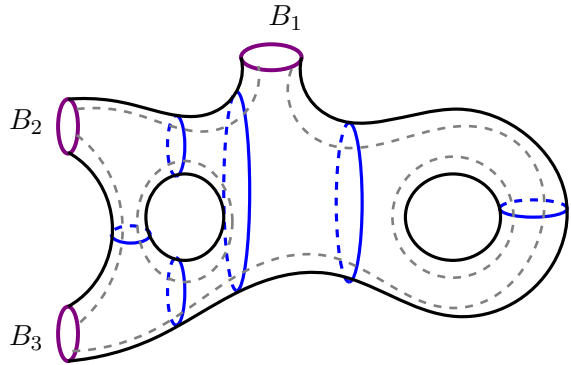
\begin{figure}
\begin{center}
\usetikzlibrary{backgrounds}
\begin{tikzpicture}[scale=1.15]
\draw[violet,line width=1.5pt] (1,.7) ellipse (.115 and .315);
\draw[violet,line width=1.5pt] (1,-1.7) ellipse (.115 and .315);
\draw[violet,line width=1.5pt] (3.35,1.5) ellipse (.35 and .15);
\draw[blue, line width=1.2pt](2.95,-1.2) arc[start angle=-90,end angle=90, x radius=.15, y radius=1.15];
\draw[blue, line width=1.2pt,style=dashed](2.95,-1.2)arc[start angle=270,end angle=90, x radius=.15, y radius=1.15];
\draw[blue, line width=1.2pt](4.25,-1.1) arc[start angle=-90,end angle=90, x radius=.15, y radius=.92];
\draw[blue, line width=1.2pt,style=dashed](4.25,-1.1)arc[start angle=270,end angle=90, x radius=.15, y radius=.92];
\draw[blue, line width=1.2pt](2.25,.125) arc[start angle=-90,end angle=90, x radius=.1, y radius=.35];
\draw[blue, line width=1.2pt,style=dashed](2.25,.125)arc[start angle=270,end angle=90, x radius=.1, y radius=.35];
\draw[blue, line width=1.2pt](2.25,-1.6) arc[start angle=-90,end angle=90, x radius=.1, y radius=.39];
\draw[blue, line width=1.2pt,style=dashed](2.25,-1.6)arc[start angle=270,end angle=90, x radius=.1, y radius=.39];
\draw[blue, line width=1.2pt,style=dashed](1.95,-.55) arc[start angle=0,end angle=180, x radius=.225, y radius=.1];
\draw[blue, line width=1.2pt](1.95,-.55)arc[start angle=360,end angle=180, x radius=.225, y radius=.1];
\draw[blue, line width=1.2pt,style=dashed](6.77,-.25)arc[start angle=0,end angle=180, x radius=.39, y radius=.1];
\draw[blue, line width=1.2pt](6.77,-.25)arc[start angle=360,end angle=180, x radius=.39, y radius=.1];
\draw[black,  very thick] (1,0.4) to[curve through={(1.25,.2)..(1.25,-1.185)}] (1,-1.385);
\draw[black,  very thick] (1,-2.015) to[curve through={(2,-1.75)..(3,-1.2)..(4,-1)..(5.5,-1.5)..(6.6,-.9)..(6.75,-.1)..(6.3,.6)..(5.5,.9)..(3.9,.9)}] (3.7,1.5);
\draw[black,  very thick] (1,1.015) to[curve through={(1.5,1)..(1.75,.95)..(2.75,.9)}] (3.0,1.5);
\draw[ line width=1.2pt](2.8,-.35) arc[start angle=0,end angle=360, x radius=.45, y radius=.5];
\draw[ line width=1.2pt](6,-.35) arc[start angle=0,end angle=360, x radius=.55, y radius=.5];
\draw[gray, line width=1pt,style=dashed](2.9,-.4) arc[start angle=0,end angle=360, x radius=.55, y radius=.7];
\draw[gray, line width=1pt,style=dashed](6.15,-.35) arc[start angle=0,end angle=360, x radius=.7, y radius=.7];
\draw[gray,  line width=1pt,style=dashed] (1.1,0.6) to[curve through={(1.45,.2)..(1.45,-1.185)}] (1.1,-1.6);
\draw[gray,  line width=1pt,style=dashed] (1.05,-1.9) to[curve through={(2,-1.65)..(3,-1)..(4,-.8)..(5.5,-1.4)..(6.4,-.7)..(6.5,-.1)..(6.2,.5)..(5.5,.7)..(3.8,.6)}](3.6,1.4) ;
\draw[gray,  line width=1pt,style=dashed] (1.1,.9) to[curve through={(1.5,.9)..(1.75,.85)..(2.75,.67)}] (3.2,1.4);

\draw (0.2,.5) node[above right] {$B_2$}  (0.2,-2) node[above right] {$B_3$}  (3.2,1.7)node [above right ] {$B_1$} ;
\end{tikzpicture}
\end{center}

\caption{A pairs of pants decomposition of a genus 2 Riemann surface with three boundaries. The  gray dashed curves provide a choice of seams. }
\label{pair of pants decomposition}
\end{figure}

 The basic idea behind the Fenchel-Nielsen construction is to decompose $\mathcal{R}$, a genus $g$ hyperbolic Riemann surface with $n$ boundaries $B_1,\cdots,B_n$ with fixed lengths, into pairs of pants using $3g-3+n$ simple closed curves (See figure (\ref{pair of pants decomposition})). Then, there are $3g-3+n$ length parameters that determine the hyperbolic structure on each pair of pants, and there are $3g-3+n$ twist parameters that determine how the pairs of pants are glued together. Taken together, these $6g-6+2n$ coordinates are the Fenchel-Nielsen coordinates \cite{Farb}. Below we describe this more precisely. \par

In order to define the Fenchel-Nielsen coordinates, we must first choose a {\it coordinate system of curves} on $\mathcal{R}$  \cite{Farb}. For this, we need to choose:
\begin{itemize}
\item a pants decomposition $\left\{ \gamma_1,\cdots,\gamma_{3g-3+n} \right\}$ of oriented simple closed geodesics (boundary curves are not included) and
\item a set $\{\beta_1,\cdots,\beta_m\}$ of {\it seams}; that is, a collection of disjoint simple closed geodesics in $\mathcal{R}$ so that the intersection of the union $\cup\beta_i$ with any pair of pants $P$ determined by the $\left\{ \gamma_j\right\}$ is a union of three disjoint arcs connecting the boundary components of $P$ pairwise. See figure (\ref{pair of pants decomposition}). 
\end{itemize}

Fix once and for all a coordinate system of curves on $\mathcal{R}$. The $3g-3+n$ number of length parameters of a point $\widetilde{\mathcal{R}}\in \mathcal{T}_{g,n}(\mathcal{R})$ are defined to be the ordered $(3g-3+n)$-tuple of positive real numbers
\begin{equation}
\left(\ell_{1}(\widetilde{\mathcal{R}}),\cdots,\ell_{3g-3+n}(\widetilde{\mathcal{R}})\right),
\end{equation} 
where $\ell_{i}(\widetilde{\mathcal{R}})$ is the hyperbolic length of $\gamma_i$ in $\widetilde{\mathcal{R}}$. The length parameters determine the hyperbolic structure of the $2g-2+n$ number of pairs of pants obtained by cutting $\mathcal{R}$ along the curves $\left\{ \gamma_1,\cdots,\gamma_{3g-3+n} \right\}$. The information about how these pants are glued together is recorded in the twist parameters $\tau_i(\widetilde{\mathcal{R}})$. \par

 \begin{figure}
\begin{center}
\usetikzlibrary{backgrounds}
\begin{tikzpicture}[scale=1.15]
\draw[violet,line width=1.5pt] (0,-1.7) ellipse (.55 and .315);
\draw[violet,line width=1.5pt] (2,-1.7) ellipse (.55 and .315);
\draw[violet,line width=1.5pt] (0,1.7) ellipse (.55 and .315);
\draw[violet,line width=1.5pt] (2,1.7) ellipse (.55 and .315);
\draw[violet, line width=1.2pt](1.7,.3) arc[start angle=360,end angle=180, x radius=.8, y radius=.15];
\draw[violet, line width=1.2pt,style=dashed](1.7,.3)arc[start angle=360,end angle=540, x radius=.8, y radius=.15];
\draw[violet, line width=1.2pt](1.7,-.3) arc[start angle=360,end angle=0, x radius=.8, y radius=.15];
\draw[red, line width=1.2pt,->](1.7,0.1) arc[start angle=360,end angle=270, x radius=.8, y radius=.15];
\draw[black,  very thick] (-0.55,-1.7) to[curve through={(-.4,-1.2)..(0,-.5)}] (0.1,-.33);
\draw[black,  very thick] (-0.55,1.7) to[curve through={(-.4,1.2)..(0,.5)}] (0.1,.3);
\draw[black,  very thick] (2.55,-1.7) to[curve through={(2.4,-1.2)..(2.0,-.65)}] (1.7,-.33);
\draw[black,  very thick] (2.55,1.7) to[curve through={(2.4,1.2)..(2,.65)}] (1.7,.3);
\draw[black,  very thick] (0.55,-1.7) to[curve through={(.95,-1)..(1.05,-1)}] (1.45,-1.7);
\draw[black,  very thick] (0.55,1.7) to[curve through={(.95,1)..(1.05,1)}] (1.45,1.7);
\draw[blue,  very thick] (0,1.4) to[curve through={(.3,.9)..(.8,.5)}] (.9,.15);
\draw[blue,  very thick] (1.9,-1.4) to[curve through={(1.5,-.9)..(1.0,-.65)}] (.9,-.45);
\draw[line width=1.2pt,->](2.5,0)--(3.5,0);
\draw (-.5,-0.25) node[above right] {$\text{twist}$}  (2.5,.15) node[above right] {$\text{glue}$}  (6.4,.15) node [above right ] {$\text{pull tight}$} ;
\draw[violet,line width=1.5pt] (4,-1.4) ellipse (.55 and .315);
\draw[violet,line width=1.5pt] (6,-1.4) ellipse (.55 and .315);
\draw[violet,line width=1.5pt] (4,1.4) ellipse (.55 and .315);
\draw[violet,line width=1.5pt] (6,1.4) ellipse (.55 and .315);
\draw[violet, line width=1.2pt](5.7,0) arc[start angle=360,end angle=180, x radius=.8, y radius=.15];
\draw[violet, line width=1.2pt,style=dashed](5.7,0)arc[start angle=360,end angle=540, x radius=.8, y radius=.15];
\draw[black,  very thick] (3.45,-1.4) to[curve through={(3.6,-.9)..(4.1,-.2)}] (4.1,-.03);
\draw[black,  very thick] (3.45,1.4) to[curve through={(3.6,.9)..(4.05,.2)}] (4.1,0);
\draw[black,  very thick] (6.55,-1.4) to[curve through={(6.4,-.9)..(5.9,-.35)}] (5.7,-.03);
\draw[black,  very thick] (6.55,1.4) to[curve through={(6.25,.7)..(5.9,.4)}] (5.7,0);
\draw[black,  very thick] (4.55,-1.4) to[curve through={(4.95,-.7)..(5.05,-.7)}] (5.45,-1.4);
\draw[black,  very thick] (4.55,1.4) to[curve through={(4.95,.7)..(5.05,.7)}] (5.45,1.4);
\draw[blue,  very thick] (3.7,1.15) to[curve through={(4.0,.6)..(4.45,.1)}] (4.5,-.15);
\draw[blue,  very thick] (5.9,-1.1) to[curve through={(5.5,-.6)..(5.0,-.35)}] (4.9,-.15);
\draw[blue,  very thick](4.5,-.125) to[curve through={(4.7,-0.155)}] (4.9,-.155);
\draw[line width=1.2pt,->](6.5,0)--(8,0);
\draw[violet,line width=1.5pt] (9,-1.4) ellipse (.55 and .315);
\draw[violet,line width=1.5pt] (11,-1.4) ellipse (.55 and .315);
\draw[violet,line width=1.5pt] (9,1.4) ellipse (.55 and .315);
\draw[violet,line width=1.5pt] (11,1.4) ellipse (.55 and .315);
\draw[violet, line width=1.2pt](10.7,0) arc[start angle=360,end angle=180, x radius=.8, y radius=.15];
\draw[violet, line width=1.2pt,style=dashed](10.7,0)arc[start angle=360,end angle=540, x radius=.8, y radius=.15];
\draw[black,  very thick] (8.45,-1.4) to[curve through={(8.6,-.9)..(9.1,-.2)}] (9.1,-.03);
\draw[black,  very thick] (8.45,1.4) to[curve through={(8.6,.9)..(9.05,.2)}] (9.1,0);
\draw[black,  very thick] (11.55,-1.4) to[curve through={(11.4,-.9)..(10.9,-.35)}] (10.7,-.03);
\draw[black,  very thick] (11.55,1.4) to[curve through={(11.25,.7)..(10.9,.4)}] (10.7,0);
\draw[black,  very thick] (9.55,-1.4) to[curve through={(9.95,-.7)..(10.05,-.7)}] (10.45,-1.4);
\draw[black,  very thick] (9.55,1.4) to[curve through={(9.95,.7)..(10.05,.7)}] (10.45,1.4);
\draw[blue,  very thick] (8.7,1.15) to[curve through={(9.0,.6)..(9.45,.1)..(9.9,-.15)..(10.5,-.6)}](10.75,-1.15);
\end{tikzpicture}
\end{center}

\caption{The effect of the twisting on geodesic arcs. If the twist parameter was zero, the geodesic arc at the end would be the union of the two geodesic arcs from the original pairs of pants.}
\label{gluing pairs of pants}
\end{figure}
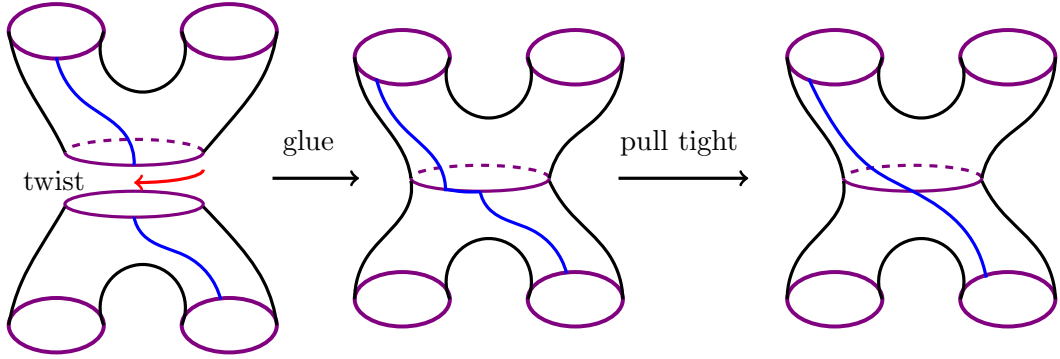
In order to understand the information recorded in the twist parameter, let us consider two hyperbolic pairs of pants with  three geodesic boundaries. If these pairs of pants have boundary components of the same  length, then we can glue them together to obtain a compact hyperbolic surface $\mathcal{R}_{0,4}$ of genus zero with four boundary components. The hyperbolic structure of $\mathcal{R}_{0,4}$ depends on how much we rotate the pairs of pants before gluing. For instance, the shortest arc connecting two boundary components of   $\mathcal{R}_{0,4}$ changes as we change the gluing instruction, (see figure (\ref{gluing pairs of pants})). Thus we have a circle's worth of choices for the hyperbolic structure of $\mathcal{R}_{0,4}$. Therefore, the twist parameters we define on the Teichm\"uller space will be real numbers, but modulo $2\pi$, they are simply recording the angles at which we glue pairs of pants. Below we explain the precise definition of the twist parameters. \par

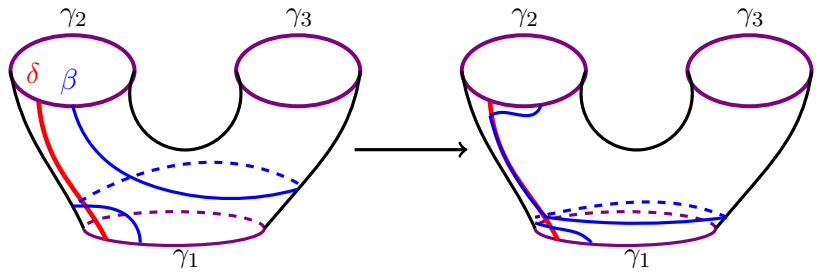
\begin{figure}
\begin{center}
\usetikzlibrary{backgrounds}
\begin{tikzpicture}[scale=1.5]
\draw[violet,line width=1.5pt] (0,1.7) ellipse (.55 and .315);
\draw[violet,line width=1.5pt] (2,1.7) ellipse (.55 and .315);
\draw[violet, line width=1.2pt](1.7,.3) arc[start angle=360,end angle=180, x radius=.8, y radius=.15];
\draw[violet, line width=1.2pt,style=dashed](1.7,.3)arc[start angle=360,end angle=540, x radius=.8, y radius=.15];
\draw[black,  very thick] (-0.55,1.7) to[curve through={(-.4,1.2)..(0,.5)}] (0.1,.3);
\draw[black,  very thick] (2.55,1.7) to[curve through={(2.4,1.2)..(2,.65)}] (1.7,.3);
\draw[black,  very thick] (0.55,1.7) to[curve through={(.95,1)..(1.05,1)}] (1.45,1.7);
\draw[red,  line width=1.75pt] (0.3,.2) to[curve through={(0.275,.25) ..(-.15,.9)}] (-.3,1.425);
\draw[blue,  very thick] (0,1.4) to[curve through={(.3,.9)..(.8,.65)}] (2,.65);
\draw[blue,  very thick,style=dashed] (2,.7) to[curve through={(1.9,.75)..(.5,.75)}] (0,.5);
\draw[blue,  very thick] (0,.5) to[curve through={(0.1,.5)..(.5,.4)}] (.6,0.175);
\draw[line width=1.2pt,->](2.5,1)--(3.5,1);
\draw[violet,line width=1.5pt] (4,1.7) ellipse (.55 and .315);
\draw[violet,line width=1.5pt] (6,1.7) ellipse (.55 and .315);
\draw[violet, line width=1.2pt](5.7,.3) arc[start angle=360,end angle=180, x radius=.8, y radius=.15];
\draw[violet, line width=1.2pt,style=dashed](5.7,.3)arc[start angle=360,end angle=540, x radius=.8, y radius=.15];
\draw[black,  very thick] (3.45,1.7) to[curve through={(3.6,1.2)..(4,.5)}] (4.1,.3);
\draw[black,  very thick] (6.55,1.7) to[curve through={(6.4,1.2)..(6,.65)}] (5.7,.3);
\draw[black,  very thick] (4.55,1.7) to[curve through={(4.95,1)..(5.05,1)}] (5.45,1.7);
\draw[red,  line width=1.75pt] (4.3,.2) to[curve through={(4.275,.25) ..(3.85,.9)}] (3.7,1.425);
\draw[blue,  very thick] (4.15,1.4) to[curve through={(4.13,1.35)..(3.8,1.33)}] (3.7,1.3);
\draw[blue,  very thick] (4.2,.4) to[curve through={(3.85,.9)}] (3.7,1.3);
\draw[blue,  very thick] (4.2,.4) to[curve through={(4.8,.35)}] (5.8,.4);
\draw[blue,  very thick,style=dashed] (5.8,.4) to[curve through={(5.7,.44)..(4.5,.48)}] (4.1,.4);
\draw[blue,  very thick] (4.1,.35) to[curve through={(4.3,.3)..(4.5,.25)}] (4.6,0.175);
\draw (-.2,2) node[above right] {$\gamma_2$}  (1.8,2) node[above right] {$\gamma_3$}  (.8,.2) node [below right ] {$\gamma_1$}  (3.8,2) node[above right] {$\gamma_2$}  (5.8,2) node[above right] {$\gamma_3$}  (4.8,.2) node [below right ] {$\gamma_1$} (-.5,1.5) node[above right] {\color{red}$\delta$}  (-.2,1.4) node [above right ] {\color{blue}$\beta$};
\end{tikzpicture}
\end{center}

\caption{Modifying an arc on a pair of pants so that it agrees with a perpendicular arc except near its endpoints.}
\label{twist parameter}
\end{figure}

Assume that  $\beta$ is an arc in a hyperbolic pair of pants $P$ connecting the boundary components $\gamma_1$ and $\gamma_2$ of $P$. Let $\delta$ be the unique shortest arc connecting $\gamma_1$ and $\gamma_2$. Let $\mathcal{N}_1$ and $\mathcal{N}_2$ be the neighbourhoods of $\gamma_1$ and $\gamma_2$. Modify $\beta$ by isotopy, leaving the endpoints fixed, so that it agrees with $\delta$ outside of $\mathcal{N}_1\cup\mathcal{N}_2$; see figure (\ref{twist parameter}). The {\it twisting number}  of $\beta$ at $\gamma_1$ is the signed horizontal displacement of the endpoints $\beta\cap\partial\mathcal{N}_1$. The orientation of $\gamma_1$ determines the sign. Similarly, the twisting number  of $\beta$ at $\gamma_2$ is the signed horizontal displacement of the endpoints $\beta\cap\partial\mathcal{N}_2$.\par

Then we define the $i^{\text{th}}$ twist parameter $\tau_i( \widetilde{\mathcal{R}})$ of a given Riemann surface $\widetilde{\mathcal{R}}\in \mathcal{T}_{g,n}(\mathcal{R})$ as follows. Assume that $\beta_j$ is one of the two seams that cross $\gamma_i$. On each side of $\gamma_i$ there is a pair of pants, and  $\beta_j$ gives an arc in each of these. Let $t_L$ and $t_R$ be the twisting numbers of each of these arcs on the left and right sides of $\gamma_i$, respectively. The $i^{\text{th}}$ twist parameter of $\widetilde{\mathcal{R}}$ is defined to be
\begin{equation}
\tau_i( \widetilde{\mathcal{R}})=2\pi\frac{t_L-t_R}{\ell_i( \widetilde{\mathcal{R}})}.
\end{equation}
Note that, there were two choices of seams $\beta_j$. It is possible to show that the twist parameters computed from the two seams are the same \cite{Farb}.

\subsection{Fundamental domain for  the MCG inside the Teichm\"uller space}

 \begin{figure}
\begin{center}
\usetikzlibrary{backgrounds}
\begin{tikzpicture}[scale=.85]
\draw[violet,line width=1.5pt] (1,.7) ellipse (.115 and .315);
\draw[violet,line width=1.5pt] (1,-1.7) ellipse (.115 and .315);
\draw[violet,line width=1.5pt] (3.35,1.5) ellipse (.35 and .15);
\draw[blue, line width=1.2pt](2.95,-1.2) arc[start angle=-90,end angle=90, x radius=.15, y radius=1.15];
\draw[blue, line width=1.2pt,style=dashed](2.95,-1.2)arc[start angle=270,end angle=90, x radius=.15, y radius=1.15];
\draw[blue, line width=1.2pt](4.25,-1.1) arc[start angle=-90,end angle=90, x radius=.15, y radius=.92];
\draw[blue, line width=1.2pt,style=dashed](4.25,-1.1)arc[start angle=270,end angle=90, x radius=.15, y radius=.92];
\draw[blue, line width=1.2pt](2.25,.125) arc[start angle=-90,end angle=90, x radius=.1, y radius=.35];
\draw[blue, line width=1.2pt,style=dashed](2.25,.125)arc[start angle=270,end angle=90, x radius=.1, y radius=.35];
\draw[blue, line width=1.2pt](2.25,-1.6) arc[start angle=-90,end angle=90, x radius=.1, y radius=.39];
\draw[blue, line width=1.2pt,style=dashed](2.25,-1.6)arc[start angle=270,end angle=90, x radius=.1, y radius=.39];
\draw[blue, line width=1.2pt,style=dashed](1.95,-.55) arc[start angle=0,end angle=180, x radius=.225, y radius=.1];
\draw[blue, line width=1.2pt](1.95,-.55)arc[start angle=360,end angle=180, x radius=.225, y radius=.1];
\draw[blue, line width=1.2pt,style=dashed](6.77,-.25)arc[start angle=0,end angle=180, x radius=.39, y radius=.1];
\draw[blue, line width=1.2pt](6.77,-.25)arc[start angle=360,end angle=180, x radius=.39, y radius=.1];
\draw[black,  very thick] (1,0.4) to[curve through={(1.25,.2)..(1.25,-1.185)}] (1,-1.385);
\draw[black,  very thick] (1,-2.015) to[curve through={(2,-1.75)..(3,-1.2)..(4,-1)..(5.5,-1.5)..(6.6,-.9)..(6.75,-.1)..(6.3,.6)..(5.5,.9)..(3.9,.9)}] (3.7,1.5);
\draw[black,  very thick] (1,1.015) to[curve through={(1.5,1)..(1.75,.95)..(2.75,.9)}] (3.0,1.5);
\draw[ line width=1.2pt](2.8,-.35) arc[start angle=0,end angle=360, x radius=.45, y radius=.5];
\draw[ line width=1.2pt](6,-.35) arc[start angle=0,end angle=360, x radius=.55, y radius=.5];
\draw (3.2,-1.7)node [below right ] {$\mathcal{R}_{1}$} ;
\draw[violet,line width=1.5pt] (10,.7) ellipse (.115 and .315);
\draw[violet,line width=1.5pt] (10,-1.7) ellipse (.115 and .315);
\draw[violet,line width=1.5pt] (12.35,1.5) ellipse (.35 and .15);
\draw[blue, line width=1.2pt](11.95,-1.2) arc[start angle=-90,end angle=90, x radius=.15, y radius=1.15];
\draw[blue, line width=1.2pt,style=dashed](11.95,-1.2)arc[start angle=270,end angle=90, x radius=.15, y radius=1.15];
\draw[blue, line width=1.2pt](13.25,-1.1) arc[start angle=-90,end angle=90, x radius=.15, y radius=.92];
\draw[blue, line width=1.2pt,style=dashed](13.25,-1.1)arc[start angle=270,end angle=90, x radius=.15, y radius=.92];
\draw[blue, line width=1.2pt](11.25,.125) arc[start angle=-90,end angle=90, x radius=.1, y radius=.35];
\draw[blue, line width=1.2pt,style=dashed](11.25,.125)arc[start angle=270,end angle=90, x radius=.1, y radius=.35];
\draw[blue, line width=1.2pt](11.25,-1.6) arc[start angle=-90,end angle=90, x radius=.1, y radius=.39];
\draw[blue, line width=1.2pt,style=dashed](11.25,-1.6)arc[start angle=270,end angle=90, x radius=.1, y radius=.39];
\draw[blue, line width=1.2pt,style=dashed](10.95,-.55) arc[start angle=0,end angle=180, x radius=.225, y radius=.1];
\draw[blue, line width=1.2pt](10.95,-.55)arc[start angle=360,end angle=180, x radius=.225, y radius=.1];
\draw[blue, line width=1.2pt](15.15,-.35)arc[start angle=360,end angle=0, x radius=.75, y radius=.65];
\draw[black,  very thick] (10,0.4) to[curve through={(10.25,.2)..(10.25,-1.185)}] (10,-1.385);
\draw[black,  very thick] (10,-2.015) to[curve through={(11,-1.75)..(12,-1.2)..(13,-1)..(14.5,-1.5)..(15.6,-.9)..(15.75,-.1)..(15.3,.6)..(14.5,.9)..(12.9,.9)}] (12.7,1.5);
\draw[black,  very thick] (10,1.015) to[curve through={(10.5,1)..(10.75,.95)..(11.75,.9)}] (12.0,1.5);
\draw[ line width=1.2pt](11.8,-.35) arc[start angle=0,end angle=360, x radius=.45, y radius=.5];
\draw[ line width=1.2pt](15,-.35) arc[start angle=0,end angle=360, x radius=.55, y radius=.5];
\draw (12.2,-1.7)node [below right ] {$\mathcal{R}_{2}$} ;
\end{tikzpicture}
\end{center}

\caption{The hyperbolic surfaces $\mathcal{R}_{1}$ and $\mathcal{R}_{2}$ are not the same point of $\mathcal{T}_{g,n}(\mathcal{R})$ due to the difference in the pants decomposition associated with them. However, both  represents the same hyperbolic Riemann surface.}
\label{different points in T}
\end{figure}
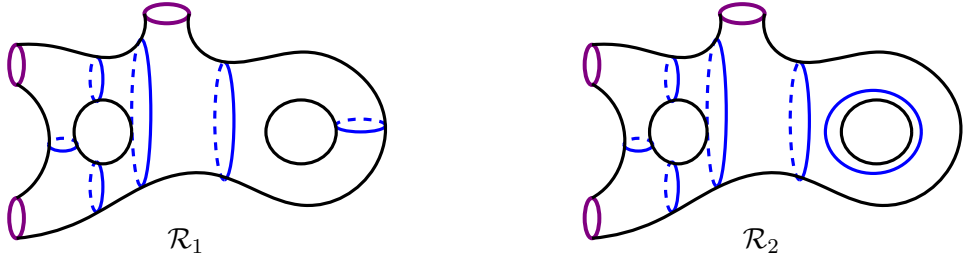

Let $\mathcal{S}$ be a  bordered Riemann surface. Denote the group of orientation-preserving diffeomorphisms of $\mathcal{S}$ that restrict to the identity on $\partial \mathcal{S}$ by $\text{Diff}^+(\mathcal{S},\partial\mathcal{S})$. Then, the {\it mapping class group} (MCG) of $\mathcal{S}$, denoted by $\text{Mod}(\mathcal{S})$, is the group
\begin{equation}
\text{MCG}(\mathcal{S})=\text{Diff}^+(\mathcal{S},\partial\mathcal{S})/\text{Diff}_0(\mathcal{S},\partial\mathcal{S}),
\end{equation}
where $\text{Diff}_0(\mathcal{S},\partial\mathcal{S})$ denotes the components of $\text{Diff}^+(\mathcal{S},\partial\mathcal{S})$ that can be continuously connected to the identity. The {\it moduli space} of hyperbolic surfaces homemorphic to $\mathcal{S}$ is defined to be the quotient space 
\begin{equation}
\mathcal{M}(\mathcal{S})=\mathcal{T}(\mathcal{S})/\text{MCG}(\mathcal{S}),
\end{equation}
where $\mathcal{T}(\mathcal{S})$ is the Teichm\"uller space of hyperbolic Riemann surfaces  homoemorphic to $\mathcal{S}$ with a specific pants decomposition associated with it. The elements in the group $\text{Mod}(\mathcal{S}$  act on $\mathcal{T}(\mathcal{S})$ simply by changing the pants decomposition. \par

    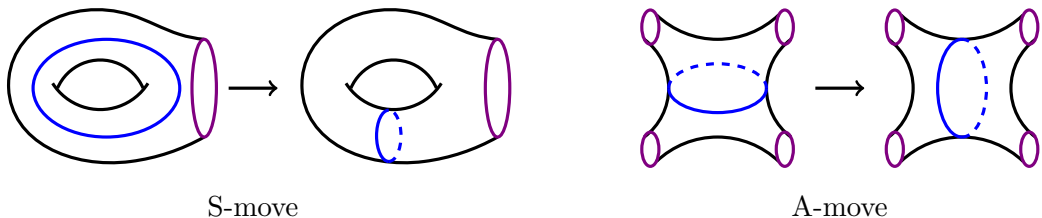
\begin{figure}
\begin{center}
\usetikzlibrary{backgrounds}
\begin{tikzpicture}[scale=.325]
\draw[black, very thick] (0,-2) to[curve through={(-3,-3)..(-8,0)..(-3,3)..(-1,2.2)}] (0,2);
\draw[black, very thick] (-2.5,0) to[curve through={(-3.5,1)..(-5,1)}] (-6,0);
\draw[black, very thick] (-2.3,.2) to[curve through={(-3.5,-.75)..(-5,-.75)}] (-6.2,.2);
\draw[violet, very thick](0,0) ellipse (.5 and 2);
\draw  node[below] at (2,-4) {S-move};
\draw[ blue, very thick](-4,0) ellipse (3 and 2);
\draw[very thick,->] (1,0)--(3,0);
\draw[black, very thick] (12,-2) to[curve through={(9,-3)..(4,0)..(9,3)..(11,2.2)}] (12,2);
\draw[black, very thick] (9.5,0) to[curve through={(8.5,1)..(7,1)}] (6,0);
\draw[black, very thick] (9.7,.2) to[curve through={(8.5,-.75)..(7,-.75)}] (5.8,.2);
\draw[violet, very thick](12,0) ellipse (.5 and 2);
\draw[ blue,very thick](7.55,-3) arc[start angle=270,end angle=90, x radius=.5, y radius=1.05];
\draw[ blue,very thick,style=dashed](7.55,-3) arc[start angle=-90,end angle=90, x radius=.5, y radius=1.05];

\draw[black,very  thick] (18,-2) to[curve through={(19,0)}] (18,2);
\draw[black,very  thick] (24,-2) to[curve through={(23,0)}] (24,2);
\draw[black,very  thick] (18.5,-3) to[curve through={(21,-2)}] (23.5,-3);
\draw[black, very thick] (18.5,3) to[curve through={(21,2)}] (23.5,3);
\draw  node[below] at (26,-4) {A-move};
\draw[ violet,very thick](18.25,-2.5) ellipse (.35 and .7);
\draw[violet, very thick](18.25,2.5) ellipse (.35 and .7);
\draw[ violet,very thick](23.75,-2.5) ellipse (.35 and .7);
\draw[violet, very thick](23.75,2.5) ellipse (.35 and .7);
\draw[ blue,very thick](19,0)  arc[start angle=180,end angle=360, x radius=2, y radius=1];
\draw[ blue,very thick,style=dashed](19,0)  arc[start angle=180,end angle=0, x radius=2, y radius=1];
\draw[ very thick,->] (25,0)--(27,0);

\draw[black,very  thick] (28,-2) to[curve through={(29,0)}] (28,2);
\draw[black, very thick] (34,-2) to[curve through={(33,0)}] (34,2);
\draw[black,very  thick] (28.5,-3) to[curve through={(31,-2)}] (33.5,-3);
\draw[black,very  thick] (28.5,3) to[curve through={(31,2)}] (33.5,3);
\draw[ violet,very thick](28.25,-2.5) ellipse (.35 and .7);
\draw[violet, very thick](28.25,2.5) ellipse (.35 and .7);
\draw[ violet,very thick](33.75,-2.5) ellipse (.35 and .7);
\draw[ violet,very thick](33.75,2.5) ellipse (.35 and .7);
\draw[ blue,very thick](31,-2) arc[start angle=270,end angle=90, x radius=1, y radius=2];
\draw[ blue,very thick,style=dashed](31,-2) arc[start angle=-90,end angle=90, x radius=1, y radius=2];

\end{tikzpicture}
\end{center}

\caption{ The S and the A moves }
\label{ A and S move}
\end{figure}
 
A presentation of the mapping class groups in terms of Dehn twists is given in  \cite{Thurs1} A pants decomposition of $\mathcal{R}$, a hyperbolic Riemann surface, decomposes  $\mathcal{R}$ into a set of pairs of pants. If $\mathcal{R}$ is not itself a pair of pants, then there are  infinitely many different isotopy classes of pants decompositions of $\mathcal{R}$. Interestingly, any two isotopy classes of pants decompositions can be joined by a finite sequence of elementary moves in which only one closed curve changes at a time \cite{Hatch1}; (see figure (\ref{ A and S move})).  The different types of elementary moves are as follows:
 \begin{itemize}
 
 \item Let $\mathcal{P}$ be a pants decomposition, and assume that one of the simple closed curve $\gamma$ of $\mathcal{P}$ is such that deleting $\gamma$ from $\mathcal{R}$ produces a one holed torus as a complementary component.This is equivalent to saying that there is a simple closed curve $\beta$ in $\mathcal{R}$ which intersects $\gamma$ in one point transversely and is disjoint from all other circles in $\mathcal{P}$. In this case, replacing $\gamma$ by $\beta$ in $\mathcal{P}$ produces a new pants decomposition $\mathcal{P}'$. This replacement is known as a simple move or S-move. 
 
 \item If $\mathcal{P}$ contains a simple closed curve $\gamma$ such that deleting $\gamma$ from $\mathcal{P}$ produces a four holed sphere as a complementary component, then we obtain a new pants decomposition $\mathcal{P}'$ by replacing $\gamma$ with a simple closed curve $\beta$ which intersects $\gamma$ transversely in two points and is disjoint from the other curves of $\mathcal{P}$. The transformation $\mathcal{P}\to \mathcal{P}'$ in this case is called an associative move or A-move.   

\end{itemize}

The inverse of an S-move is again an S-move, and the inverse of an A-move is again an A-move. Unfortunately, the presentation so obtained is rather complicated, and stands in need of considerable simplification before much light can be shed on the structure of the MCG. As a result, {\it in the generic situation, it seems hopeless to obtain the explicit description of a fundamental domain  for the action of the MCG on the Teichm\"uller space \cite{penner1,penner2}}. 

\section{The Mirzakhani-McShane identity}\label{MMidentity}

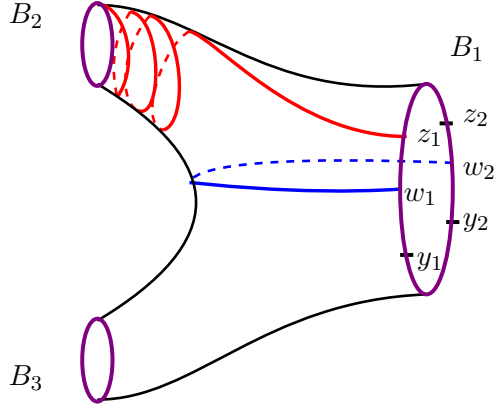
\begin{figure}
\begin{center}
\begin{tikzpicture}[scale=1.75]
\draw[line width=1pt] (1,1) .. controls (1.75,1) and (2.25,.25)  ..(3.5,.4);
\draw[line width=1pt] (1,-2) .. controls(1.75,-2) and (2.25,-1.25)  ..(3.5,-1.2);
\draw[line width=1pt,style=dashed,blue] (3.7,-.2) .. controls (2.5,-.15) and (1.75,-.2)  ..(1.7,-.35);
\draw[line width=1.3pt,blue] (3.3,-.4) .. controls (2.5,-.45) and (1.75,-.35)  ..(1.7,-.35);
\draw[line width=1.3pt,red] (3.35,0) .. controls (2.5,0) and (1.9,.75)  ..(1.7,.8);
\draw[line width=1pt,style=dashed,red] (1.7,.8) .. controls (1.5,.7) and (1.3,0.1)  ..(1.5,0.05);
\draw[line width=1.3pt,red] (1.4,.92) .. controls (1.65,.85) and (1.7,0.1)  ..(1.5,0.05);
\draw[line width=1pt,style=dashed,red] (1.4,.92) .. controls (1.2,.7) and (1.15,0.3)  ..(1.3,0.2);
\draw[line width=1.3pt,red] (1.25,.95) .. controls (1.5,.9) and (1.5,0.1)  ..(1.3,0.2);
\draw[line width=1pt,style=dashed,red] (1.25,.95) .. controls (1.1,.7) and (1.11,0.5)  ..(1.15,0.3);
\draw[line width=1.3pt,red] (1.05,1) .. controls (1.25,.9) and (1.3,0.5)  ..(1.15,0.3);
\draw[line width=1pt] (1,.4) .. controls(2,-.2)  and (2,-.8) ..(1,-1.4);
\draw[line width=1.5pt] (3.6,.1)--(3.7,.1);
\draw[line width=1.5pt] (3.65,-.65)--(3.75,-.65);
\draw[line width=1.5pt] (3.3,-.9)--(3.4,-.9);
\draw[violet, line width=1.5pt] (1,.7) ellipse (.115 and .315);
\draw[violet,line width=1.5pt] (1,-1.7) ellipse (.115 and .315);
\draw[violet, line width=1.5pt] (3.5,-.4) ellipse (.2 and .8);
\draw (0.25,.75) node[above right] {\color{black}$B_2$}  (0.25,-2) node[above right] {\color{black}$B_3$}  (3.6,.5)node [above right ] {\color{black}$B_1$}  (3.35,-.15)  node[above right] {$z_1$}(3.7,0)  node[above right] {$z_2$} (3.7,-.8)  node[above right] {$y_2$} (3.35,-1.1)  node[above right] {$y_1$} (3.25,-.6)  node[above right] {$w_1$}  (3.7,-.4)node [above right ] {$w_2$} ;
\end{tikzpicture}

\end{center}

\caption{The complete geodesics that are disjoint from $B_2,B_3$ and orthogonal to $B_1$.}
\label{complete geodesics}
\end{figure}

 Before stating the Mirzakhni-McShane identity, let us discuss some aspects of infinite simple geodesic rays on a hyperbolic pair of pants. Consider $P(x_1,x_2,x_3)$, the unique hyperbolic pair of pants  with the  geodesic boundary curves $(B_1,B_2,B_3)$ such that $$l_{B_i}(P)=x_i\qquad i=1,2,3.$$  $P(x_1,x_2,x_3)$ is constructed by pasting two copies  of the (unique) right hexagons along the three  edges. The edges of the right hexagons are  the geodesics  that meet perpendicularly with the non-adjacent sides of length $\frac{x_1}{2},\frac{x_2}{2}$ and $\frac{x_3}{2}$. Consequently, $P(x_1,x_2,x_3)$ admits a reflection  involution symmetry $\mathcal{J}$ which interchanges the two hexagons. Such a hyperbolic pair of pants has five complete geodesics disjoint form $B_1,B_2$ and $B_3$. Two of these geodesics meet the border $B_1$, say at positions $z_1$ and $z_2$, and spiral around the border $B_2$, see figure (\ref{complete geodesics}).  Similarly, the other two geodesics meet the border $B_1$, say at $y_1$ and $y_2$, and spiral around the border $B_3$. The  simple geodesic that emanates perpendicularly from  the border $B_1$ to itself meet the border  $B_1$ perpendicularly at two points, say at  $w_1$ and $w_2$, is the fifth complete geodesic. The involution symmetry $\mathcal{J}$ relates the two points $w_1$ and $w_2$, i.e. $\mathcal{J}(w_1)=w_2$. Likewise,  $\mathcal{J}(z_1)=z_2$ and $\mathcal{J}(y_1)=y_2$.  The geodesic length  of the interval   between $z_1$ and $z_2$ along the boundary  $\beta_1$ containing $w_1$ and $w_2$ is given by \cite{Mirzakhani:2006fta}
    \begin{equation}\label{rinterval}
    \mathcal{D}\left(x_1,x_2,x_3\right)=x_1-\mathrm{ln}\left(\frac{\mathrm{cosh}(\frac{x_2}{2})+\mathrm{cosh}\left(\frac{x_1+x_3}{2}\right)}{\mathrm{cosh}\left(\frac{x_2}{2}\right)+\mathrm{cosh}\left(\frac{x_1-x_3}{2}\right)}\right).
    \end{equation}
    Twice of the geodesic distance between the two geodesics that are perpendicular to the boundary $B_1$ and spiralling around the boundary $B_2$ and the boundary $B_3$ is given by \cite{Mirzakhani:2006fta}
     \begin{equation}\label{einterval}
    \mathcal{E}\left(x_1,x_2,x_3\right)=2~\mathrm{ln}\left(\frac{e^{\frac{x_1}{2}}+e^{\frac{x_2+x_3}{2}}}{e^{-\frac{x_1}{2}}+e^{\frac{x_2+x_3}{2}}}\right).
    \end{equation}
 
\begin{figure}
\begin{center}
\usetikzlibrary{backgrounds}
\begin{tikzpicture}[scale=1]
\draw[line width=1pt] (1,1) .. controls (1.75,1) and (2.25,.75)  ..(2.75,.2);
\draw[line width=1pt] (1,-2) .. controls(1.75,-2) and (2.25,-1.75)  ..(3,-1);
\draw[line width=1pt] (1,.4) .. controls(1.7,0)  and (1.7,-1) ..(1,-1.4);
\draw[violet,line width=1pt] (1,.7) ellipse (.115 and .315);
\draw[violet,line width=1pt] (1,-1.7) ellipse (.115 and .315);
\draw[line width=1pt] (3,-1) .. controls(3.3,-.75) and (3.75,-.75) ..(4,-1);
\draw[line width=1pt] (4,0.2) .. controls(5,1) and (6,1) ..(7,0);
\draw[line width=1pt] (2.75,0.2) .. controls(3,0.05) and (3.1,0.2) ..(3,1.5);
\draw[line width=1pt] (4,0.2) .. controls(3.8,0.05) and (3.6,0.2) ..(3.7,1.5);
\draw[line width=1pt] (4,-1) .. controls(5,-2) and (6,-2) ..(7,-1);
\draw[line width=1pt] (7,0) .. controls(7.2,-.15)  ..(8,-.15);
\draw[line width=1pt] (7,-1) .. controls(7.2,-.85)  ..(8,-.85);
\draw[violet, line width=1pt] (8,-.5) ellipse (.15 and .35);
\draw[blue, line width=1pt] (3.35,1.5) ellipse (.35 and .15);
\draw[line width=1pt] (4.75,-.75) .. controls(5.25,-1.1) and (5.75,-1.1) ..(6.25,-.75);
\draw[line width=1pt] (4.75,-.35) .. controls(5.25,0.1) and (5.75,0.1) ..(6.25,-.35);
\draw[line width=1pt] (4.75,-.35) .. controls(4.65,-.475) and (4.65,-.625) ..(4.75,-.75);
\draw[line width=1pt] (6.25,-.35) .. controls(6.35,-.475) and (6.35,-.625) ..(6.25,-.75);
\draw[red, line width=1pt, style=dashed] [rotate around={160:(2.8,.17)}]  (2.8,.17) arc[start angle=360,end angle=180, x radius=1, y radius=.15];
		\draw[red, line width=1.3pt, style] [rotate around={160:(2.8,.17)}]  (2.8,.17)arc[start angle=0,end angle=180, x radius=1, y radius=.15];
\draw[line width=1pt] (2.2,-.5) ellipse (.25 and .4);
\draw[red, line width=1pt, style=dashed] (5.5,.76) arc[start angle=450,end angle=270, x radius=.15, y radius=.4];
		\draw[red, line width=1.3pt, style] (5.5,.76)arc[start angle=90,end angle=270, x radius=.15, y radius=.4];
\draw[line width=1pt] (2.2,-.5) ellipse (.25 and .4);

\draw (-0.25,.5) node[above right] {\color{black}$L_2$}  (-0.25,-2) node[above right] {\color{black}$L_3$} (8.2,-.2)node [below right ] {\color{black}$L_4$}  (3.2,1.5)node [above right ] {\color{black}$L_1$} (3.5,-.3) node  [below right ] {$\alpha_1$} (5.55,.5) node  [below right ] {$\alpha_2$};
\end{tikzpicture}
\end{center}

\caption{Scissoring  the surface along the curve $\alpha_1+\alpha_2$ produces a pair of pants and genus 1 surface with 5 borders.}
\label{cutting2}
\end{figure}
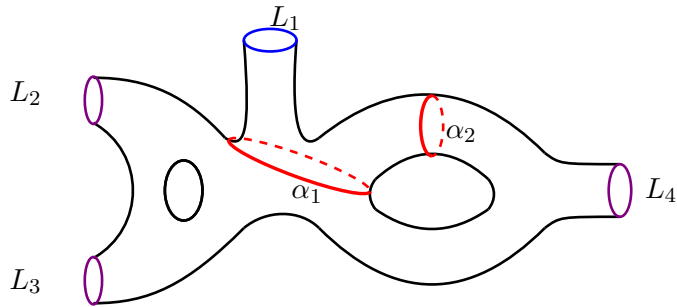

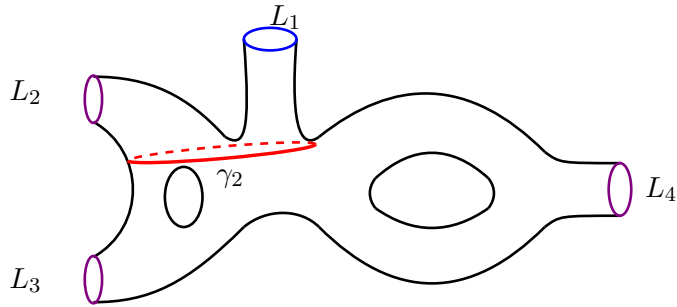
\begin{figure}
\begin{center}
\usetikzlibrary{backgrounds}
\begin{tikzpicture}[scale=1]
\draw[line width=1pt] (1,1) .. controls (1.75,1) and (2.25,.75)  ..(2.75,.2);
\draw[line width=1pt] (1,-2) .. controls(1.75,-2) and (2.25,-1.75)  ..(3,-1);
\draw[line width=1pt] (1,.4) .. controls(1.7,0)  and (1.7,-1) ..(1,-1.4);
\draw[violet, line width=1pt] (1,.7) ellipse (.115 and .315);
\draw[violet, line width=1pt] (1,-1.7) ellipse (.115 and .315);
\draw[line width=1pt] (3,-1) .. controls(3.3,-.75) and (3.75,-.75) ..(4,-1);
\draw[line width=1pt] (4,0.2) .. controls(5,1) and (6,1) ..(7,0);
\draw[line width=1pt] (2.75,0.2) .. controls(3,0.05) and (3.1,0.2) ..(3,1.5);
\draw[line width=1pt] (4,0.2) .. controls(3.8,0.05) and (3.6,0.2) ..(3.7,1.5);
\draw[line width=1pt] (4,-1) .. controls(5,-2) and (6,-2) ..(7,-1);
\draw[line width=1pt] (7,0) .. controls(7.2,-.15)  ..(8,-.15);
\draw[line width=1pt] (7,-1) .. controls(7.2,-.85)  ..(8,-.85);
\draw[violet,line width=1pt] (8,-.5) ellipse (.15 and .35);
\draw[blue, line width=1pt] (3.35,1.5) ellipse (.35 and .15);
\draw[line width=1pt] (4.75,-.75) .. controls(5.25,-1.1) and (5.75,-1.1) ..(6.25,-.75);
\draw[line width=1pt] (4.75,-.35) .. controls(5.25,0.1) and (5.75,0.1) ..(6.25,-.35);
\draw[line width=1pt] (4.75,-.35) .. controls(4.65,-.475) and (4.65,-.625) ..(4.75,-.75);
\draw[line width=1pt] (6.25,-.35) .. controls(6.35,-.475) and (6.35,-.625) ..(6.25,-.75);
\draw[red, line width=1.3pt, style]  [rotate around={5:(3.95,.1)}]  (3.95,.1) arc[start angle=360,end angle=180, x radius=1.25, y radius=.08];
		\draw[red, line width=1pt, style=dashed]  [rotate around={5:(3.95,.1)}]  (3.95,.1)arc[start angle=0,end angle=180, x radius=1.25, y radius=.08];
\draw[line width=1pt] (2.2,-.6) ellipse (.25 and .4);

\draw (-0.25,.5) node[above right] {\color{black}$L_2$}  (-0.25,-2) node[above right] {\color{black}$L_3$} (8.2,-.2)node [below right ] {\color{black}$L_4$}  (3.2,1.5)node [above right ] {\color{black}$L_1$} (2.5,-.1) node  [below right ] {$\gamma_2$};
\end{tikzpicture}
\end{center}

\caption{Scissoring  the surface along the curve $\gamma_2$ produces a pair of pants and a genus 2 surface with 3 borders.}
\label{cutting4}
\end{figure}

  We say that  three isotopy classes of the connected simple closed curves $(\alpha_1,\alpha_2,\alpha_3)$ on $\mathcal{R}$, a genus $g$ hyperbolic Riemann surface with $n$ borders $\mathbf{L}=(L_1,\cdots,L_n)$ having lengths $(l_1,\cdots,l_n)$, bound a pair of pants if there exists an embedded pair of pants $P\subset \mathcal{R}$ such that $\partial P=\{\alpha_1,\alpha_2,\alpha_3\}$. The  boundary curves can have vanishing lengths, which turns a boundary to a puncture. Following definitions are useful:

  \begin{itemize}
  
  \item For $1\leq i\leq n$, we define $\mathcal{F}_i$ be the set of unordered pairs of isotopy classes of the simple closed curves $\{\alpha_1,\alpha_2\}$ bounding a pairs of pants with $L_i$ such that $\alpha_1,\alpha_2\notin \partial(\mathcal{R})$ (see figure (\ref{cutting3}) and  (\ref{cutting2})). 
  
  \item For $1\leq i\neq j \leq n$, we define $\mathcal{F}_{i,j}$ be the set of isotopy classes of the simple closed curves $\gamma$ bounding a pairs of pants containing the borders $L_i$ and  $L_j$ (see figure (\ref{cutting4})). 
  
  \end{itemize}

 \begin{figure}
\begin{center}
\usetikzlibrary{backgrounds}
\begin{tikzpicture}[scale=1]
\draw[line width=1pt] (1,1) .. controls (1.75,1) and (2.25,.75)  ..(2.75,.2);
\draw[line width=1pt] (1,-2) .. controls(1.75,-2) and (2.25,-1.75)  ..(3,-1);
\draw[line width=1pt] (1,.4) .. controls(1.7,0)  and (1.7,-1) ..(1,-1.4);
\draw[violet, line width=1pt] (1,.7) ellipse (.115 and .315);
\draw[violet,line width=1pt] (1,-1.7) ellipse (.115 and .315);
\draw[ line width=1pt] (3,-1) .. controls(3.3,-.75) and (3.75,-.75) ..(4,-1);
\draw[line width=1pt] (4,0.2) .. controls(5,1) and (6,1) ..(7,0);
\draw[line width=1pt] (2.75,0.2) .. controls(3,0.05) and (3.1,0.2) ..(3,1.5);
\draw[line width=1pt] (4,0.2) .. controls(3.8,0.05) and (3.6,0.2) ..(3.7,1.5);
\draw[line width=1pt] (4,-1) .. controls(5,-2) and (6,-2) ..(7,-1);
\draw[line width=1pt] (7,0) .. controls(7.2,-.15)  ..(8,-.15);
\draw[line width=1pt] (7,-1) .. controls(7.2,-.85)  ..(8,-.85);
\draw[violet, line width=1pt] (8,-.5) ellipse (.15 and .35);
\draw[blue, line width=1pt] (3.35,1.5) ellipse (.35 and .15);
\draw[line width=1pt] (4.75,-.75) .. controls(5.25,-1.1) and (5.75,-1.1) ..(6.25,-.75);
\draw[line width=1pt] (4.75,-.35) .. controls(5.25,0.1) and (5.75,0.1) ..(6.25,-.35);
\draw[line width=1pt] (4.75,-.35) .. controls(4.65,-.475) and (4.65,-.625) ..(4.75,-.75);
\draw[line width=1pt] (6.25,-.35) .. controls(6.35,-.475) and (6.35,-.625) ..(6.25,-.75);
	\draw[red, line width=1pt, style=dashed] (2.8,.13) arc[start angle=450,end angle=270, x radius=.15, y radius=.65];
		\draw[red, line width=1.3pt, style] (2.8,.13) arc[start angle=90,end angle=270, x radius=.15, y radius=.65];
			
	\draw[red, line width=1pt, style=dashed] (4.15,.26) arc[start angle=450,end angle=270, x radius=.15, y radius=.7];
		\draw[red, line width=1.3pt, style] (4.15,.26) arc[start angle=90,end angle=270, x radius=.15, y radius=.7];

\draw[line width=1pt] (2.2,-.5) ellipse (.25 and .4);

\draw (0,.5) node[above right] {\color{black}$L_2$}  (0,-2) node[above right] {\color{black}$L_3$} (8.2,-.2)node [below right ] {\color{black}$L_4$}  (3.2,1.5)node [above right ] {\color{black}$L_1$} (2,-1.5) node  [above right ] {$\alpha_1$} (4.3,-1.5) node  [above right ] {$ \alpha_2$};
\end{tikzpicture}
\end{center}

\caption{Scissoring  the surface along the curve $\alpha_1+\alpha_2$ produces a pair of pants, a genus 1 surface with 3 borders and a genus 1 surface with 2 borders.}
\label{cutting3}
\end{figure}
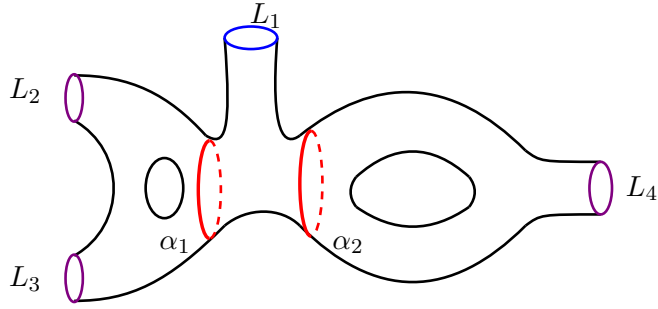
  
  Let us state the Mirzakhani-McShane identity for  hyperbolic bordered surfaces. For any genus $g$ hyperbolic Riemann surface $\mathcal{R}\in \mathcal{T}_{g,n}(l_1,\cdots,l_n)$ with $n$ borders $L_1,\cdots,L_n$ having lengths $l_1,\cdots,l_n$, satisfying $3g-3+n>0$, we have 
  \begin{equation}\label{gmidentity}
  \sum_{\{\alpha_1,\alpha_2\}\in \mathcal{F}_1}\mathcal{D}(l_1,l_{\alpha_1}(\mathcal{R}),l_{\alpha_2}(\mathcal{R}))+\sum_{i=2}^n\sum_{\gamma\in \mathcal{F}_{1,i}}\mathcal{E}(l_1,l_i,l_{\gamma}(\mathcal{R}))=l_1.
  \end{equation}
  
  We use this identity write down a decomposition of  unity:
   \begin{equation}\label{gmidentity11}
  1=\sum_{k}\sum_{\{\alpha^k_{1},\alpha^k_{2}\}\in \mathcal{F}^k_1}\frac{\mathcal{D}(l_1,l_{\alpha^k_1}(\mathcal{R}),l_{\alpha^k_2}(\mathcal{R}))}{l_1}+\sum_{i=2}^n\sum_{\gamma\in \mathcal{F}_{1,i}}\frac{\mathcal{E}(l_1,l_i,l_{\gamma}(\mathcal{R}))}{l_1}.
  \end{equation}
Here, we decomposed  the summation over each simple closed geodesics that are disjoint from the boundaries into a sum over a set of simple closed geodesic that are related each other by the action of MCG and a sum over a discrete variable which differentiate the class of simple closed geodesics that are not related by the action of MCG.

  \section{The Luo-Tan dilogarithm identity}\label{LuoTan}

The most important ingredient for obtaining an effective integral for an integral that involves integration over the moduli space of Riemann surfaces is the identity (\ref{mcshane}).  The Luo-Tan idenity for simple closed geodesics on hyperbolic Riemann surfaces with or without borders \cite{LT01, HT01} provides an example such a decomposition of unity. In this section, we describe this identity in detail.

\subsection{Properly and quasi-properly  embedded geometric surfaces}

    \begin{figure}
\begin{center}
\usetikzlibrary{backgrounds}
\begin{tikzpicture}[scale=.6]

\draw[line width=1.2pt, black] (27.5,-4) to[curve through={(25,-2.5)..(20,0)..(25,2.5)}] (27.5,4);
\draw[line width=1.2pt,black] (25.5,0) to[curve through={(24.5,.5)..(23,.5)}] (22,0);
\draw[line width=1.2pt,black] (25.7,.2) to[curve through={(24.5,-.5)..(23,-.5)}] (21.8,.2);
\draw[line width=1.2pt,black] (28,-2.5) to[curve through={(27.5,-1)..(27.5,1)}] (28,2.5);
\draw[line width=1.4pt,violet](27.75,3.25) ellipse (.5 and .85);
\draw[line width=1.4pt,violet](27.75,-3.25) ellipse (.5 and .85);
\draw  node[below] at (19.5,0.5) {$ \mathcal{C}_2$};
\draw  node[below] at (26.5,-3.5) {$\mathcal{C}_1$};
\draw  node[below] at (29,-3) {$ B_2$};
\draw  node[below] at (29,3.5) {$B_1$};
\draw  node[below] at (24,-1) {$ P_2$};
\draw  node[below] at (26.5,0) {$P_1$};
\draw[red,  line width=1.2pt,style=dashed](22,0) arc[start angle=360,end angle=180, x radius=1, y radius=.6];
\draw[red,  line width=1.5pt](22,0) arc[start angle=0,end angle=180, x radius=1, y radius=.6];
\draw[blue, line width=1.2pt,style=dashed](26.5,3.1) arc[start angle=450,end angle=270, x radius=.5, y radius=3.1];
\draw[blue, line width=1.5pt](26.5,3.1)arc[start angle=90,end angle=270, x radius=.5, y radius=3.1];
\end{tikzpicture}
\end{center}

\caption{The pair of pants $P_1$ with boundaries $\partial P_1=\left\{ \mathcal{C}_1, B_1,B_2\right\}$ is a properly embedded pair of pants on the torus with two borders $B_1$ and $B_2$. The pair of pants $P_2$ with boundaries $\partial P_2=\left\{ \mathcal{C}_1, \mathcal{C}_2,\mathcal{C}_2\right\}$ is a quasi properly embedded pair of pants on the torus.}
\label{proper and quasi embedding}
\end{figure}
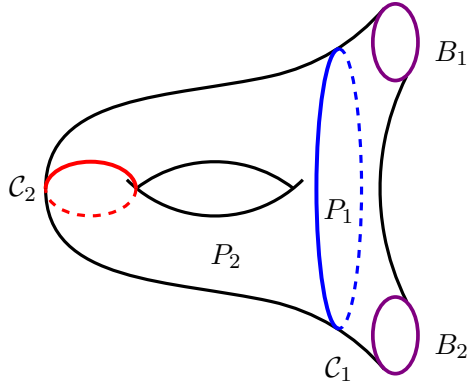

Consider $\mathcal{R}_{g,n}^r$, a genus $g$ hyperbolic Riemann surface  with $n$ geodesic boundary components and $r$ cusps (punctures). We say that a compact embedded subsurface $\mathcal{R}_{g_1,n_1}^{r_1}\subset \mathcal{R}_{g,n}^{r}$ with $g_1\leq g,n_1\leq n,r_1\leq r$ is {\it geometric}, if the boundaries of $\mathcal{R}_{g_1,n_1}^{r_1}$ are geodesics. We say $\mathcal{R}_{g_1,n_1}^{r_1}$ {\it proper}, if the inclusion map 
\begin{equation}
\imath:\mathcal{R}_{g_1,n_1}^{r_1}\to \mathcal{R}_{g,n}^{r},
\end{equation} 
is injective. Therefore, the  subsurface is $\mathcal{R}_{g_1,n_1}^{r_1}$ is said to be a {\it properly  embedded geometric surface}, if its boundaries are geodesics on $\mathcal{R}_{g,n}^{r}$ and the inclusion map is one-to one. \par

For example, each  pair of pants in the   pants decomposition of a genus zero hyperbolic Riemann surface using $n-3$ non-homotopic disjoint simple closed geodesics is a properly embedded geometric surface. However, it is impossible to obtain the   pants decomposition of genus $1$ hyperbolic Riemann surface with one border using properly embedded geometric pair of pants. This is due to the following fact. We obtain a torus with one border  from a hyperbolic pair of pants $P$ by identifying the two boundaries of it. The identification of the boundary makes the inclusion map 
\begin{equation}
\imath:P\to \mathcal{R}_{1,1}^{0},
\end{equation}
 non-injective. Therefore, the only properly embedded hyperbolic surface inside a  hyperbolic torus with one border is the surface itself. \par
 
 We define a {\it quasi-embedded geometric} pair of pants   $P$  in $\mathcal{R}_{g,n}^{r}$ to be an immersion $P$ into $\mathcal{R}_{g,n}^{r}$ which is injective on the interior $\text{int}(P)$ of $P$ such that the boundaries are mapped to geodesics, but two of the boundaries are mapped to the same geodesic. Hence, a quasi-embedded geometric pair of pants is contained in a unique embedded geometric 1-holed torus. Conversely, every embedded geometric 1-holed torus together with a non-trivial simple closed geodesic on the torus that is not parallel to its boundary geodesic determines a quasi-embedded geometric pair of pants (see figure (\ref{proper and quasi embedding})).

\subsection{The Roger's dilogarithm functions}

The dilogarithm function $\text{Li}_2$ is defined for $z\in \mathbb{C}$ with   $\left |z\right |<1$ by the following Taylor series
\begin{equation}\label{dilog1}
\text{Li}_2(z)=\sum_{n=1}^{\infty}\frac{z^n}{n^2}.
\end{equation}
It is straightforward to verify that, for $x\in \mathbb{R}$ with $\left |x\right |<1$, the dilogarithm function has the following expression
\begin{equation}\label{dilog2}
\text{Li}_2(x)=-\int_0^x \frac{\text{ln}\left(1-z\right)}{z}dz.
\end{equation}
The Roger's dilogarithm function $\mathcal{L}$ is defined by
\begin{equation}\label{rogerdilog1}
\mathcal{L}(x)\equiv\text{Li}_2(x)+\frac{1}{2}\text{ln}\left(\left|x\right|\right)\text{ln}\left(1-x\right).
\end{equation}
The Roger's $\mathcal{L}$-function has the following special values
\begin{align}\label{specialvaluerogerl}
\mathcal{L}(0)&=0, \nonumber\\
 \mathcal{L}(\frac{1}{2})&=\frac{\pi^2}{12}, \nonumber\\
  \mathcal{L}(1)&=\frac{\pi^2}{6}, \nonumber\\
   \mathcal{L}(-1)&=-\frac{\pi^2}{12}.
\end{align}
The first derivative of the Roger's $\mathcal{L}$-function has the following simple form
\begin{equation}\label{fdrogersdilog}
\mathcal{L}'(z)=-\frac{1}{2}\left( \frac{\text{ln}(1-z)}{z}+\frac{\text{ln}(z)}{1-z}\right).
\end{equation}
The $\mathcal{L}$-function satisfies the  following Euler relations 
\begin{equation}\label{ELidenity1}
\mathcal{L}(x)+\mathcal{L}(1-x)=\frac{\pi^2}{6},
\end{equation}
for $x\in (0,1)$, and 
\begin{equation}\label{ELidenity2}
\mathcal{L}(-x)+\mathcal{L}(-\frac{1}{x})=2\mathcal{L}(-1)=-\frac{\pi^2}{6},
\end{equation}
for $x>0$. It satisfies the Landen's identity given by
\begin{equation}\label{Landen}
\mathcal{L}\left(\frac{x}{x-1} \right)=-\mathcal{L}\left(x \right),
\end{equation}
for $x\in (0,1)$. If we define $y\equiv \frac{x}{x-1}$, then from the Landen's identity, we get:
\begin{equation}\label{xtoinfty}
\text{lim}_{y\to \infty}\mathcal{L}(y)=-\mathcal{L}(1)=-\frac{\pi^2}{6}.
\end{equation}
It also satisfies the following pentagon relation, for $x,y\in [0,1]$ and $xy\neq 1$
\begin{equation}\label{Lpentidenity}
\mathcal{L}(x)+\mathcal{L}(y)+\mathcal{L}(1-xy)+\mathcal{L}\left(\frac{1-x}{1-xy} \right)+\mathcal{L}\left(\frac{1-y}{1-xy} \right)=\frac{\pi^2}{2}.
\end{equation} 
Finally, we define the Lasso function $La(x,y)$ in terms of  the Roger's $\mathcal{L}$-function as follows
\begin{equation}\label{Lasso}
La(x,y)\equiv\mathcal{L}(y)+\mathcal{L}\left(\frac{1-y}{1-xy}\right)-\mathcal{L}\left(\frac{1-x}{1-xy}\right).
\end{equation}

\subsection{The length invariants of a pair of pants and a 1-holed tori}

\begin{figure}
\begin{center}
\usetikzlibrary{backgrounds}
\begin{tikzpicture}[scale=.6]
\draw[line width=1.2pt,black] (27.75,1.25) to[curve through={(26.5,0)}] (27.75,-1.25);
\draw[line width=1.2pt,black] (27.5,5) to[curve through={(21.5,1.9)}] (20.5,2.1);
\draw[line width=1.2pt,black] (27.5,-5) to[curve through={(21.5,-1.75)}] (20.5,-1.95);
\draw[line width=1.5pt,violet](27.75,3.25) ellipse (.5 and 2);
\draw[line width=1.5pt,violet](27.75,-3.25) ellipse (.5 and 2);
\draw[line width=1.5pt,blue] (27.35,1.75) to[curve through={(25.75,0)}] (27.35,-1.75);
\draw[line width=1.5pt,blue] (27.25,3.5) to[curve through={(21.5,1.3)}] (20.1,1.5);
\draw[line width=1.5pt,blue] (27.25,-3.5) to[curve through={(21.5,-1.3)}] (20.2,-1.5);
\draw[red, line width=1.2pt,style=dashed](26.5,0) arc[start angle=0,end angle=82, x radius=6.5, y radius=.7];
\draw[red, line width=1.5pt](26.5,0)arc[start angle=360,end angle=270, x radius=6.5, y radius=.7];
\draw  node[below] at (19.5,0.5) {$ B_1$};
\draw  node[below] at (26,-1) {$M_1$};
\draw  node[below] at (25.5,2.25) {$M_2$};
\draw  node[below] at (24,-.5) {$M_3$};
\draw  node[below] at (21.5,0.25) {$D_1$};
\draw  node[below] at (29,-3) {$ B_3$};
\draw  node[below] at (29,3.5) {$B_2$};
\draw[violet, line width=1.2pt,style=dashed](20.5,2.1) arc[start angle=450,end angle=270, x radius=.5, y radius=2];
\draw[violet, line width=1.5pt](20.5,2.1)arc[start angle=90,end angle=270, x radius=.5, y radius=2];
\end{tikzpicture}
\end{center}

\caption{The geodesic arcs connecting the boundaries of the pair of pants.}
\label{arcs on pair of pants}
\end{figure}
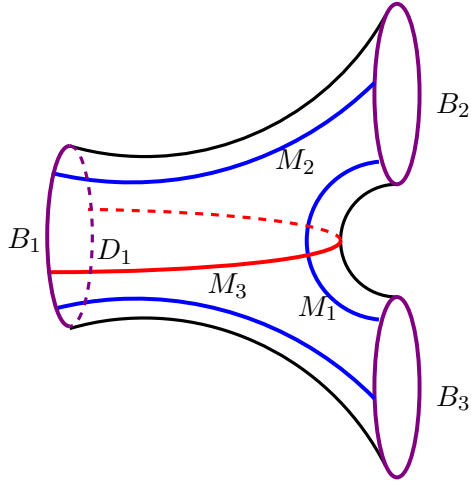

Consider a hyperbolic pair of pants $P$ with geodesic boundaries $B_1,B_2,B_3$ having lengths $l_1,l_2,l_3$ respectively. Let $M_i$ be the shortest geodesic arc between $B_j$ and $B_k$ having length $m_i$, for $\left \{ i,j,k\right\}=\left\{1,2,3 \right\}$. Denote by $D_i$, the shortest non-trivial geodesic arc from $B_i$ to itself having length $p_i$. The hyperbolic metric on the pair of pants makes the geodesic arcs $M_i$ and $D_i$ orthogonal to $\partial P$, the boundaries of the pair of pants $P$  (see figure (\ref{arcs on pair of pants})). \par

By cutting  along the geodesic arcs $M_i,~i=1,2,3$, we can decompose the pair of pants $P$ into two right-angled hyperbolic hexagons with cyclicly ordered side-lengths $\left\{  \frac{l_1}{2},m_3,\frac{l_2}{2},m_1,\frac{l_3}{2},m_2\right\}$. Then, using the following sine and cosine rules for the right angled hexagons and pentagons \begin{align}\label{sine}
\frac{\text{sinh}~m_i}{\text{sinh}\left(\frac{l_i}{2}\right)}&=\frac{\text{sinh}~m_j}{\text{sinh}\left(\frac{l_j}{2}\right)}=\frac{\text{sinh}~m_k}{\text{sinh}\left(\frac{l_k}{2}\right)},\nonumber\\
\text{cosh}~m_i~\text{sinh}\left(\frac{l_j}{2}\right)\text{sinh}\left(\frac{l_k}{2}\right)&=\text{cosh}\left(\frac{l_i}{2}\right)+\text{cosh}\left( \frac{l_j}{2}\right)\text{cosh}\left( \frac{l_k}{2}\right),\nonumber\\
\text{cosh}\left(\frac{p_k}{2}\right)&=\text{sinh}\left(\frac{l_i}{2}\right)\text{sinh}~m_j.
\end{align}
We can express the lengths of the geodesics arcs $D_1,D_2,D_3,M_1,M_2,M_3$ in terms of the lengths of the boundary geodesics $B_1,B_2,B_3$:
\begin{align}\label{sine}
\text{cosh}~m_i &=\text{cosh}\left( \frac{l_i}{2}\right)\text{cosech}\left(\frac{l_j}{2}\right)\text{cosech}\left(\frac{l_k}{2}\right)+\text{coth}\left( \frac{l_j}{2}\right)\text{coth}\left( \frac{l_k}{2}\right),\nonumber\\
\text{cosh}\left(\frac{p_k}{2}\right)&=\text{sinh}\left(\frac{l_i}{2}\right)\text{sinh}~m_j.
\end{align}

    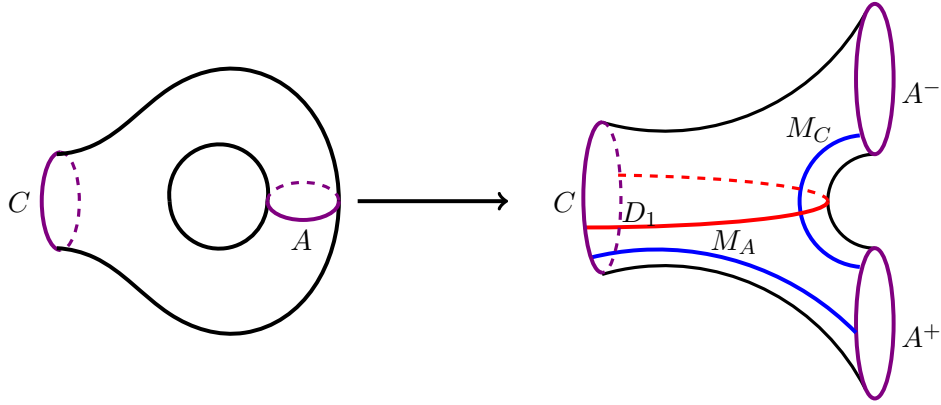
\begin{figure}
\begin{center}
\usetikzlibrary{backgrounds}
\begin{tikzpicture}[scale=.5]
\draw[violet, line width=1.2pt,style=dashed](6.1,1.3) arc[start angle=450,end angle=270, x radius=.5, y radius=1.3];
\draw[violet, line width=1.5pt](6.1,1.3)arc[start angle=90,end angle=270, x radius=.5, y radius=1.3];
\draw[line width=1.5pt,black] (6,1.25) to[curve through={(7,1.5)..(9,3)..(11,3.5)..(13.5,0)..(11,-3.5)..(9,-3)..(7,-1.5)}] (6,-1.25);
\draw[line width=1.5pt,black] (9,0) to[curve through={(10.5,1.5)}] (11.6,0);
\draw[line width=1.5pt,black] (9,0) to[curve through={(10.5,-1.25)}] (11.6,0);
\draw[violet, line width=1.2pt,style=dashed](11.6,0) arc[start angle=180,end angle=0, x radius=.95, y radius=.5];
\draw[violet, line width=1.5pt](11.6,0)arc[start angle=180,end angle=360, x radius=.95, y radius=.5];
\draw[line width=1.5pt,black,->] (14,0)--(18,0);
\draw[line width=1.2pt,black] (27.75,1.25) to[curve through={(26.5,0)}] (27.75,-1.25);
\draw[line width=1.2pt,black] (27.5,5) to[curve through={(21.5,1.9)}] (20.5,2.1);
\draw[line width=1.2pt,black] (27.5,-5) to[curve through={(21.5,-1.75)}] (20.5,-1.95);
\draw[line width=1.5pt,violet](27.75,3.25) ellipse (.5 and 2);
\draw[line width=1.5pt,violet](27.75,-3.25) ellipse (.5 and 2);
\draw[line width=1.5pt,blue] (27.35,1.75) to[curve through={(25.75,0)}] (27.35,-1.75);
\draw[line width=1.5pt,blue] (27.25,-3.5) to[curve through={(21.5,-1.3)}] (20.2,-1.5);
\draw[red, line width=1.2pt,style=dashed](26.5,0) arc[start angle=0,end angle=82, x radius=6.5, y radius=.7];
\draw[red, line width=1.5pt](26.5,0)arc[start angle=360,end angle=270, x radius=6.5, y radius=.7];
\draw  node[below] at (5,0.5) {$ C$};
\draw  node[below] at (12.5,-0.5) {$ A$};
\draw  node[below] at (19.5,0.5) {$ C$};
\draw  node[below] at (26,2.5) {$M_C$};
\draw  node[below] at (24,-.5) {$M_A$};
\draw  node[below] at (21.5,0.25) {$D_1$};
\draw  node[below] at (29,-3) {$ A^+$};
\draw  node[below] at (29,3.5) {$A^-$};
\draw[violet, line width=1.2pt,style=dashed](20.5,2.1) arc[start angle=450,end angle=270, x radius=.5, y radius=2];
\draw[violet, line width=1.5pt](20.5,2.1)arc[start angle=90,end angle=270, x radius=.5, y radius=2];
\end{tikzpicture}
\end{center}

\caption{The image of geodesic arcs on 1 holed torus connecting the boundaries of the pair of pants in the pair of pants obtained by cutting along the geodesic $A$.}
\label{1 holed torus to pair of pants}
\end{figure}
Let us discuss the length invariants of $T$,  a hyperbolic 1-holed torus with boundary component $C$.  Consider an arbitrary closed simple geodesic $A$ that is not parallel to the boundary of $T$. We obtain a hyperbolic pair of pants $P_A$ with boundary geodesics $C,A^+$ and $A^-$ by cutting $T$ along the geodesic $A$.  The pair of pants  $P_A$ is a quasi-properly embedded geometric surface. We denote the length of the  geodesics $C$ and  $A$ by $c$ and $a$ respectively. The length of the shortest geodesic $M_A$ between $C$ and $A^+$ in $P_A$ (or $A^-$) by $m_A$. Finally, $p_A$ denotes the length of the shortest non-trivial geodesic arc from $C$ to $C$ in $P_A$ and $q_A$ denotes the length of the shortest non-trivial geodesic arc from $A^+$ to $A^-$ in $P_A$. Again, using the sine and cosine rules for the hexagons and the pentagons correspond to the hyperbolic pair of pants $P_A$  associated with  $T$, we can express the lengths $m_A,p_A$ and $q_A$ of geodesic arcs in terms of the lengths $c$ and $a$ of boundary geodesics as follows
\begin{align}\label{sinePA}
\text{cosh}~m_A &=\text{cosh}\left( \frac{a}{2}\right)\text{cosech}\left(\frac{a}{2}\right)\text{cosech}\left(\frac{c}{2}\right)+\text{coth}\left( \frac{a}{2}\right)\text{coth}\left( \frac{c}{2}\right),\nonumber\\
\text{cosh}~q_A &=\text{cosh}\left( \frac{c}{2}\right)\text{cosech}\left(\frac{a}{2}\right)\text{cosech}\left(\frac{c}{2}\right)+\text{coth}\left( \frac{a}{2}\right)\text{coth}\left( \frac{a}{2}\right),\nonumber\\
\text{cosh}\left(\frac{p_A}{2}\right)&=\text{sinh}\left(\frac{a}{2}\right)\text{sinh}~m_A.
\end{align} 

\subsection{The Luo-Tan identity}

Consider $\mathcal{R}_{g,n}^{0}$, a hyperbolic Riemann surface of genus $g$ with $n$ geodesic boundary components and no cusps. On $\mathcal{R}_{g,n}^{0}$ there are two types of geometric pairs of pants: quasi-properly embedded geometric pairs of pants and three kinds of properly embedded geometric pairs of pants. Below, we list some useful   functions associated  which each type of geometric pairs of pants. \\

\noindent{\bf Properly embedded geometric pairs of pants}\par
\begin{enumerate}
\item Assume that $P_3$ is a properly embedded geometric pair of pants with no boundaries in common with that of the surface $\mathcal{R}_{g,n}^{0}$. Let the length invariants of $P_3$ be $l_i,m_i$ and $p_i$ for $i=1,2,3$. We define the function $\mathcal{K}_3(P_3)$ in terms of the length invariants of $P_3$ as follows
\begin{equation}\label{f1}
\mathcal{K}_3(P_3)\equiv 4\pi^2-8\left\{\sum_{i=1}^3\left( \mathcal{L}\left(\text{sech}^2\left(\frac{m_i}{2}\right) \right)+\mathcal{L}\left(\text{sech}^2\left(\frac{p_i}{2}\right) \right)\right)+\sum_{i\neq j} La\left( e^{-l_i},\text{tanh}^2\left(\frac{m_i}{2}\right)\right)\right\}.
\end{equation}
\item Assume that $P_2$ is a properly embedded geometric pair of pants with only one boundary in common with the boundaries of the surface $\mathcal{R}_{g,n}^{0}$. Let $L_1$ be the boundary that is also the boundary of  $\mathcal{R}_{g,n}^{0}$ and the length invariants of $P_2$ be $l_i,m_i$ and $p_i$ for $i=1,2,3$. We define the function $\mathcal{K}_2(P_2)$ in terms of the length invariants of $P_2$ as follows
\begin{equation}\label{f1}
\mathcal{K}_2(P_2)\equiv \mathcal{K}_3(P_2)+8\left\{ \mathcal{L}\left(\text{sech}^2\left(\frac{p_1}{2}\right) \right)+La\left( e^{-l_2},\text{tanh}^2\left(\frac{m_3}{2}\right)\right)+La\left( e^{-l_3},\text{tanh}^2\left(\frac{m_2}{2}\right)\right)\right\}.
\end{equation}
\item Assume that $P_1$ is a properly embedded geometric pair of pants with only two boundaries in common with the boundaries of the surface $\mathcal{R}_{g,n}^{0}$. Let $L_1$ and $L_2$ be the boundaries that are also the boundaries of  $\mathcal{R}_{g,n}^{0}$ and the length invariants of $P_1$ be $l_i,m_i$ and $p_i$ for $i=1,2,3$. We define the function $\mathcal{K}_1(P_1)$ in terms of the length invariants of $P_1$ as follows
\begin{align}\label{f1}
\mathcal{K}_1(P_1)\equiv \mathcal{K}_3(P_1)&+8\Big\{\mathcal{L}\left(\text{sech}^2\left(\frac{p_1}{2}\right) \right)+ \mathcal{L}\left(\text{sech}^2\left(\frac{p_2}{2}\right) \right)+\mathcal{L}\left(\text{sech}^2\left(\frac{m_3}{2}\right) \right)\nonumber\\ 
&+La\left( e^{-l_2},\text{tanh}^2\left(\frac{m_3}{2}\right)\right)+La\left( e^{-l_3},\text{tanh}^2\left(\frac{m_2}{2}\right)\right)\nonumber\\ 
&+La\left( e^{-l_3},\text{tanh}^2\left(\frac{m_1}{2}\right)\right)+ La\left( e^{-l_1},\text{tanh}^2\left(\frac{m_3}{2}\right)\right)\Big\}.
\end{align}
\end{enumerate}
 \noindent{\bf Quasi-Properly Embedded Geometric Pairs of Pants}\\
Assume that $P_T$ is a quasi-properly embedded geometric pair of pants in a 1-holed torus $T$ which is a properly geometric embedded surface on $\mathcal{R}_{g,n}^{0}$. Let $C$ is the boundary of $T$ and $A$ be the simple closed geodesic  along  which we cut $T$ to obtain $P_T$. Assume that the length invariants of $T$ be $c,a,m_A,q_A$ and $p_A$ as explained in the previous subsection. We define the function $\mathcal{K}_T(P_T)$ in terms of the length invariants of $P_T$ as follows
\begin{align}\label{g}
\mathcal{K}_T(P_T)&\equiv 8\Big\{ \mathcal{L}\left(\text{tanh}^2\left(\frac{q_A}{2}\right) \right)+2\mathcal{L}\left(\text{tanh}^2\left(\frac{m_A}{2}\right) \right)-\mathcal{L}\left(\text{sech}^2\left(\frac{p_A}{2}\right) \right)\nonumber\\ 
&-2La\left( e^{-a},\text{tanh}^2\left(\frac{m_A}{2}\right)\right)-2 La\left( e^{-\frac{c}{2}},\text{tanh}^2\left(\frac{m_A}{2}\right)\right)\Big\}.
\end{align}
Now, we are in a position to state the Luo-Tan dilogarithm identity. \par

\noindent{\bf The Luo-Tan Identity}: {\it Let $\mathcal{R}_{g,n}^{0}$ be a hyperbolic Riemann surface with $n$ boundaries. Then the functions $\mathcal{K}_1(P_1),\mathcal{K}_2(P_2),\mathcal{K}_3(P_3)$ and $\mathcal{K}_T(P_T)$ satisfies the identity given by
\begin{equation}\label{LuoTanidentity}
\sum_{P_1}\mathcal{K}_1(P_1)+\sum_{P_2}\mathcal{K}_2(P_2)+\sum_{P_3}\mathcal{K}_3(P_3)+\sum_{P_T}\mathcal{K}_T(P_T)=4\pi^2(2g-2+n),
\end{equation}
where the first sum is over all properly embedded geometric pairs of pants sum is over all properly embedded geometric pair of pants $P_1\subset \mathcal{R}_{g,n}^{0}$ with exactly two boundary component in $\partial \mathcal{R}_{g,n}^{0}$,   the second sum is over all properly embedded geometric pairs of pants $P_2\subset \mathcal{R}_{g,n}^{0}$ with exactly one boundary component in $\partial \mathcal{R}_{g,n}^{0}$, the third sum is over all properly embedded geometric pairs of pants $P_3\subset \mathcal{R}_{g,n}^{0}$ such that $\partial P_3\cap \partial \mathcal{R}_{g,n}^{0}=\emptyset$,  and the fourth sum is over all quasi-properly embedded geometric pairs of pants $P_T\subset \mathcal{R}_{g,n}^{0}$. Moreover, if the lengths of $r$ boundary components of $ \mathcal{R}_{g,n}^{0}$ tends to zero, then, we obtain the identity for the hyperbolic surfaces  $\mathcal{R}_{g,n-r}^{r}$ of genus $g$ with $n-r$ geodesic boundary and $r$ cusps. }\par

Let us elaborate on the Luo-Tan identity. A properly embedded geometric pair of pants   inside  $\mathcal{R}_{g,n}^{0}$ can be understood as a set of three simple closed geodesics on $\mathcal{R}_{g,n}^{0}$ that bound $P$.  Therefore, we can replace the sum over all properly embedded geometric pair of pants $P_1$ with the sum over all tuple of simple closed geodesics $\left\{ L_i,L_j,A_{i,j}\right\}$ on $\mathcal{R}_{g,n}^{0}$ that bound the pair of pants $P_1$. Here, $L_i$ and $L_j$  are two boundaries of $\mathcal{R}_{g,n}^{0}$ for $i,j=1,\cdots,n$ and $A_{i,j}$ is a simple closed geodesic which is disjoint from the boundaries of $\mathcal{R}_{g,n}^{0}$. The sum over all properly embedded geometric pair of pants $P_2$ can be replaced with the sum over all tuple of simple closed geodesics $\left\{ L_i,B_i,C_i \right\}$ on $\mathcal{R}_{g,n}^{0}$ that bound the pair of pants $P_2$, for $i=1,\cdots,n$. Here $B_i$ and $C_i$ are two simple closed geodesic disjoint from the boundaries.  The sum over all properly embedded geometric pair of pants $P_3$ can be replaced with the sum over all tuple of simple closed geodesics $\left\{ X,Y,Z \right\}$ on $\mathcal{R}_{g,n}^{0}$ that bound the pair of pants $P_3$. Here, $X,Y$ and $Z$ are three simple closed geodesic disjoint from the boundaries. Similarly, the sum over all quasi-properly embedded pairs of pants $P_T$ can be replaced by the sum over pairs of simple closed geodesics $\left\{ U,V \right\}$ on $\mathcal{R}_{g,n}^{0}$ that bound $P_T$, i.e. $\partial P_T=\{U,V,V\}$. Then we can express the Luo-Tan identity as a decomposition of unity, as follows:
\begin{align}\label{LuoTanidentity1}
1=&\sum_{i< j; i,j =1}^n\sum_{A_{i,j}\in\mathcal{G}_{i,j}}\frac{\mathcal{K}_1\left(L_i,L_j,A_{i,j}\right)}{4\pi^2(2g-2+n)}+\sum_{i =1}^n\sum_k\sum_{\{B_{i,k},C_{i,k}\}\in\mathcal{G}^k_{i}}\frac{\mathcal{K}_2\left(L_i,B_{i,k},C_{i,k}\right)}{4\pi^2(2g-2+n)}\nonumber\\ &+\sum_{q,m}\sum_{\{X_q,Y_m,Z_{q,m}\}\in\mathcal{G}^{q,m}}\frac{\mathcal{K}_3\left(X_{q},Y_m,Z_{q,m}\right)}{4\pi^2(2g-2+n)} +\sum_r\sum_{\{U_r,V_r\}\in\mathcal{G}^{r}_T}\frac{\mathcal{K}_T(U_r,V_r)}{4\pi^2(2g-2+n)}.
\end{align}
Here, we decomposed  the summation over  simple closed geodesics $\gamma$ that are disjoint from the boundaries into a summation over a set of simple closed geodesics that are related each other by the action of elements in ${\text{Mod}}_{g,n}/\text{Stab}(\gamma)$ and a summation over a discrete variables which differentiate the class of simple closed geodesics $\gamma$ that are not related by the action of elements in ${\text{Mod}}_{g,n}/\text{Stab}(\gamma)$. For the first term in the right hand side of (\ref{LuoTanidentity1}) $\gamma=A_{i,j}$, for the second term $\gamma=B_{i,k}+C_{i,k}$, for the third term $\gamma=X_q+Y_m+Z_{q,m}$ and for the last term $\gamma=U_r+V_r$. An arbitrary element in the set $\mathcal{G}_{i,j}$ together with the boundaries $L_i$ and $L_j$ of $\mathcal{R}^0_{g,n}$ form tuple of simple closed geodesics that can be identified as the boundary of a properly embedded geometric pairs of pants inside $\mathcal{R}^0_{g,n}$. An arbitrary element in the set $\mathcal{G}_{i}^k$, which is a pair of closed simple geodesics disjoint from the boundary, together with the boundary $L_i$  of $\mathcal{R}^0_{g,n}$ form tuple of simple closed geodesics that can be identified as the boundary of a properly embedded geometric pairs of pants inside $\mathcal{R}^0_{g,n}$. An arbitrary element in the set $\mathcal{G}^{q,m}$, which is a tuple of simple closed geodesics disjoint from the boundary, can be identified as the boundary of a properly embedded geometric pair of pants with respect to $\mathcal{R}^0_{g,n}$. Finally, an arbitrary element in the set $\mathcal{G}^{r}_T$, which is a pair of simple closed geodesics disjoint from the boundary, can be identified as the boundary of a quasi-properly embedded geometric pair of pants inside $\mathcal{R}^0_{g,n}$. \par

The functions $\mathcal{K}_1,\mathcal{K}_2,\mathcal{K}_3,\mathcal{K}_T$ appearing in the Luo-Tan identity (\ref{LuoTanidentity}) have the following important property:
\begin{equation}\label{f012Tproperty}
\lim_{l_i\to \infty} \mathcal{K}_I(P_I)=\lim_{l_i\to \infty}\mathcal{O}\left( e^{-l_{i}}\right)=0,\qquad I\in \left\{ 1,2,3,T\right\}, \qquad i\in\left\{1,2,3 \right\},
\end{equation}
where $l_i$ is the length of the $i^{\text{th}}$ boundary of the pair of pants $P_I$. Let us  verify this.  The function $\mathcal{K}_3(P_3)$ can be written as 
\begin{equation}\label{f0P0}
\mathcal{K}_3(P_3)=4\sum_{i\neq j}\left\{ 2\mathcal{L}\left( \frac{1-x_i}{1-x_iy_j}\right)-2\mathcal{L}\left( \frac{1-y_j}{1-x_iy_j}\right)-\mathcal{L}(y_j)-\mathcal{L}\left( \frac{(1-y_j)^2x_i}{(1-x_i)^2y_j}\right)\right\},
\end{equation}
where 
\begin{equation}
x_i\equiv e^{-l_i}, \qquad y_i \equiv \text{tanh}^2\left(\frac{m_i}{2} \right),
\end{equation}
with $m_i$ given by
\begin{equation}
\text{cosh}~m_i=\frac{\text{cosh}\left(\frac{l_i}{2}\right)+\text{cosh}\left( \frac{l_j}{2}\right)\text{cosh}\left( \frac{l_k}{2}\right)}{\text{sinh}\left(\frac{l_j}{2}\right)\text{sinh}\left(\frac{l_k}{2}\right)}.
\end{equation}
For $i\neq j, k$ and $\left\{ i,j,k\right\}=\left\{ 1,2,3\right\}$, it is straightforward to derive that 
\begin{align}\label{limity}
\lim_{l_i\to \infty}y_i&=1, \nonumber\\
\lim_{l_i\to \infty}y_j&=x_k, \nonumber\\
\lim_{l_i\to \infty}y_k&=x_j.
\end{align}
Then using (\ref{limity}), (\ref{specialvaluerogerl}) and (\ref{ELidenity1}), we can  show that $\lim_{l_i\to \infty} \mathcal{K}_3(P_3)=0$ for $i\in \left\{ 1,2,3\right\}$. Again, by repeating the same analysis, we can see that  $\lim_{l_i\to \infty} \mathcal{K}_I(P_I)=0$ for $i\in \left\{ 1,2,3\right\}$ and $I \in \left\{ 1,2,T\right\}$.

\end{document}